\documentclass[pra,letterpaper,twocolumn,aps,footinbib,superscriptaddress,showpacs,babel]{revtex4-2}

\usepackage[utf8]{inputenc}
\usepackage{amsmath} 
\usepackage{mathtools} 
\usepackage{amsfonts} %
\usepackage{amssymb} %
\usepackage{mleftright}
\usepackage{hhline}
\usepackage{color}
\usepackage{graphics}
\usepackage[colorlinks=true,linkcolor=red,citecolor=green,urlcolor=blue]{hyperref} 
\usepackage{bbold} 
\usepackage{bm} 

\newcommand*{\diff}{\mathrm{d}}
\newcommand*{\transpose}{\mathrm{T}}
\newcommand*{\trace}{\mathrm{Tr}}

\newcommand*{\mrm}[1]{\mathrm{#1}}

\newcommand*{\mcl}[1]{\mathcal{#1}}
\newcommand*{\expo}[1]{\mathrm{e}^{#1}}

\newcommand*{\Dop}[1]{\mcl{D}[#1]} 
\newcommand*{\Dopdag}[1]{\mcl{D}^{\dag}[#1]} 
\newcommand*{\Hop}[1]{\mcl{H}[#1]} 
\newcommand*{\ito}{\textbf{(I)}} 
\newcommand*{\bito}{\textbf{(BI)}} 
\newcommand*{\dspop}{\op{D}} 

\newcommand*{\op}[1]{\hat{#1}}
\newcommand*{\mat}[1]{#1} 
\newcommand*{\vect}[1]{\bm{\mrm{#1}}}
\newcommand*{\oprho}{\rho}
\newcommand*{\opE}{\op{E}}

\newcommand*{\purity}{\mcl{P}}
\newcommand*{\asympt}{\infty} 
\newcommand*{\weyl}{\mcl{W}}
\newcommand*{\cmlnt}{\kappa}

\newcommand*{\freqmech}{\Omega_{\mrm{m}}}
\newcommand*{\freqeff}{\Omega_{\mrm{eff}}}
\newcommand*{\freqcav}{\omega_{\mrm{c}}}
\newcommand*{\freqdrive}{\omega_{0}}
\newcommand*{\freqLO}{\omega_{\mrm{lo}}}
\newcommand{\ket}[1]{| #1 \rangle}
\newcommand{\bra}[1]{\langle #1 |}
\newcommand{\braket}[2]{\langle #1 | #2 \rangle}

\newcommand{\eg}{e.\,g.}

\newcommand{\ie}{i.\,e.}

\begin{document}

\title{Quantum retrodiction in Gaussian systems and applications in optomechanics}

\author{Jonas Lammers}
\author{Klemens Hammerer}
\email{Klemens.Hammerer@itp.uni-hannover.de}
\affiliation{Institute of Theoretical Physics, Leibniz Universität Hannover, Appelstraße 2, 30167 Hannover, Germany}

\date{\today}

\begin{abstract}
What knowledge can be obtained from the record of a continuous measurement about the quantum state the measured system was in at the beginning of the measurement? The task of quantum state retrodiction, the inverse of the more common state prediction, is rigorously and elegantly addressed in quantum measurement theory through retrodictive Positive Operator Valued Measures. This article provides an introduction to this general framework, presents its practical formulation for retrodicting Gaussian quantum states using continuous-time homodyne measurements, and applies it to optomechanical systems. We identify and characterise achievable retrodictive POVMs in common optomechanical operating modes with resonant or off-resonant driving fields and specific choices of local oscillator frequencies in homodyne detection. In particular, we demonstrate the possibility of a near-ideal measurement of the quadrature of the mechanical oscillator, giving direct access to the position or momentum distribution of the oscillator at a given time. This forms the basis for complete quantum state tomography, albeit in a destructive manner.
\end{abstract}

\maketitle

\mathtoolsset{centercolon} 


\section{Introduction}

Continuous measurements \cite{Wiseman2010,Jacobs2014,Barchielli2009} are a powerful tool for the preparation and control of quantum states in open systems and as such are of great importance for studies of fundamental physics and applications in quantum technology. Based on a continuous measurement record, it is possible to track the quantum trajectory of a system in its Hilbert space in real time, as demonstrated in circuit QED systems~\cite{Weber2016,HacohenGourgy2020}, atomic ensembles~\cite{Geremia2003,Kong2020}, and in optomechanics~\cite{Hofer2017} with micromechanical oscillators~\cite{Iwasawa2013,Wieczorek2015,Rossi2018,Thomas2020,Meng2022} and levitated nanoparticles~\cite{Setter2018,Liao2019,Liao2019,Magrini2021}. Determining the conditional quantum state formally requires to solve the stochastic Schrödinger or master equation~\cite{Wiseman2010,Jacobs2014,Barchielli2009}, which is a daunting task in general. For the important case of linear quantum systems, which includes most applications in optomechanics and atomic ensembles, the integration of the Schrödinger equation simplifies greatly and turns out to be equivalent to classical Kalman filtering~\cite{Zhang2022}. For this reason, these well-established and powerful tools of classical estimation and control theory are increasingly finding application in quantum science~\cite{Ma2022} and are becoming a well-accepted technique for preparing quantum states.

\begin{figure}
	\centering
	\def\svgwidth{\columnwidth}
	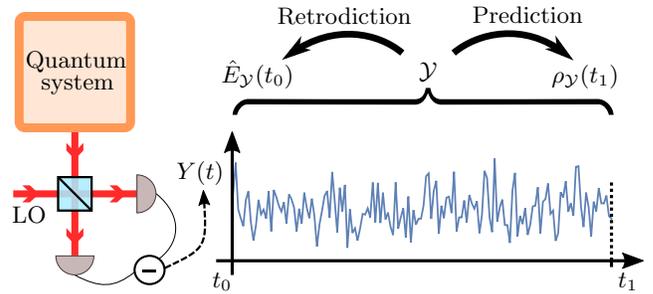
	\caption{Schematic of a continuously monitored quantum system:	The output of field a quantum system is combined with a strong local oscillator (LO) to perform homodyne detection from time $t_{0}$ to $t_{1}$, producing some measurement record $\mcl{Y}$. Starting from a known initial state $\oprho(t_{0})$ this record can be used by integrating a stochastic master, Eq.~\eqref{eq:ItoSimpleMasterEquation}, to predict the system state $\oprho_{\mcl{Y}}(t_{1})$ conditioned on the record $\mcl{Y}$. Alternatively, the measurement record can be used to retrodict an effect operator $\opE_{\mcl{Y}}(t_{0})$, cf. Eq.~\eqref{eq:SimpleEffectEquation}, that characterizes an effective POVM measurement on the initial state $\oprho(t_{0})$.
	\label{fig:PreVsRetrodiction}}
\end{figure}

Like any measurement in quantum mechanics, continuous measurements not only determine the state of the system post measurement, but also provide information about its initial state prior. The dual use of continuous measurements for predictive preparation and retrospective analysis of quantum states, cf. Fig.~\ref{fig:PreVsRetrodiction}, as well as their combination in what is referred to as quantum state smoothing, has received considerable attention in the theoretical literature, for a review see~\cite{Chantasri2021}. Retrospective state analysis and smoothing have been investigated experimentally in cavity- and circuit QED~\cite{Rybarczyk2015,Tan2015,Tan2016,Foroozani2016,Tan2017}, atomic ensembles~\cite{Bao2020,Bao2020a}, and optomechanics~\cite{Rossi2018,Thomas2020,Kohler2020}. However, compared to quantum state preparation by filtering, the applications of these concepts for state readout seem to be less known, although they represent powerful tools for quantum state verification and tomography.

Here we aim to give a self-contained and accessible introduction to the theory of quantum state retrodiction based on continuous measurements and its formulation for linear quantum systems. The main equations of this theory have been derived before in the context of quantum state smoothing in~\cite{Huang2018,Zhang2017,Warszawski2020}. We focus our presentation here on the aspect of state retrodiction and aim to provide operational recipes for this. The general formalism is applied to optomechanical systems, for which we identify and characterize the retrodictive measurements achievable there in terms of their Positive Operator Valued Measures (POVMs). In particular, we consider the common regimes of driving the optomechanical cavity on resonance or on its red or blue mechanical sidebands and discuss the role of the local oscillator frequency in homodyne detection. In each case we determine the realized POVM and compare to what is achieved in state filtering in the same configuration. As a main finding, we show that red-detuned driving in the resolved-sideband limit allows for an almost perfect quadrature measurement, which is back-action free but completely destructive. Our treatment accounts for imperfections due to thermal noise and detection inefficiencies, and studies requirements on the quantum cooperativity for performing efficient state readout. In particular, we determine the concrete filter functions that are necessary for the post-processing of the photocurrent in order to realize certain POVMs.

The article is organized as follows: In Sec.~\ref{sec:StatePreparation} we recapitulate the description of conditional state preparation through continuous measurement based on stochastic master equations and the equivalent Kalman filter, emphasizing the operational interpretation of the central formulas. In close analogy we introduce in Sec.~\ref{sec:StateVerification} the formalism of retrodictive POVMs and its application to linear quantum systems, where the POVM consists of Gaussian effect operators conveniently characterized by their first and second moments. In Sec.~\ref{sec:DecayingCavity} we illustrate the application of this formalism to the simple case of a decaying cavity. Finally, in Sec.~\ref{sec:RealisticOM} we provide a rather detailed modelling of an optomechanical system and derive the retrodictive POVMs in various parameter regimes.

\section{Conditional State Preparation Through Continuous Measurements}\label{sec:StatePreparation}

\subsection{Conditional Master Equation}\label{sec:ForwardMasterEquation}
To set the scene and introduce some notation, we start with an overview of the concept of conditional (stochastic) master equations, referring to~\cite{Wiseman2010,Jacobs2014} for detailed derivations. These describe the evolution of continuously monitored quantum systems, and are used to prepare \textit{conditional} (or \textit{filtered}) quantum states.

We consider an open quantum system governed by Hamiltonian $\op{H}$ and coupled to a Markovian bath via jump operator $\op{L}$. This gives rise to a quantum master equation \cite{Wiseman2010,Jacobs2014} for the system's density operator $\oprho(t)$,
\begin{align}\label{eq:UnobservedMasterEquation}
	\begin{split}
		\diff \oprho (t) & = -i[\op{H},\oprho(t)]\diff t + \Dop{\op{L}}\oprho(t)\diff t,
	\end{split}
\end{align}
with the usual Lindblad superoperator $\Dop{\op{L}}\oprho = \op{L}\oprho\op{L}^\dag - (\op{L}^\dag\op{L}\oprho + \oprho \op{L}^\dag \op{L})/2$. The generalization to multiple jump operators is straightforward. We will designate all operators (except density operators) by caret superscripts. The increment $\diff \oprho(t) := \oprho(t+\diff t) - \oprho(t)$ propagates the state by an infinitesimal amount forward in time. Integrating this equation of motion yields a trace-preserving completely-positive map $\mcl{N}_{t_{0},t}$ which takes an initial state $\oprho(t_{0})$ to a corresponding state at a later time, $\oprho(t) = \mcl{N}_{t_{0},t}[\oprho(t_{0})]$ \cite{Nielsen2010}.
with the usual Lindblad superoperator $\Dop{\op{L}}\oprho = \op{L}\oprho\op{L}^\dag - (\op{L}^\dag\op{L}\oprho + \oprho \op{L}^\dag \op{L})/2$. The generalization to multiple jump operators is straightforward, see Eq.~\eqref{eq:ItoSimpleMasterEquation}. We will designate all operators (except density operators) by caret superscripts. The increment $\diff \oprho(t) := \oprho(t+\diff t) - \oprho(t)$ propagates the state by an infinitesimal amount forward in time. Integrating this equation of motion yields a trace-preserving completely-positive map $\mcl{N}_{t_{0},t}$ which takes an initial state $\oprho(t_{0})$ to a corresponding state at a later time, $\oprho(t) = \mcl{N}_{t_{0},t}[\oprho(t_{0})]$ \cite{Nielsen2010}.

Further information about the state can be gained by monitoring the bath to which the system is coupled~\cite{Wiseman2010,Jacobs2014,Barchielli2009}. In that case conditioning the state on the knowledge gained from these indirect measurements is known as \textit{filtering}~\cite{Bouten2007}. We only consider the case of homodyne (and later heterodyne) measurements, as we are ultimately interested in \emph{linear} dynamics. Other measurement schemes, such as photon counting, would take the conditional dynamics out of this regime. A continuous homodyne detection of the outgoing mode, as sketched in Fig.~\ref{fig:PreVsRetrodiction}, yields a stochastic photocurrent $I(t)$. This can be normalized, $Y(t) := I(t)/\alpha$, with some $\alpha\in\mathbb{R}$ so that for vacuum input its increment $\delta  Y(t) = Y(t + \delta  t) - Y(t)$ has the variance of white noise, $\overline{\delta  Y(t)^2} = \overline{\delta  I(t)^2}/\alpha^2 \equiv \delta  t$ where the bar denotes an ensemble average~\footnote{Hence $\alpha = (\overline{\delta  I(t)^2}/\delta  t)^{1/2}$ with $\delta  t$ chosen as small as possible without violating the assumption that the noise is indeed white, \ie, has independent increments from one moment to the next.}. The measured signal can be decomposed into a deterministic and stochastic part as
\begin{align}\label{eq:MeasurementWienerRelation}
	\ito\ \diff Y(t) & = \langle \op{C} + \op{C}^\dag \rangle_{\rho(t)}\diff t + \diff W(t).
\end{align}
Here, $\op{C}=\sqrt{\eta}\expo{i\theta}\op{L}$ denotes the measurement operator, which includes imperfect detection efficiency $\eta\in[0,1]$ and the local oscillator phase $\theta$. Angled brackets denote an expectation value, $\langle \op{C} \rangle_{\rho} := \trace\{ \op{C} \oprho\}$, and $\diff W$ is a stochastic Wiener increment satisfying the It\^{o} relation $(\diff W)^2 = \diff t$. Equation~\eqref{eq:MeasurementWienerRelation} is a stochastic It\^{o} equation \cite{Mikosch1998,Gardiner2009} denoted by the $\ito$ in front. Depending on the measurement results, the system satisfies the conditional master equation
\begin{align}\label{eq:ItoSimpleMasterEquation}
	\begin{split}
		\ito\  \diff \oprho (t) & = -i[\op{H},\oprho(t)]\diff t + \Dop{\op{L}}\oprho(t)\diff t \\
	 & \qquad + \Hop{\op{C}}\oprho(t) \diff W(t),
	\end{split}
\end{align}
with superoperator $\Hop{\op{C}}\oprho := (\op{C}-\langle \op{C} \rangle_{\rho})\oprho + \oprho(\op{C}^\dag - \langle \op{C}^\dag \rangle_{\rho})$. Assume the system has evolved from $t_{0}$ to $t_{1}$ and produced some measurement record $\mcl{Y} = \{ Y(s) , t_{0} \leq s < t_{1} \}$, as depicted in Fig.~\ref{fig:PreVsRetrodiction}. By integrating the master equation from $t_{0}$ to $t_{1}$ we obtain a conditional (or \textit{filtered}) state $\oprho_{\mcl{Y}}(t_{1}) = \mcl{N}_{t_{0},t_{1}|\mcl{Y}}[\oprho(t_{0})]$ dependent on the initial state $\oprho(t_{0})$ and conditioned on the record $\mcl{Y}$.

The conditional master equation Eq.~\eqref{eq:ItoSimpleMasterEquation} can be generalized to $N_{L}$ Markovian baths and $N_{C}$ monitored channels,
\begin{align}\label{eq:ItoMasterEquation}
	\begin{split}
		\ito\  \diff \oprho (t) & = -i[\op{H},\oprho(t)]\diff t + \sum_{j=1}^{N_L}\Dop{\op{L}_{j}}\oprho(t)\diff t \\
		 & \qquad + \sum_{k=1}^{N_C} \Hop{\op{C}_{k}}\oprho(t) \diff W_{k}(t).
	\end{split}
\end{align}
The measurement operators $\op{C}_{k}$ do not necessarily correspond one-to-one to the jump operators $\op{L}_{j}$ as before, and we will see an example in Sec.~\ref{sec:RealisticOM} where effectively $N_{C} > N_{L}$. However, since any information recorded by the observer must have previously leaked from the system it holds that $\sum\nolimits_{j}\op{L}_{j}^\dagger\op{L}_{j} - \sum\nolimits_{k}\op{C}_{k}^\dag \op{C}_{k}\geq 0$. The $\diff W_{j}$ are mutually independent Wiener increments satisfying the It\^{o} relation
\begin{align}
	\diff W_{j}(t)\diff W_{k}(t) = \delta_{jk}\diff t,
\end{align}
and each $\diff W_{j}$ is related to a corresponding homodyne measurement increment $\diff Y_{j}$ as
\begin{align}
	\ito\ \diff Y_{j}(t) & = \langle \op{C}_{j} + \op{C}_{j}^\dag \rangle_{\rho(t)}\diff t + \diff W_{j}(t).
\end{align}
For details and derivations of this general formalism for describing quantum dynamics conditioned on continuous homodyne detection, we refer once more to~\cite{Wiseman2010,Jacobs2014}.

\subsection{Linear Dynamics}\label{sec:LinearDynamics}

\subsubsection{Linear systems}
We now apply these concepts to linear systems with Gaussian states governed by the general master equation Eq.~\eqref{eq:ItoMasterEquation}. We consider a bosonic quantum system with $M$ modes and $2M$ associated canonical operators $\op{r}_j$ which we collect into a vector $\vect{\op{r}}=(\op{r}_j)_{j=1,\dots,2M}$. The $\op{r}_{j}$ satisfy canonical commutation relations
\begin{align}\label{eq:SigmaDefinition}
	i\sigma_{jk} & := [\op{r}_{j},\op{r}_{k}],
\end{align}
giving rise to a skew-symmetric matrix $\mat{\sigma}\in\mathbb{R}^{2M\times 2M}$. For example, the usual choice for an oscillator with $M$ modes would be $\vect{\op{r}} = [ \vect{\op{x}}^\transpose, \vect{\op{p}}^\transpose ]^\transpose = [ \op{x}_{1}, \dots, \op{x}_{M},\op{p}_{1},\dots,\op{p}_{M} ]^\transpose,$ which entails
\begin{align}
	\mat{\sigma} & = \begin{bmatrix} \mathbb{0}_{M} & \mathbb{1}_{M} \\ -\mathbb{1}_{M} & \mathbb{0}_{M} \end{bmatrix}.
\end{align}
In a linear system the Hamiltonian is at most quadratic in the canonical operators while the jump and measurement operators are at most linear. $\op{H}$ can be expressed as
\begin{align}
	\op{H} = \frac{1}{2} \vect{\op{r}}^\transpose \mat{H} \vect{\op{r}},
\end{align}
with a symmetric matrix $\mat{H}\in\mathbb{R}^{2M\times 2M}$. Without loss of generality, we assume that $\op{H}$ does not contain terms linear in $\vect{\op{r}}$~\footnote{Such terms can always be set to zero by a suitable shift $\vect{\op{r}}\rightarrow \vect{\op{r}}+\vect{\xi}$ where $\vect{\xi}\in\mathbb{R}^{2M}$.}. We write the $N_{C}$ linear measurement operators as
\begin{align}
	\vect{\op{C}} = (\mat{A} + i\mat{B} )\vect{\op{r}},
\end{align}
with $\mat{A},\mat{B} \in\mathbb{R}^{N_{C} \times 2M}$, and $N_{L}$ jump operators as
\begin{align}
	\vect{\op{L}} & = \mat{\Lambda}\vect{\op{r}},\\
	\mat{\Lambda}^\dag \mat{\Lambda} & =: \mat{\Delta} + i\mat{\Omega},
\end{align}
with complex $\mat{\Lambda}\in\mathbb{C}^{N_{L}\times 2M}$ and $\mat{\Delta},\mat{\Omega}\in\mathbb{R}^{2M\times 2M}$ symmetric and skew-symmetric respectively.

\subsubsection{Gaussian states}\label{sec:GaussianStates}
A Gaussian state $\oprho$ \cite{Olivares2012,Wang2008,Adesso2014,Genoni2016,Weedbrook2012} is, by definition, any state with a Gaussian phase-space distribution. Gaussian states are \textit{fully determined} by their first- and second-order cumulants \cite{Ivan2012}, namely a vector of means
\begin{align}
	\vect{r}_{\rho} := \langle \vect{\op{r}} \rangle_{\rho} : = \trace\{\vect{\op{r}}\oprho\} \in\mathbb{R}^{2M}
\end{align}
and a symmetric covariance matrix
\begin{align}\label{eq:CovarianceMatrixDefinition}
	\mat{V}_{jk}^{\rho} := \langle \{ \op{r}_{j} - r_{j}^{\rho}, \op{r}_{k} - r_{k}^{\rho} \} \rangle_{\rho}\in\mathbb{R}^{2M\times 2M}. 
\end{align}
All higher-order cumulants are identically zero, so knowing $\vect{r}_{\rho}$ and $\mat{V}_{\rho}$ determines the full Wigner function of $\oprho$ and thus also $\oprho$ itself. Note that the normalization of $\mat{V}_{\rho}$ chosen in Eq.~\eqref{eq:CovarianceMatrixDefinition} means that diagonal elements correspond to twice the variance, \eg, $V_{jj}^{\rho} = 2(\langle \op{r}_{j}^{2}\rangle - \langle \op{r}_{j} \rangle^{2})$. 

The assumption of a Gaussian initial state $\oprho(t_{0})$ is both convenient and reasonable. Since Gaussian operators have the tremendously useful property to remain Gaussian under linear dynamics they are easy to work with. Additionally, considering only Gaussian states is justified since Gaussian measurements \cite{Jacobs2006,VanHandel2009} and Gaussian baths \cite{Zurek1993} tend to ``Gaussify'' the state of the system. Mathematically this means that if we start with an arbitrary initial state $\oprho(t_{0})$, higher-order cumulants of order $\geq 3$ are damped by the dynamics. 
Depending on how slowly this damping happens, if our linear system is initially prepared in a non-Gaussian state these higher orders may need to be taken into account, which we do in App.~\ref{sec:EvolutionOfGeneralQuantumStates}. But for now we focus on the case of Gaussian initial states only.

It is known \cite{Barnett1997,Zhang2017} that a master equation for $\oprho$ can be directly translated into differential equations for the means and covariance matrix, as detailed in App.~\ref{sec:Appendix:CumulantEoMs}. For a Gaussian state one finds
\begin{align} \label{eq:FwdWienerMeanEoM}
	\ito\ \diff\vect{r}_{\rho}(t) & = \mat{Q}\vect{r}_{\rho}(t)\diff t + \bigl(\mat{V}_{\rho}(t)\mat{A}^\transpose - \mat{\sigma}\mat{B}^\transpose\bigr)\diff \vect{W}(t), 
\end{align}
with the drift matrix
\begin{align}
	\mat{Q} & := \mat{\sigma}(\mat{H} + \mat{\Omega}),
\end{align}
comprising unitary and dissipative terms. If we reintroduce the actually measured homodyne signal
\begin{align}
	\ito\ \diff \vect{Y}(t) = 2\mat{A}\vect{r}_{\rho}(t)\diff t + \diff \vect{W}(t),
\end{align}
we can write
\begin{align} \label{eq:FwdMeanEquationOfMotion}
	\ito\ \diff\vect{r}_{\rho}(t)	& = \mat{M}_{\rho}(t)\vect{r}_{\rho}(t)\diff t + \bigl(\mat{V}_{\rho}(t)\mat{A}^\transpose - \mat{\sigma}\mat{B}^\transpose\bigr)\diff \vect{Y}(t), 
\end{align}
with the \textit{conditional drift matrix}
\begin{align} \label{eq:FwdConditionalDriftMatrix}
	\mat{M}_{\rho}(t) & := \mat{Q}+2\mat{\sigma}\mat{B}^\transpose\mat{A} - 2\mat{V}_{\rho}(t)\mat{A}^\transpose\mat{A}. 
\end{align}
In Eq.~\eqref{eq:FwdMeanEquationOfMotion} the measurement current $\diff \vect{Y}(t)$ enters the evolution of the conditional means only through multiplication with the measurement matrices $\mat{A}$ and $\mat{B}$. Hence, reducing the detection efficiency which corresponds to $\mat{A},\,\mat{B}\to 0$ causes the stochastic increment to disappear as it should. Note that the covariance matrix $\mat{V}_{\rho}(t)$ twice enters Eq.~\eqref{eq:FwdMeanEquationOfMotion}, once through the drift matrix $\mat{M}_{\rho}(t)$ and once directly coupled to $\diff\vect{Y}(t)$. The latter term has the effect that a large variance, which corresponds to large uncertainty about the state, boosts the effect each bit of gathered information has on the evolution of the conditional means.

The covariance matrix satisfies the deterministic equation
	\begin{align}\label{eq:FwdCovarianceEquationOfMotion}
		\begin{split}
		\frac{\diff \mat{V}_{\rho}(t)}{\diff t} & = \mat{M}_{\rho}(t)\mat{V}_{\rho}(t) + \mat{V}_{\rho}(t)\mat{M}_{\rho}^\transpose(t) \\
		& \quad + \mat{D} + 2\mat{V}_{\rho}(t)\mat{A}^\transpose\mat{A}\mat{V}_{\rho}(t). 
		\end{split}
	\end{align}
with \textit{diffusion matrix}
\begin{align}
	\mat{D} & := 2\mat{\sigma}\bigl(\mat{\Delta} - \mat{B}^\transpose\mat{B}\bigr)\mat{\sigma}^\transpose. 
\end{align}
The evolution of $\mat{V}_{\rho}(t)$ is independent of the means $\vect{r}_{\rho}(t)$ or any other cumulants, which is a peculiarity of Gaussian dynamics. However, while it is independent of the measurement record and not a stochastic equation, it does depend on the measurement device through matrices $\mat{A},\,\mat{B}$. This is reasonable since the information gained from observations of the system conditions the state, reducing its uncertainty.

In the following we assume stable dynamics, which 
makes the covariance matrix collapse to some steady state matrix $\mat{V}_{\rho}(t) \to \mat{V}_{\rho}^{\asympt}$ asymptotically for $t\to\infty$ from any initial $\mat{V}_{\rho}(t_{0})$. We find $\mat{V}_{\rho}^{\asympt}$ by solving the Riccati equation $\dot{\mat{V}}_{\rho} = 0$ which implies
\begin{align} \label{eq:StableFwdDynamicsRiccati}
	\mat{M}_{\rho}^{\asympt}\mat{V}_{\rho}^{\asympt} + \mat{V}_{\rho}^{\asympt}(\mat{M}_{\rho}^{\asympt})^\transpose & = - \mat{D} - 2\mat{V}_{\rho}^{\asympt}\mat{A}^\transpose\mat{A}\mat{V}_{\rho}^{\asympt}, 
\end{align}
where $\mat{M}_{\rho}^{\asympt}$ is just $\mat{M}_{\rho}(t)$ with $\mat{V}_{\rho}(t) \mapsto \mat{V}_{\rho}^{\asympt}$. The right-hand side is negative definite and the covariance matrix is positive definite for proper quantum states, so $\mat{M}_{\rho}^{\asympt}$ only has eigenvalues with negative real part. Now if the experiment has been running sufficiently long we can simply plug $\mat{V}_{\rho}^{\asympt}$ and $\mat{M}_{\rho}^{\asympt}$ into Eq.~\eqref{eq:FwdMeanEquationOfMotion} to find the evolution of the means,
\begin{multline} \label{eq:FwdIntegratedMeans}
	\ito\ \vect{r}_{\rho}(t) = \expo{(t-t_{0})\mat{M}_{\rho}^{\asympt}}\vect{r}_{\rho}(t_{0}) \\
	+ \int_{t_{0}}^{t}\expo{(t-\tau)\mat{M}_{\rho}^{\asympt}} \bigl(\mat{V}_{\rho}^{\asympt}\mat{A}^\transpose - \mat{\sigma}\mat{B}^\transpose \bigr)\diff \vect{Y}(\tau). 
\end{multline}
Because $\mat{M}_{\rho}^{\asympt}$ is stable (all eigenvalues have non-positive real parts) we see that the initial condition $\vect{r}_{\rho}(t_{0})$ is damped exponentially, as is the integrand in the second line.

Here we see that the means (and thus the whole state) do not depend on the entire continuous measurement record $\mcl{Y}$ as such, but only on the It\^{o}-integral in the second line of Eq.~\eqref{eq:FwdIntegratedMeans}, which is a simple vector of $2M$ real numbers for a system composed of $M$ subsystems. Thus, the integral kernel $\expo{\mat{M}_{\rho}^{\asympt}(t-\tau)} \bigl(\mat{V}_{\rho}^{\asympt}\mat{A}^\transpose - \mat{\sigma}\mat{B}^\transpose \bigr)$ 
actually picks out a set of $2M$ temporal modes of the monitored fields. Each of these (not necessarily orthogonal) modes of the light fields provides an estimate for one of the $2M$ phase space variables of the system. We will elaborate on this aspect further in Sec.~\ref{sec:ModeFunctions}. For the particular case of a freely decaying monitored cavity this fact was pointed out already by Wiseman \cite{Wiseman1996}, and in Sec.~\ref{sec:DecayingCavity} we will treat this cavity as an illustrative example of the formalism developed here, reproducing the results of \cite{Wiseman1996}.

\subsection{Interpretation and Discussion}

\subsubsection{Conditional quantum states}
We now want to remind the reader of how the conditional Gaussian quantum state has to be interpreted, and what its preparation via continuous measurements means from an operational perspective.

The means $\vect{r}_\rho(t)$ and covariance matrix $V_\rho(t)$ determined from Eq.~\eqref{eq:FwdMeanEquationOfMotion} and Eq.~\eqref{eq:FwdCovarianceEquationOfMotion} fully determine the density matrix for the conditional state. It is instructive to note that the Gaussian density matrix is always of the form \cite{Giedke2001,Fiurasek2007}
\begin{align}\label{eq:DOGaussianState}
  \rho(t)&\propto \dspop(\vect{r}_{\rho}(t))\exp\Bigl[-\vect{\op{r}}^\transpose\Gamma_\rho(t)\vect{\op{r}}\Bigr]\dspop^\dagger(\vect{r}_\rho(t)).
\end{align}
Here $\dspop(\vect{q})=\exp(-i \vect{q}\mat{\sigma}\vect{\op{r}})$ is a displacement operator in phase space, and the matrix $\Gamma_\rho(t)$ is a simple functional of the covariance matrix \footnote{Let $V=S^T(T\oplus T)S$ be the symplectic diagonalization of the covariance matrix $V$, that is $S$ is a symplectic matrix with $S^T\sigma S=\sigma$ and $T$ is a diagonal matrix with entries $\tau_k$. Then $\Gamma=S^{-1}(K\oplus K)(S^{-1})^T$ where $K$ is a diagonal matrix with entries $\kappa_k$ such that $\tau_k=(1-e^{-\kappa_k})^{-1}$.}. The shape of the Gaussian wave packet in phase space is determined by the middle term on the right hand side, which evolves deterministically and is independent of the measurement results. The wave packet's position in phase space is set by the displacement operators, and does depend on the photocurrent via Eq.~\eqref{eq:FwdIntegratedMeans}.

Prediction of a conditional quantum state based on a continuous measurement during time interval $[t_0,t]$ starting from a known Gaussian initial state therefore simply means to calculate the means according to Eq.~\eqref{eq:FwdIntegratedMeans}. Knowing those numbers the prediction is that a hypothetical projective measurement of canonical operators at time $t$ will give results with these same averages, and second moments according to the covariance matrix $V_\rho(t)$ which depends only on the initial condition. Statistics of any other measurement can be determined from Eq.~\eqref{eq:DOGaussianState}. For stable dynamics dependencies on initial conditions will disappear in the long run, and the covariance matrix will become time-independent. The wave packet will then have a fixed shape and undergo stochastic motion in phase space with positions known from the photocurrent.

The quality of the conditional preparation can be judged from the purity $\purity (\oprho) = \trace\{\oprho^2\} \leq 1$ of the conditional state. It tells us how close it is to a pure state and thus quantifies the amount of classical uncertainty in $\oprho$. For a Gaussian state with $M$ modes it is given by \cite{Paris2003}
\begin{align}\label{eq:Purity}
	\purity (\oprho) = 1/\sqrt{\det(\mat{V_{\rho}})}. 
\end{align}
Unobserved dissipation tends to reduce the purity while monitoring the dynamics and conditioning the state increases the purity. Ideally, perfect detection allows to prepare pure states, which are the only states  with $\purity (\oprho) = 1$. The bound $\purity \leq 1$ implies $\det(\mat{V_{\rho}}) \geq 1$ 
which is also imposed by Heisenberg's uncertainty relation. We briefly recall prototypical pure Gaussian states of a single mode. Coherent states $|\alpha\rangle$ have equal variances $V_{xx}=V_{pp}=1$ 
and vanishing covariance. The vacuum $|0\rangle$ is a special coherent state with vanishing means. Squeezed states \cite{Barnett1997} have the variance in one quadrature reduced below \textit{shot noise}, that is below $1$ (the variance of vacuum). The conjugate quadrature is then necessarily anti-squeezed to satisfy Heisenberg's uncertainty relation. An important class of non-pure Gaussian states are thermal states. These have vanishing covariance and equal variance $V_{xx}=V_{pp}=2\bar{n}+1$ 
, where $\bar{n}\geq 0$ is the mean number of excitations. Importantly, $\purity = 1/(2\bar{n}+1)$ 
decreases as $\bar{n}$ grows.

\subsubsection{Mode functions}\label{sec:ModeFunctions}
We mentioned at the end of Sec.~\ref{sec:GaussianStates} that the kernels in the forward and backward integrals of the means in Eqs.~\eqref{eq:FwdIntegratedMeans} and Eq.~\eqref{eq:BwdIntegratedMeans} each pick out sets of temporal modes. Recall that the means $\vect{r}_{\rho}(t)=(r_j(t))_{j=1,\dots,2M}$ in Eq.~\eqref{eq:FwdIntegratedMeans} depend on the measurement currents $\vect{Y}(\tau)=(Y_k(\tau))_{k=1,\dots,N_{C}}$ only through integration 
with respect to the functions
\begin{align}
	f_{jk}^{\rho}(t,\tau) := \bigl[ \expo{(t-\tau)\mat{M}_{\rho}^{\asympt}} \bigl(\mat{V}_{\rho}^{\asympt}\mat{A}^\transpose - \mat{\sigma}\mat{B}^\transpose \bigr) \bigr]_{jk}. 
\end{align}
Each (unnormalized) temporal mode function $f_{jk}^{\rho}(t,\tau)$ is integrated with a corresponding signal $Y_{k}(\tau)$,
\begin{align}
	X_{j}(t) & := \int_{t_{0}}^{t} f_{jk}^{\rho}(t,\tau) \diff Y_{k}(\tau),
\end{align}
to  enter the evolution of $r_{j}(t)$. A measurement current $Y_{k}(\tau)$ results from a quadrature measurement of some outgoing light field, $Y_{k}(\tau) \propto \langle \op{x}^{out}_{k}(\tau) \rangle$. Thus integration with $f_{jk}^{\rho}$ effectively corresponds to the measurement of a quadrature operator of a certain temporal mode $\op{X}_{j}(t) \propto \int^{t} f_{jk}^{\rho}(t,\tau) \op{x}_{k}(\tau) \diff \tau$ with result $X(t)$. Of course the mode functions $f_{jk}^{\rho}$, and thus the new optical modes, will generally not be orthogonal.

\section{State Verification Using Retrodictive POVMs}\label{sec:StateVerification}

\subsection{Retrodictive POVMs}
In the previous section we have seen how to use continuous monitoring for the preparation of conditional states (\textit{filtering}). We are now going to show how to interpret the measurement record instead as an instantaneous \textit{Positive-Operator Valued Measure (POVM)} \cite{Nielsen2010,Jacobs2014,Wiseman2010} measurement. To fully appreciate this result let us first remind the reader of a few facts about POVMs and general measurements in quantum mechanics.

\subsubsection{Positive-operator valued measures}
A general measurement of a given quantum state $\oprho$ is always composed of (i) possible measurement outcomes $x\in\mcl{X}$, (ii) probabilities for those outcomes to occur $P(x|\oprho)$, and (iii) the effect that obtaining some outcome $x$ has on the system, \ie, the post-measurement state $\oprho_{x}\propto \op{M}_{x}\oprho\op{M}_{x}^\dag$ where $\op{M}_{x}$ incorporates the measurement back action on the state. The probability for a particular $x$ to be measured is given by
\begin{align} \label{eq:POVMProbability}
	P(x|\oprho) & = \trace\{ \op{M}_{x} \oprho \op{M}_{x}^\dag \} = \trace\{ \op{M}_{x}^\dag \op{M}_{x}\oprho \} = \trace\{ \opE_{x} \oprho \}
\end{align}
with the positive \textit{effect operator} $\opE_{x} := \op{M}_{x}^\dag \op{M}_{x}$. Because $\sum_{x}P(x|\oprho) = 1$ must hold for any $\oprho$, the operators $\opE_{x}$ must resolve the identity $\sum_{x}\opE_{x}=\op{1}$. Without reference to the $\op{M}_{x}$ any collection of positive self-adjoint operators $\{\opE_{x},x\in\mcl{X}\}$ which resolve the identity is called a \textit{Positive-Operator Valued Measure (POVM)}.

\subsubsection{Continuous monitoring as POVM measurement}\label{sec:ContinuousMonitoringAsPOVMMeas}
To see how to reinterpret the measurement record let us again consider the simple system governed by the master equation \eqref{eq:ItoSimpleMasterEquation}, and an evolution from $t_0$ to $t_1$ that produced some record $\mcl{Y} = \{ Y(s) , t_{0} \leq s < t_{1} \}$. Note that Eq.~\eqref{eq:ItoSimpleMasterEquation} is nonlinear in $\oprho$ in order to yield a trace-preserving map $\mcl{N}_{t_{0},t_{1}|\mcl{Y}}$. If instead we consider the linear equation \cite{Wiseman1996}
\begin{align}\label{eq:ItoLinearMasterEquation}
	\begin{split}
		\ito\ \diff \tilde{\oprho} (t) & = -i[\op{H},\tilde{\oprho}(t)]\diff t + \Dop{\op{L}}\tilde{\oprho}(t)\diff t \\
		 & \qquad + \bigl(\op{C}\tilde{\oprho}(t) + \tilde{\oprho}(t)\op{C}^\dagger\bigr) \diff Y(t),
	\end{split}
\end{align}
we find it generates equivalent but non--trace-preserving dynamics,
\begin{align}\label{eq:UnnormalizedState}
	\tilde{\oprho}_{\mcl{Y}}(t_{1}) & = \tilde{\mcl{N}}_{t_{0},t_{1}|\mcl{Y}}[\oprho(t_{0})],
\end{align}
denoted by a tilde. The trace of the conditional state now carries additional information, namely the probability for $\mcl{Y}$ to have occurred given an initial $\oprho(t_{0})$,
\begin{align}
	P(\mcl{Y}|\oprho(t_{0})) & = \trace\{ \tilde{\oprho}_{\mcl{Y}}(t_{1}) \}.
\end{align}
If we plug Eq.~\eqref{eq:UnnormalizedState} into this expression and include an identity operator $\op{1}$, we can write
\begin{align}
		P(\mcl{Y}|\oprho(t_{0})) & = \trace\{ \op{1} \tilde{\mcl{N}}_{t_{0},t_{1}|\mcl{Y}}[\oprho(t_{0})] \}\\
		& = \trace\{ \tilde{\mcl{N}}_{t_{0},t_{1}|\mcl{Y}}^\dag[\op{1}] \oprho(t_{0}) \}, \label{eq:AdjointChannelOnIdentity}
\end{align}
where $\tilde{\mcl{N}}_{t_{0},t_{1}|\mcl{Y}}^\dag$ is the Hilbert-Schmidt adjoint channel of $\tilde{\mcl{N}}_{t_{0},t_{1}|\mcl{Y}}$ that acts on $\op{1}$. We now  define
\begin{align}\label{eq:DefinitionEffectOp}
	\opE_{\mcl{Y}}(t_{0}) := \tilde{\mcl{N}}_{t_{0},t_{1}|\mcl{Y}}^\dag[\op{1}],
\end{align}
 which will play a crucial role throughout this article. With this definition Eq.~\eqref{eq:AdjointChannelOnIdentity} can be rewritten as
\begin{align}\label{eq:ContMeasAsPOVMMeas}
	P(\mcl{Y}|\oprho(t_{0})) & = \trace\{ \opE_{\mcl{Y}}(t_{0}) \oprho(t_{0}) \}. 
\end{align}
Comparing Eq.~\eqref{eq:ContMeasAsPOVMMeas} to Eq.~\eqref{eq:POVMProbability} shows that $\{\opE_{\mcl{Y}}(t_{0}),\mcl{Y}\in\mathfrak{Y} \}$ indeed constitutes a POVM on the initial state $\oprho(t_{0})$. Here, the ``outcomes'' $x\equiv \mcl{Y}\in\mathfrak{Y}$ comprise all possible records one could observe, and $\sum_{Y\in\mcl{Y}} P(\mcl{Y}|\oprho(t_{0})) = 1$ because the sum corresponds to averaging over (\ie, ignoring) the observations which yields the unconditional trace-preserving evolution \eqref{eq:UnobservedMasterEquation}. As in filtering, we will show later that the effect operators actually depend only on certain weighted integrals of the measurement record $\mcl{Y}$, and not on the whole continuous record as such.

\subsection{Backward effect equation}\label{sec:BackwardEffectEquation}
Just as the conditional quantum state, the effect operators $\opE_{\mcl{Y}}(t)$ themselves can be considered as dynamical quantities obeying a certain (stochastic) equation of motion. In open but unobserved systems Barnett, Pegg, and Jeffers \cite{Barnett2000,Barnett2001,Pegg2002} derived a deterministic differential equation describing the propagation \textit{backwards in time} of effect operators to yield effective POVMs at past times. Tsang \cite{Tsang2009a,Tsang2010,Tsang2010a} and M\o lmer et al.~\cite{Gammelmark2013,Zhang2017} incorporated continuous observations into Bayesian updates of past measurement results, which (arguably, see \cite{Guevara2015}) extends classical \textit{smoothing} to the quantum domain, and results in a stochastic differential equation for $\opE_{\mcl{Y}}(t)$.

For a given system dynamics the effect operators are backpropagated by a channel adjoint to that of the state. More specifically, for continuously monitored systems governed by conditional master equation \eqref{eq:ItoLinearMasterEquation}, the adjoint \textit{conditional effect equation}, which takes some effect operator $\opE(t)$ from the future to the past reads \cite{Tsang2009a,Tsang2010,Gammelmark2013}
\begin{align}\label{eq:SimpleEffectEquation}
	\begin{split}
		\bito\  {-\diff}\opE(t) & := \op{E}(t-\diff t) - \op{E}(t) \\
		& = i[\op{H},\opE(t)]\diff t + \Dopdag{\op{L}}\opE(t)\diff t\\
		&\qquad + \bigl(\op{C}^\dag\opE(t) + \opE(t)\op{C}\bigr)\diff Y(t),
	\end{split}
\end{align}
with adjoint Lindblad superoperator $\mcl{D}^\dag[\op{L}]\opE := \op{L}^\dag \opE \op{L} - \frac{1}{2}\op{L}^\dag \op{L}\opE -\frac{1}{2} \opE\op{L}^\dagger\op{L}$. The $\bito$ indicates that the equation has to be treated as a backward It\^{o} equation. In App.~\ref{sec:Appendix:DerivationEffectEquation} we give a detailed derivation of this equation, and comment further on its interpretation as a differential equation for propagation backwards in time. Note that we defined the increment $\diff\opE$ with an explicit minus sign. This differs from the convention of M\o lmer et al.~\cite{Gammelmark2013,Zhang2017}, but follows the convention of Tsang \cite{Tsang2009a,Tsang2010}. 

Comparing the effect equation \eqref{eq:SimpleEffectEquation} to the forward master equation \eqref{eq:ItoLinearMasterEquation} we observe the following differences. The sign of the Hamiltonian changes, which we expect from the usual time-reversal in closed systems. The Lindblad superoperator $\mcl{D}$ is replaced by its adjoint $\mcl{D}^\dag$ which is no longer trace-preserving but vanishes when applied to the identity. The measurement operator $\op{C}$ is replaced by its adjoint.

Solving Eq.~\eqref{eq:SimpleEffectEquation} for $\opE(t)$ for $t\leq t_1$ requires a certain final condition $\opE(t_1)$. We have motivated the definition \eqref{eq:DefinitionEffectOp} of the effect operator by means of the final condition $\opE(t_1)=\op{1}$. This can be interpreted as describing a situation where at time $t_1$ a certain $\{\opE_{x},x\in\mcl{X}\}$ is performed on the system but the outcome $x$ is not registered. If the outcome $x$ is registered, and we want to describe a dynamics post-selected on it, one needs to replace the identity in Eq.~\eqref{eq:AdjointChannelOnIdentity} by $\opE_{x}(t_{1})$ to obtain an effective $\opE_{x,\mcl{Y}}(t_{0})$. This general observation-assisted backpropagation is what we refer to as \textit{retrodiction}. It is remarkable that a non-trivial POVM can also be retrodicted starting from the trivial effect operator $\opE(t_{1})=\op{1}$ and using nothing but knowledge of the system's dynamics and continuous observations. In fact, in many relevant cases the final condition on $\opE$ will be damped out in the long run, just as initial conditions for the forward propagated density matrix become irrelevant for stable dynamics. This point will be addressed more rigorously further below. 

The unnormalized effect equation generalizing Eq.~\eqref{eq:SimpleEffectEquation} to multiple observed and unobserved channels reads
\begin{align} \label{eq:BackwardItoEffectEquation}
\begin{split}
	\bito\  {-\diff} \opE(t) & = i[\op{H},\opE(t)]\diff t + \sum_{j=1}^{N_{L}}\mcl{D}^\dag[\op{L}_{j}]\opE(t)\diff t \\
	& \quad + \sum_{k=1}^{N_{C}}\bigl(\op{C}_{k}^\dag\opE(t) + \opE(t)\op{C}_{k}\bigr) \diff Y_{k} (t).
\end{split}
\end{align}
Since we only consider conditional dynamics from now on, we will drop the subscript $\mcl{Y}$ and remember that both $\oprho$ and $\opE$ depend on respective parts of the monitoring record.

\subsection{Linear Dynamics and Gaussian POVMs}
As in Sec.~\ref{sec:LinearDynamics} we focus our approach on linear systems. Like density operators we can represent effect operators in terms of phase space distributions, which lets us translate the effect equation into differential equations for the cumulants. One only needs to be more careful with the definition of statistical quantities, as $\opE$ does not have unit trace (or may not be trace class at all). For example the means and covariance matrix are given by
\begin{align}
	r_{j}^{E} & := \langle \op{r}_{j} \rangle_{E} := \frac{\trace\{\op{r}_{j}\opE\}}{\trace\{\opE\}}, \\
	V_{jk}^{E} & = \langle \{ \op{r}_{j} - r_{j}^{E}, \op{r}_{k} - r_{k}^{E} \} \rangle_{E}, 
\end{align}
where the expectation value $\langle \cdot \rangle_{E}$ is explicitly normalized, and defined as long as $\trace\{\opE\}$ exists.

Gaussian effect operators and their time dynamics have been treated recently by Zhang and M{\o}lmer~\cite{Zhang2017}, Huang and Sarovar~\cite{Huang2018}, and Warszawski et al.~\cite{Warszawski2020}. Since we aim to keep our treatment self-contained we reproduce a number of results (in particular on the Gaussian equations of motion of the effect operator) presented there. Our derivation and presentation is meant to complement these previous ones by further details and background. In particular it was not apparent to us if the restriction to Gaussian effect operators is justified as it is for quantum states (cf.\ the discussion in Sec.~\ref{sec:GaussianStates}). In App.~\ref{sec:Appendix:BackwardAddition} we consider the evolution of general effect operators, and show that it is very similar to that of general quantum states. Hence one can apply a notion of backward stability analogous to that of quantum states.

To obtain the evolution of the means and covariance matrix associated with $\opE$ from the corresponding equations for $\vect{r}_{\rho}$ and $\mat{V}_{\rho}$ let us rewrite the effect equation \eqref{eq:BackwardItoEffectEquation} as
\begin{align}
	\begin{split}
		\bito\  {-\diff} \opE(t) & = -i[-\op{H},\opE(t)]\diff t + \sum_{j=1}^{N_{L}} \mcl{D}[\op{L}_{j}]\opE(t)\\
		& \quad + \sum_{j=1}^{N_{L}} \bigl(\op{L}_{j}^\dag \opE(t)\op{L}_{j} - \op{L}_{j} \opE(t)\op{L}_{j}^\dag\bigr)\diff t\\
		& \quad + \sum_{k=1}^{N_{C}}\bigl(\op{C}_{k}^\dag\opE(t) + \opE(t)\op{C}_{k}\bigr) \diff Y_{k} (t),
	\end{split}
\end{align}
where the second line compensates for the replacement of $\mcl{D}^\dag$ by $\mcl{D}$. We see that this equation is structurally very similar to the unnormalized master equation \eqref{eq:ItoMasterEquation}, so Eqs.~\eqref{eq:FwdMeanEquationOfMotion} and \eqref{eq:FwdCovarianceEquationOfMotion} for $\diff \vect{r}_{\rho}$ and $\dot{\mat{V}}_{\rho}$ serve as a good starting point with the following changes: (i) Time-reversal requires us to treat them as backward It\^{o} equations, cf.\@ App.~\ref{sec:Appendix:BackwardItoIntegration}. (ii) The sign flip of $\op{H}$ causes $\mat{H}\mapsto -\mat{H}$ and replacing the measurement operators $\op{C}_{k}$ by their adjoint entails $\mat{B} \mapsto -\mat{B}$. (iii) Working out the change stemming from the sandwich terms in the second line we find in App.~\ref{sec:Appendix:BackwardAddition} that it contributes terms ${-2}\mat{\sigma}\mat{\Omega}\vect{r}_{E}$ and $-\mat{\sigma}\mat{\Omega}\mat{V}_{E} - (\mat{\sigma}\mat{\Omega}\mat{V}_{E})^\transpose$ 
to the evolution of the means and covariance matrix, respectively. Together with $\mat{H}\mapsto -\mat{H}$ this simply changes the sign of the unconditional drift matrix $\mat{Q}\mapsto -\mat{Q}$. Hence the backward It\^{o} equation for the means reads
\begin{align}\label{eq:BwdMeanEquationOfMotion}
\begin{split}
		&\bito\  {-\diff\vect{r}_{E}(t)}  := \vect{r}_{E}(t-\diff t) - \vect{r}_{E}(t) \\
		&\qquad = \mat{M}_{E}(t)\vect{r}_{E}(t)\diff t + \bigl( 2\mat{V}_{E}(t)\mat{A}^\transpose + \mat{\sigma}\mat{B}^\transpose \bigr) \diff\vect{Y}(t),
\end{split}
\end{align}
with the conditional backward drift matrix
\begin{align} \label{eq:BwdConditionalDriftMatrix}
	\mat{M}_{E}(t) & := -\mat{Q}-2\mat{\sigma}\mat{B}^\transpose\mat{A} - 2\mat{V}_{E}(t)\mat{A}^\transpose\mat{A}. 
\end{align}

The deterministic backward Riccati equation for the covariance matrix is similar to Eq.~\eqref{eq:FwdCovarianceEquationOfMotion},
\begin{align} \label{eq:BwdCovarianceEquationOfMotion}
\begin{split}
	{-\frac{\diff \mat{V}_{E}(t)}{\diff t}} & := \mat{V}_{E}(t-\diff t)-\mat{V}_{E}(t) \\
	& = \mat{M}_{E}(t)\mat{V}_{E}(t) + \mat{V}_{E}(t)\mat{M}_{E}^\transpose(t) \\
	& \quad + \mat{D} - 2\mat{V}_{E}(t)\mat{A}^\transpose\mat{A}\mat{V}_{E}(t), 
\end{split}
\end{align}
and clearly shows the importance of continuous observations for retrodiction. Without observations (\ie, when $\mat{A}=\mat{B}= 0$) the drift matrices would be equal up to a sign, $\mat{M}_{\rho}(t)=-\mat{M}_{E}(t)=\mat{Q}$. At the same time the quadratic Riccati equations for the respective covariance matrices would turn into linear Lyapunov equations. Assuming stable forward dynamics with a positive steady state solution $\mat{V}_{\rho}^{\asympt}>0$ of Eq.~\eqref{eq:StableFwdDynamicsRiccati},
\begin{align}
	\mat{Q}\mat{V}_{\rho}^{\asympt} + \mat{V}_{\rho}^{\asympt}\mat{Q}^\transpose & = - \mat{D},
\end{align}
would preclude stable backward dynamics: there cannot simultaneously be a positive asymptotic covariance matrix $\mat{V}_{E}^{\asympt}>0$ for $t\to -\infty$ that satisfies
\begin{align}
	-\mat{Q}\mat{V}_{E}^{\asympt} - \mat{V}_{E}^{\asympt}\mat{Q}^\transpose & = - \mat{D}.
\end{align}
Only the presence of a sufficiently large quadratic $\mat{A}^\transpose\mat{A}$-term in Eq.~\eqref{eq:BwdCovarianceEquationOfMotion}, corresponding to sufficiently efficient observations, allows us to find an asymptotic solution $\mat{V}_{E}^{\asympt}>0$. Analogous to Eq.~\eqref{eq:StableFwdDynamicsRiccati} this implies an asymptotic drift matrix $\mat{M}_{E}^{\asympt}$ whose eigenvalues have negative real parts.

Assuming stable backward dynamics that make any Gaussian effect operator with $\mat{V}_{E}(t_{1})$ collapse to $\mat{V}_{E}^{\asympt}$ as $t\to -\infty$, we can plug the asymptotic solution $\mat{V}_{E}^{\asympt}$ into the equation for the means. Similar to the forward solution in Eq.~\eqref{eq:FwdIntegratedMeans} we find
\begin{multline} \label{eq:BwdIntegratedMeans}
	\bito\  \vect{r}_{E}(t) = \expo{(t_{1}-t)\mat{M}_{E}^{\asympt}}\vect{r}_{E}(t_{1}) \\
	+ \int_{t}^{t_{1}}\expo{(\tau-t)\mat{M}_{E}^{\asympt}}\bigl(\mat{V}_{E}^{\asympt}\mat{A}^\transpose + \mat{\sigma}\mat{B}^\transpose\bigr)\diff \vect{Y}(\tau), 
\end{multline}
where the integral is a backward It\^{o} integral as explained in Appendix \ref{sec:Appendix:BackwardItoIntegration}. The negative eigenvalues of $\mat{M}_{E}^{\asympt}$ again cause exponential damping of the final condition $\vect{r}_{E}(t_{1})$ and of the integrand, which picks out a different set of modes compared to the forward integral in Eq.~\eqref{eq:FwdIntegratedMeans}, cf.~Sec.~\ref{sec:ModeFunctions}.

\subsection{Interpretation of retrodictive POVMs}
In analogy to Eq.~\eqref{eq:DOGaussianState}, the POVM realized at time $t$ in retrodiction based on continuous homodyne detection during some time interval $[t,t_1]$ can be written as
\begin{align}\label{eq:POVMOps}
  \Bigl\{\dspop(\vect{r}_{E}(t))\opE_{0}(t)\dspop^\dagger(\vect{r}_{E}(t))\Bigr\}.
\end{align}
Here $\opE_0(t)=\exp\bigl[-\vect{\op{r}}^\transpose\mat{\Gamma}_E(t)\vect{\op{r}}\bigr]$ is independent of the means, since $\mat{\Gamma}_E(t)$ is determined by the covariance matrix $\mat{V}_E(t)$ as explained below Eq.~\eqref{eq:DOGaussianState}. Means $\vect{r}_{E}(t)$ and covariance matrix $\mat{V}_{E}(t)$ are determined by \eqref{eq:BwdMeanEquationOfMotion} and \eqref{eq:BwdCovarianceEquationOfMotion}. The POVM elements all correspond to displaced versions of the operator $\opE_0(t)$. The shape of $\opE_0(t)$ determines the resolution in phase space achieved by the POVM in retrodiction. It is again useful to consider the purity of a Gaussian effect operator $\opE$ with covariance matrix $\mat{V}_{E}$ which is computed as in Eq.~\eqref{eq:Purity}. Unit purity means the given POVM actually corresponds to projections onto pure states, constituting a quantum-limited measurement. Pure POVM elements with equal variances then indicate a projection onto coherent states, which corresponds to a heterodyne measurement of both quadratures \cite{Wiseman2010}, that is $\left\{|\alpha\rangle\langle\alpha|/\pi\right\}$. Asymmetric variances, on the other hand, indicate squeezed projectors that correspond, in the ideal limit of infinite squeezing, to a measurement of only one quadrature, $\{|x\rangle\langle x|\}$, where $|x\rangle$ denotes a quadrature eigenstate.

Reduced purity means additional uncertainty and thus lower resolution of the measurement. We will see in the examples in Sections \ref{sec:DecayingCavity} and \ref{sec:RealisticOM} that the purity of retrodicted effect operators decreases quickly when the detection efficiency is low or there is coupling to unobserved baths. Quite generally, for systems subject to continuous time measurements with a  measurement rate $\Gamma$ (including losses) competing with other decoherence processes happening at rate $\gamma$ one will find that the dynamics of both conditional density and effect matrix crucially depend on a quantum cooperativity parameter $C_{q}=\Gamma/\gamma$. The regime $C_q>1$ signifies the possibility to produce quantum limited POVMs in retrodiction just as it allows pure conditional quantum states in prediction. In Sec.~\ref{sec:RealisticOM} we will prove this statement in great detail for continuous measurements on optomechanical systems.

While it is possible to perform quantum limited POVMs corresponding to projections on pure states it cannot be used as a means for \textit{preparation} of pure quantum states. The ``collapsed'' posterior state is physically not realized since retrodictive POVMs are destructive: once all information necessary for realizing the POVM has been gathered the system's state has already evolved into something different, whose best description is just the conditional quantum state. It does not make sense to consider the posterior state after the measurement in a similar way as it is useless to ask for the state of a photon after photo-detection.

Repeated measurement of a POVM \eqref{eq:POVMOps} on identically prepared systems in state $\oprho_0$ will map out the probability distribution
\begin{align*}
  P(\vect{r}|\oprho_0)=\trace\{ D(\vect{r})\opE_0 D^\dagger(\vect{r}) \oprho_0 \}.
\end{align*}
This is the information on the state $\oprho_0$ which is directly accessible via retrodictive POVMs. Other relevant aspects regarding the quantum state may be inferred from such information, possibly collected for different POVMs by changing the dynamics -- and with it the equations of motion for $\opE(t)$ -- of the system.

One may, for example, be interested in reconstructing the density matrix $\oprho_0$ itself which corresponds to the problem of quantum state tomography. Recapitulating the methods available to perform this task is beyond the scope of this article, and we refer to the literature in this field \cite{Lvovsky2009,Paris2004}. We just state two particularly simple cases: The heterodyne POVM directly provides the Mandel $Q$-function of the quantum state, $Q(\alpha)=\bra{\alpha}\oprho_0\ket{\alpha}$. If $\oprho_0$ was known to be Gaussian this POVM will directly give the correct means and (co)variances with one unit of added quantum noise in each quadrature. A POVM corresponding to an infinitely squeezed state will give access directly to the marginal distribution in the respective quadrature $\bra{x}\oprho_0\ket{x}$.

It is worth emphasizing that the Gaussian POVMs realized by linear dynamics considered here may well be applied to non-Gaussian states. No assumption on the initial \emph{state} $\oprho_0$ went into the derivation of the equations of motion \eqref{eq:BwdMeanEquationOfMotion} and \eqref{eq:BwdCovarianceEquationOfMotion} for the Gaussian operator $\opE(t)$. Provided the measurement delivers sufficient resolution in phase space the tools of retrodictive Gaussian POVMs can therefore well be used for verification of non-Gaussian states which have been created initially by some different means. (Of course those initial states cannot emerge as conditional states from Gaussian dynamics and homodyne detections alone.) Along these lines, preparation and verification of Fock states in macroscopic mechanical oscillators have been discussed by Khalili et al. \cite{Khalili2010}, see also \cite{Miao2010}.


\section{Basic examples}
\label{sec:DecayingCavity}

In this section we will consider two basic but illustrative examples of the formalism developed so far which will provide a firm basis for the more serious application to optomechanical systems in Sec.~\ref{sec:RealisticOM}.

\subsection{Monitoring a decaying cavity}

\begin{figure}
	\centering
	\def\svgwidth{.8\columnwidth}
	\input{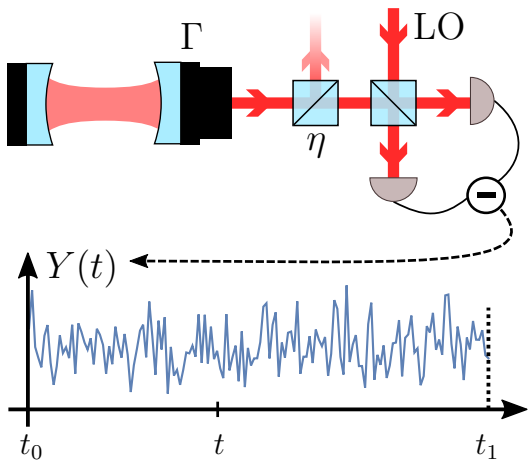}
	\caption{Schematic of a freely decaying cavity monitored from time $t_{0}$ to $t_{1}$. Light leaving the cavity at rate $\Gamma$ is superposed on a balanced beam-splitter with a strong local oscillator (LO). Two photodetectors monitor the output ports and their photocurrents are subtracted to yield a time-continuous homodyne measurement signal $Y(t)$. Imperfect detection is modeled as photon loss induced by a second beam splitter which only transmits a fraction $\eta$ of the signal light. \label{fig:SimpleCavity}}
\end{figure}

Let us start with the simple example of a decaying cavity undergoing homodyne detection, depicted in Fig.~\ref{fig:SimpleCavity}. This example was used by Wiseman \cite{Wiseman1996} to illustrate the interpretation of quantum trajectories in measurement theory as retrodictive POVM elements. Using operator algebra he showed that with an ideal detector and infinite observation time one can perform a projective measurement of the initial state of the cavity onto a quadrature eigenstate. Using the formalism developed in the previous sections we will treat the same setup here for homodyne detection with efficiency $\eta$. For ideal detection $\eta\rightarrow 1$ recover the result of Wiseman.

\subsubsection{Conditional state evolution}

We consider an ideal freely damped cavity with $\op{H}=0$ and decay rate $\Gamma$. The output is mixed with a strong local oscillator 
with adjustable relative phase $\phi$ to perform homodyne detection with efficiency $\eta\in [0,1]$. For later reference we first study the corresponding stochastic master equation for the conditional state of the intra-cavity field \cite{Wiseman1996}. In a frame rotating at the cavity frequency this is
\begin{align}\label{eq:SimpleCavityME}
	\ito\ \diff \oprho(t) & = \Gamma\Dop{\op{a}}\oprho(t)\diff t + \sqrt{\eta \Gamma}\Hop{\expo{-i\phi}\op{a}}\oprho(t) \diff W(t),
\end{align}
where $\op{a}^\dag,\op{a}$ are the cavity creation and annihilation operators (CAOs). The canonical quadrature operators are $\op{x} = (\op{a} + \op{a}^\dag)/\sqrt{2}$ and $\op{p} = -i(\op{a} - \op{a}^\dag)/\sqrt{2}$ which we collect into a vector $\op{\vect{r}} = \begin{bmatrix} \op{x} & \op{p} \end{bmatrix}^\transpose$. Then the Wiener increment $\diff W(t)$ is related to the detector output $\diff Y(t)$ as
\begin{align}
	\diff Y(t) = \sqrt{2\eta\Gamma}\langle \op{x}_{\phi} \rangle_{\rho(t)}\diff t + \diff W(t),
\end{align}
with $\op{x}_{\phi} := \cos(\phi)\op{x} + \sin(\phi)\op{p}$. Due to the symmetry of the problem we choose $\phi = 0$ without loss of generality, observing only the $\op{x}$-quadrature of the cavity.

Spelling out Eqs.~\eqref{eq:FwdWienerMeanEoM} and \eqref{eq:FwdCovarianceEquationOfMotion} for $\diff \vect{r}_{\rho}$ and $\dot{\mat{V}}_{\rho}$ we find
\begin{subequations}\label{eq:SimpleCavityMeans}
\begin{align}
	\ito\ \diff x_{\rho}(t) & = -\frac{\Gamma}{2}x_{\rho} \diff t + \sqrt{\frac{\eta \Gamma}{2}}(V_{xx}^{\rho}(t)-1)\diff W(t),\\ 
	\ito\ \diff p_{\rho}(t) & = -\frac{\Gamma}{2}p_{\rho} \diff t + \sqrt{\frac{\eta \Gamma}{2}}V_{xp}^{\rho}(t)\diff W(t), 
\end{align}
\end{subequations}
and
\begin{subequations}\label{eq:SimpleCavityCovarianceMatrix}
\begin{align}
	\dot{V}_{xx}^{\rho} & = - (1-2\eta)\Gamma V_{xx}^{\rho} + (1-\eta)\Gamma - \eta \Gamma (V_{xx}^{\rho})^2,\\ 
	\dot{V}_{xp}^{\rho} & = - (1-\eta)\Gamma V_{xp}^{\rho} - \eta\Gamma V_{xx}^{\rho}V_{xp}^{\rho},\\ 
	\dot{V}_{pp}^{\rho} & = - \Gamma V_{pp}^{\rho} + \Gamma - \eta \Gamma (V_{xp}^{\rho})^2. 
\end{align}
\end{subequations}
The steady-state covariance matrix $\mat{V}_{\rho}^{\asympt}$ satisfying the Riccati equation $\dot{\mat{V}}_{\rho}=0$ is given by the variances and covariance
\begin{align}
	V_{xx}^{\rho} & = V_{pp}^{\rho} = 1,\qquad V_{xp}^{\rho} = 0. 
\end{align}

Computing the purity $\purity (\oprho) = 1 / \sqrt{\det(\mat{V}_{\rho}^{\asympt})} = 1$ 
shows that the prepared steady state is pure, which together with equal variances implies it is a coherent state. However, plugging $\mat{V}_{\rho}^{\asympt}$ into Eqs.~\eqref{eq:SimpleCavityMeans} for the means makes $\diff W$ drop out, so the asymptotic forward evolution does not depend on the monitored output. In the long term both mean values decay exponentially, affirming the expected result that for long times a decaying cavity will simply collapse to the vacuum state, $\oprho_{\asympt} = \ket{0}\bra{0}$.

This insight is important. It shows that the covariance matrix and purity alone do not let us judge the effectiveness of a given preparation (or retrodiction) scheme. If the unconditional dynamics produce some mixed steady state we can increase our knowledge by monitoring the output. At long times the conditional dynamics will produce a state with fixed covariance matrix and measurement-dependent means that move around phase space, such that the averaged conditional dynamics agree with the unconditional dynamics. However, if the unconditional dynamics already yield a quantum-limited state (such as the vacuum in the present example) then there is nothing to be gained from observing the output. These statements apply to both quantum states and effect operators.

While the observations cannot aid the (long-term) state preparation we will now see how they let us infer information about the \textit{initial} state of the cavity.

\subsubsection{Retrodiction of POVM elements}
The equation adjoint to Eq.~\eqref{eq:SimpleCavityME} for the backwards-propagating POVM element $\opE$ reads
\begin{align}\label{eq:EffectOpEomCavity}
	\begin{split}
		\bito\ {-\diff} \opE(t) & = \Gamma\Dopdag{\op{a}}\opE(t)\diff t \\
		& \quad + \sqrt{\eta \Gamma}(\op{a}^\dag\opE(t) + \opE(t)\op{a}) \diff Y(t).
	\end{split}
\end{align}
In \cite{Wiseman1996} Wiseman essentially constructed an operator solution of this equation for a unit-efficiency measurement and showed that the corresponding POVM corresponds to a projection on quadrature eigenstates. Restricting ourselves to Gaussian POVMs (cf.\@ App.~\ref{sec:Appendix:BackwardAddition}) we instead directly write down the (normalized) equations of motion of means and covariance matrix, Eqs.~\eqref{eq:BwdMeanEquationOfMotion} and \eqref{eq:BwdCovarianceEquationOfMotion},
\begin{subequations}\label{eq:SimpleCavity:BackwardMeans}
\begin{align}
	\bito\ {-\diff} x_{E}(t) & = \frac{\Gamma}{2}x_{E} \diff t + \sqrt{\frac{\eta \Gamma}{2}}(V_{xx}^{E}(t)+1)\diff W(t),\\ 
	\bito\ {-\diff} p_{E}(t) & = \frac{\Gamma}{2}p_{E} \diff t + \sqrt{\frac{\eta \Gamma}{2}}V_{xp}^{E}(t)\diff W(t), 
\end{align}
\end{subequations}
and
\begin{subequations}\label{eq:SimpleCavity:BackwardCovMat}
\begin{align}
	-\dot{V}_{xx}^{E} & = (1-2\eta)\Gamma V_{xx}^{E} + (1-\eta)\Gamma - \eta \Gamma (V_{xx}^{E})^2,\\ 
	-\dot{V}_{xp}^{E} & = (1-\eta)\Gamma V_{xp}^{E} - \eta\Gamma V_{xx}^{E}V_{xp}^{E},\\ 
	-\dot{V}_{pp}^{E} & = \Gamma V_{pp}^{E} + \Gamma - \eta \Gamma (V_{xp}^{E})^2. 
\end{align}
\end{subequations}
We solve $\dot{V}_{xx}^{E} = 0$ to find the asymptotic solution
\begin{align}
	V_{xx}^{E} & = \frac{1-\eta}{\eta}, 
\end{align}
which entails constant covariance, $\dot{V}_{xp}^{E} \equiv 0$, independent of its current value. Note that the asymptotic $\op{x}$-variance vanishes, $V_{xx}^{E}\to 0$, as $\eta\to 1$ which shows that the corresponding effect operator measures $\op{x}$ with arbitrary precision. The effect operator will be squeezed in $\op{x}$ (\ie, $V_{xx}^{E}<1$) 
for any $\eta > \frac{1}{2}$. On the other hand, $V_{xx}^{E}\to \infty$ as $\eta\to 0$ emphasizing that retrodiction crucially depends on observations. When attempting to solve $\dot{V}_{pp}^{E} = 0$ we find that there is no finite asymptotic solution, $V_{xp}^{E}$ and $V_{pp}^{E}$, which simultaneously satisfies $V_{pp}^{E} \geq 0$ and $\det[\mat{V}_{E}]\geq 0$, which are necessary requirements for a proper covariance matrix. Thus $V_{pp}^{E}(t)$ grows beyond all bounds as time runs backwards, which is in line with the fact that our setup only gathers information about $\op{x}$. Thus, retrodiction allows to effectively perform a projective  measurement of a quadrature operator on the initial state of the cavity. By changing the homodyne angle analogous results can be obtained for any quadrature $\op{x}_{\phi}$. This agrees with the finding of \cite{Wiseman1996} derived using completely different methods exploiting operator algebra. One can check by direct computation (paying attention to detail \footnote{When computing $\diff \opE(t)$ of the effect operator in \cite{Wiseman1996} it is important to reintroduce the initial time which was set to zero by Wiseman, and to take the derivative with respect to this time.})
that the effect operator constructed in \cite{Wiseman1996} indeed satisfies the equation of motion \eqref{eq:EffectOpEomCavity}.

We can now also derive the filter functions or temporal modes which have to be extracted from the photocurrent. Plugging the asymptotic variance $V_{xx}^{E}$ into the equation for $x_{E}$ we find
\begin{align}
	\bito\ {-\diff} x_{E}(t) & = \frac{\Gamma}{2}x_{E} \diff t + \sqrt{\frac{\Gamma}{2\eta}}\diff W(t)\\
	& = -\frac{\Gamma}{2}x_{E} \diff t + \sqrt{\frac{\Gamma}{2\eta}}\diff Y(t).
\end{align}
The solution to this equation is given by
\begin{align}
	\begin{split}
		\bito\ x_{E}(t) & = \expo{-\Gamma(t_{1}-t)/2} x_{E}(t_{1}) \\
		& \qquad + \sqrt{\frac{\Gamma}{2\eta}}\int_{t}^{t_{1}} \expo{-\Gamma(t_{1}-\tau)/2} \diff Y(t),
	\end{split}
\end{align}
for $t\leq t_{1}$ so the final value $x_{E}(t_{1})$ is exponentially damped, and far into the past the mean $\op{x}$-quadrature of the retrodicted effect operator will depend only on the integrated measurement current. The temporal mode to be extracted from the continuous measurement is an exponentially decaying function in time with width $\Gamma/2$ set by the cavity decay rate.

\subsection{Beam-splitter vs.~squeezing interaction}
We will now examine why we can prepare only a coherent state (the vacuum) but are able to measure squeezed states. This is due to the beam-splitter (BS) coupling between the cavity and the field outside,
\begin{align}
	\op{H}_{\mrm{int}}^{\mrm{BS}} & = \Gamma (\op{a}\op{c}_{\mrm{out}}^\dag + \op{a}^\dag \op{c}_{\mrm{out}}),
\end{align}
where $\op{c}_{\mrm{out}}^\dag,\op{c}_{\mrm{out}}$ are the CAOs corresponding to the outgoing mode being measured. This interaction causes a state swap between the intracavity and outside fields.
To illustrate this further let us replace the beam-splitter coupling by a two-mode squeezing (TMS) interaction,
\begin{align}
	\op{H}_{\mrm{int}}^{\mrm{TMS}} & = \Gamma (\op{a}^\dag\op{c}_{\mrm{out}}^\dag + \op{a}\op{c}_{\mrm{out}}).
\end{align}
For our simple cavity this is obviously unrealistic but we will encounter the TMS interaction again in optomechanical systems, so it is worthwhile understanding the effect this has on the dynamics. $\op{H}_{\mrm{int}}^{\mrm{TMS}}$ creates entangled pairs of photons so detecting the outgoing light will reveal information about the current state of the cavity but not about what it was before the interaction. The corresponding master equation reads
\begin{align}
	\ito\  \diff \oprho(t) & = \Gamma\Dop{\op{a}^\dag}\oprho(t)\diff t + \sqrt{\eta \Gamma}\Hop{\op{a}^\dag}\oprho(t) \diff W(t).
\end{align}
This yields equations of motion for the means and (co)variances of the conditional state,
\begin{align}
	\ito\ \diff x_{\rho}(t) & = \frac{\Gamma}{2}x_{\rho} \diff t + \sqrt{\frac{\eta \Gamma}{2}}(V_{xx}^{\rho}(t)+1)\diff W(t),\\ 
	\ito\ \diff p_{\rho}(t) & = \frac{\Gamma}{2}p_{\rho} \diff t + \sqrt{\frac{\eta \Gamma}{2}}V_{xp}^{\rho}(t)\diff W(t), 
\end{align}
and
\begin{align}
	\dot{V}_{xx}^{\rho} & = (1-2\eta)\Gamma V_{xx}^{\rho} + (1-\eta)\Gamma - \eta \Gamma (V_{xx}^{\rho})^2,\\ 
	\dot{V}_{xp}^{\rho} & = (1-\eta)\Gamma V_{xp}^{\rho} - \eta\Gamma V_{xx}^{\rho}V_{xp}^{\rho},\\ 
	\dot{V}_{pp}^{\rho} & = \Gamma V_{pp}^{\rho} + \Gamma - \eta \Gamma (V_{xp}^{\rho})^2, 
\end{align}
which are exactly the same as the backward Eqs.~\eqref{eq:SimpleCavity:BackwardMeans} and \eqref{eq:SimpleCavity:BackwardCovMat} for the BS interaction. So while $V_{pp}^{\rho}(t)$ will grow beyond all bounds asymptotically, we can condition the $\op{x}$-quadrature to arbitrary precision limited only by our detection efficiency $\eta$, meaning we can prepare arbitrarily squeezed states. Similarly, the situation is also reversed for the backward effect equation, yielding equations for means and covariance matrix given by Eqs.~\eqref{eq:SimpleCavityMeans} and \eqref{eq:SimpleCavityCovarianceMatrix}. This means the effect operators will become independent of the photocurrent in the long-time limit and project only on the vacuum state. 
We summarize the effect of each coupling on the performance of both pre- and retrodiction in Table~\ref{table}.

\begin{center}
\begin{table}[t]
	\begin{tabular}{|c||c|c|}\hline
			& Prediction $\oprho$	& Retrodiction $\opE$	\\\hhline{|=#=|=|}
	beam splitter interaction	& Coherent		& Squeezed	\\ \hline
	two-mode squeezing int.	& Squeezed		& Coherent \\ \hline
	\end{tabular}
\caption{Summary of the predicted quantum state and the retrodictive POVM realized for a single mode coupled via a beam splitter or a two-mode squeezing interaction to the monitoring field.}\label{table}
\end{table}
\end{center}


\section{Conditional state preparation and verification in optomechanics}
\label{sec:RealisticOM}

\begin{figure}
	\centering
	\def\svgwidth{.8\columnwidth}
	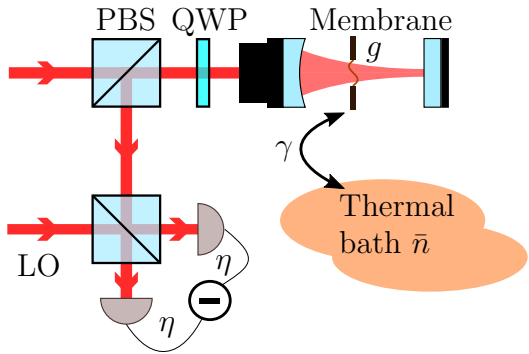
	\caption{Schematic of a micromechanical membrane coupled to a driven cavity with coupling strength $g$. Before entering the cavity the linearly polarized driving field is transmitted through a polarizing beam-splitter (PBS) and quarter-wave plate (QWP). After interaction with the cavity and membrane the outgoing light again passes the QWP, such that it becomes orthogonally polarized to the incoming light. It is reflected off the PBS and superposed on a second beam splitter with a strong local oscillator (LO) to perform homodyne or heterodyne detection with detection efficiency $\eta$. The membrane is additionally coupled to a thermal bath with rate $\gamma$ and mean phonon number $\bar{n}$. \label{fig:Membrane}}
\end{figure}

The illustrative examples studied in the previous sections provide the background for the main application of the formalism to time continuous measurements on optomechanical systems \cite{Chen2013,Aspelmeyer2014}. The system of interest is a single mode of a mechanical oscillator, such as a membrane depicted in Fig.~\ref{fig:Membrane}, which couples to the light field inside a resonantly driven cavity. The light escaping the cavity is then mixed with a local oscillator to perform heterodyne detection. We will be interested in the weak coupling limit of optomechanics, where the cavity can be adiabatically eliminated, and the time continuous measurement effectively concerns the mechanical system only. It is important to note that this weak coupling limit does not exclude the regime of strong quantum cooperativity where the measurement back action noise process effectively becomes stronger than all other noise processes acting on the oscillator. Quantum cooperativities on the order of 100 have been demonstrated in recent optomechanical systems \cite{Rossi2018}. It is clear that the tools of quantum state pre- and retrodiction become especially powerful in such a regime.

The adiabatic limit of the conditional optomechanical master equation has been treated in great detail in \cite{Hofer2015}. We summarize here the main aspects, and then apply it to discuss pre-  and retrodiction.

\subsection{Optomechanical setup}
We consider a mechanical mode with frequency $\freqmech$ coupled to a cavity with resonance frequency $\freqcav$, driven by a strong coherent field with frequency $\freqdrive$.
We move to a rotating frame with respect to the drive $\freqdrive$, and assume the generated intracavity amplitude is large so we can linearize the radiation pressure interaction. Following standard treatment \cite{Aspelmeyer2014} this yields
\begin{subequations}\label{eq:LinearHamiltonian}
\begin{align}
	\op{H}_{\mrm{lin}} & = \op{H}_{0} + g\bigl( \op{a} + \op{a}^\dag \bigr) \bigl( \op{a}_{\mrm{c}} + \op{a}_{\mrm{c}}^\dag \bigr), \\
	\op{H}_{0} & = \freqmech \op{a}^\dag \op{a} - \Delta_{\mrm{c}}\op{a}_{\mrm{c}}^\dag \op{a}_{\mrm{c}},
\end{align}
\end{subequations}
where $\op{H}_{0}$ comprises the local Hamiltonians of cavity and mechanics with $\Delta_{\mrm{c}} = \freqdrive-\freqcav$,
and $g \propto g_{0}$ 
is the cavity-enhanced optomechanical coupling strength. $\op{a}$ and $\op{a}_{\mrm{c}}$ are the annihilation operators of the mechanical and cavity mode, respectively.

The cavity field leaks out at a full width at half maximum (FWHM) decay rate $\kappa$. The (unconditional) master equation of the joint state $\oprho_{\mrm{mc}}$ of mechanical and cavity mode reads
\begin{align}\label{eq:UnconditionalMasterEquation}
	\begin{split}
		\dot{\oprho}_{\mrm{mc}}(t) & = -i[\op{H}_{\mrm{lin}},\oprho_{\mrm{mc}}(t)] + \kappa \Dop{\op{a}_{\mrm{c}}}\oprho_{\mrm{mc}}(t) \\
		& \quad + \mcl{L}_{\mrm{th}}\oprho_{\mrm{mc}}(t),
	\end{split}
\end{align}
where we have also included a thermal bath,
\begin{align}\label{eq:ThermalBath}
	\begin{split}
		\mcl{L}_{\mrm{th}}\oprho_{\mrm{mc}}(t) & = \gamma(\bar{n}+1) \Dop{\op{a}}\oprho_{\mrm{mc}}(t)\\
		&\quad + \gamma\bar{n} \Dop{\op{a}^\dag}\oprho_{\mrm{mc}}(t),
	\end{split}
\end{align}
with mean phonon number $\bar{n}$ which couples to the mechanical oscillator at rate $\gamma$ (FWHM of the mechanical mode).

We monitor the field that leaks from the cavity using homodyne or heterodyne detection. As usual the outgoing field is combined on a balanced beam splitter with a strong local oscillator, and the difference of the measured intensities in the two output beams is the measurement current, depicted in the bottom left of Fig.~\ref{fig:Membrane}. As compared to the conditional master equation Eq.~\eqref{eq:SimpleCavityME} studied in Sec.~\ref{sec:DecayingCavity} on the decaying cavity we consider here a slightly more general setup where the local oscillator frequency $\freqLO$ may be detuned from the driving frequency $\freqdrive$, captured by $\Delta_{\mrm{lo}} = \freqLO - \freqdrive$. This realizes a measurement of the outgoing field quadrature operator $\op{a}_{\text{out}}(t)\expo{-i\Delta_{\mrm{lo}}t + i\phi_{\mrm{lo}}} + \op{a}_{\text{out}}^\dag(t)\expo{i\Delta_{\mrm{lo}}t - i\phi_{\mrm{lo}}}$, where $\phi_{\mrm{lo}}$ is the tunable phase of the local oscillator. This yields a conditional master equation for cavity and mechanics,
\begin{align}
	\begin{split} \label{eq:ConditionalMasterEquation}
		\ito\ \diff \oprho(t) & = -i[\op{H}_{\mrm{lin}},\oprho_{\mrm{mc}}(t)]\diff t + \kappa \Dop{\op{a}_{\mrm{c}}}\oprho_{\mrm{mc}}(t)\diff t \\
		& \quad + \mcl{L}_{\mrm{th}}\oprho_{\mrm{mc}}(t)\diff t \\
		& \quad + \sqrt{\eta\kappa}\mcl{H}\bigl[ \op{a}_{\mrm{c}} \expo{i(\Delta_{\mrm{c}} + \Delta_{\mrm{lo}})t - i\phi_{\mrm{lo}}} \bigr]\diff W(t),
	\end{split}
\end{align}
where $\eta\in[0,1]$ is the detection efficiency.

We would like an effective master equation for the mechanics alone. To this end one can start from the combined master equation Eq.~\eqref{eq:ConditionalMasterEquation} and move to an interaction picture with respect to $\op{H}_{0}$.
Assuming the cavity field decays fast on the time-scale set by the optomechanical interaction, $g/\kappa \ll 1$, one can adiabatically eliminate the cavity dynamics from the description. For details of this procedure see \cite{Hofer2015,Hofer2017}. But before we state the result let us take a closer look at the optomechanical interaction.

\subsection{Optomechanical interaction}\label{sec:OptomechanicalInteraction}
\begin{figure}
	\centering
	\def\svgwidth{\columnwidth}
	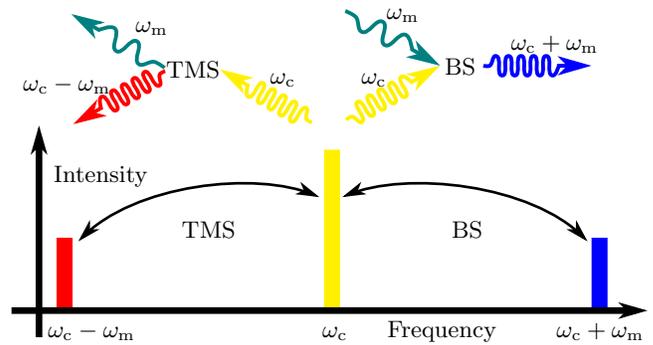
	\caption{Top: schematic conversion processes occurring in the optomechanical setup depicted in Fig.~\ref{fig:Membrane}, that scatter cavity photons at frequency $\freqcav$ into the outgoing sidebands while creating or annihilating a mechanical phonon at frequency $\freqmech$. Bottom: the spectrum of the outgoing light (not to scale). As discussed in Sec.~\ref{sec:OptomechanicalInteraction} the linearized optomechanical interaction facilitates two processes:
	The beam splitter (BS) interaction converts a cavity photon into a phonon and an outgoing photon in the lower (red) sideband at $\freqcav - \freqmech$. Two-mode squeezing (TMS) combines a cavity photon and a phonon to produce an outgoing photon in the upper (blue) sideband at $\freqcav + \freqmech$.
	\label{fig:Sidebands}}
\end{figure}

The linearized radiation-pressure interaction is given by the last term in Eq.~\eqref{eq:LinearHamiltonian}. The interaction decomposes into two terms: (i) a beam-splitter (BS) coupling $g(\op{a}\op{a}_{\mrm{c}}^\dag + \op{a}^\dag \op{a}_{\mrm{c}})$ and (ii) a two-mode squeezing (TMS) part $g(\op{a}\op{a}_{\mrm{c}} + \op{a}^\dag \op{a}_{\mrm{c}}^\dag)$. These give rise to Stokes and anti-Stokes scattering processes depicted in Fig.~\ref{fig:Sidebands}.
If we work in an interaction picture with respect to $\op{H}_{0}$
these terms oscillate at frequencies $\freqmech \pm \Delta_{\mrm{c}}$. For a \textit{red-detuned} drive, $\Delta_{\mrm{c}} = -\freqmech$, the BS interaction becomes resonant and is thus enhanced while the TMS interaction oscillates quickly at $2\freqmech$ and is suppressed. For a \textit{blue-detuned} drive, $\Delta_{\mrm{c}} = \freqmech$, the situation is reversed so the TMS interaction is enhanced and the BS interaction suppressed. For a resonant drive, $\Delta_{\mrm{c}} = 0$, both processes contribute equally.

As we have seen in the initial example in Section \ref{sec:DecayingCavity}, the entangling TMS interaction enhances our ability to prepare a conditional mechanical state. Because the outgoing light is entangled with the mechanics, performing a quantum-limited squeezed detection will also project the oscillator onto a squeezed state. On the other hand, the BS interaction generates light with the mechanical state swapped onto it. Observing it lets us determine what the state was before the interaction but will not enable the preparation of squeezed states. For retrodiction the situation is reversed. Extracting information about the system in the past from BS light produces squeezed effect operators (sharp measurements) on the past state, while entangled TMS light lets us retrodict coherent effect operators at best. Thus TMS (blue drive) enhances our ability to prepare while the BS interaction (red drive) enhances our ability to retrodict.

\subsection{Master equation of the mechanics}\label{sec:AdiabaticElimination}
In \cite{Hofer2015,Hofer2017} the master equation Eq.~\eqref{eq:ConditionalMasterEquation} is turned into an effective evolution equation for the mechanical state $\oprho_{\mrm{m}} \equiv \oprho$ through adiabatic elimination of the cavity mode. Since the result is not a proper Lindblad master equation one needs to perform a rotating wave approximation for which we integrate the dynamics over a short time,
\begin{align}
	\ito\ \delta\oprho(t) := \int_{t}^{t+\delta t} \diff\oprho(\tau).
\end{align}
We are interested here in the case of mechanical oscillators with high quality factors $Q=\freqmech/\gamma$ where $\freqmech$ is much larger than other system frequencies set by the optomechanical interaction and decoherence, \ie, $\freqmech \gg g^2/\kappa,\ \bar{n}\gamma$. In fact, we assume $\freqmech$ is so much larger that we can choose $\delta t$ such that $\freqmech \gg 1/\delta t \gg g^2/\kappa,\ \bar{n}\gamma$, which allows us to pull $\oprho(t)$ out of all deterministic time integrals since it is approximately constant on this time-scale. Note that this requires $Q\gg\bar{n}$ to be fulfilled with a safe margin. We emphasize that sideband resolution ($\freqmech\gg\kappa$) is not required for the following. We can then perform the rotating wave approximation by dropping all resonant terms oscillating at $\pm 2\freqmech$. Choosing the right phase $\phi_{\mrm{lo}}$ and quadrature frame, we find
\begin{subequations}\label{eq:AveragedConditionalMasterEquation}
\begin{align}
	\begin{split}
		\ito\ \delta \oprho(t) & = \Gamma_{-}\mcl{D}[\op{a}]\oprho(t)\delta t + \Gamma_{+}\mcl{D}[\op{a}^\dag]\oprho(t)\delta t \\
		& \quad + \sqrt{\eta} \int_{t}^{t+\delta t} \mcl{H}\bigl[ \op{C}(\tau;\Delta_{\mrm{lo}}) \bigr]\oprho(\tau)\diff W(\tau)\\
		& \quad + \mcl{L}_{\mrm{th}}\oprho(t)\delta t.
	\end{split}
\end{align}
with the time-dependent measurement operator
\begin{align}
\begin{split}
	\op{C}(\tau;\Delta_{\mrm{lo}}) & := \sqrt{\Gamma_{-}}\op{a}\expo{-i(\freqeff-\Delta_{\mrm{lo}})\tau } \\
	& \quad + \sqrt{\Gamma_{+}}\op{a}^\dag\expo{i(\freqeff+\Delta_{\mrm{lo}})\tau}.
\end{split}
\end{align}
\end{subequations}
The effective mechanical frequency
\begin{subequations}\label{eq:EffectiveFrequency}
\begin{align}
	\freqeff & := \freqmech - \sqrt{2}g^2(\beta_{+}+\beta_{-}),\\
	\beta_{\pm} & := \frac{\Delta_{\mrm{c}} \pm \freqmech}{(\kappa/2)^2 + (\Delta_{\mrm{c}} \pm \freqmech)^2}
\end{align}
\end{subequations}
results from a shift of $\freqmech$ due to the optical spring effect, and the rates
\begin{align}\label{eq:StokesRates}
	\Gamma_{\pm} & := \frac{g^2 \kappa}{(\kappa/2)^2 + (-\Delta_{\mrm{c}} \pm \freqmech)^2} 
\end{align}
are the usual Stokes and anti-Stokes rates known from sideband cooling.
From these we can define two effective cooperativities
\begin{align}\label{eq:QuantumCooperativities}
	C_{\pm} := \Gamma_{\pm}/\gamma = C_{\mrm{cl}} \frac{\kappa^2}{\kappa^2 + 4(-\Delta_{\mrm{c}} \pm \freqmech)^2},
\end{align}
in terms of the classical cooperativity
\begin{align}
	C_{\mrm{cl}} & = \frac{4g^2}{\kappa\gamma}.
\end{align}
Each $C_{\pm}$ compares the rate of the respective (anti-)Stokes process to the incoherent coupling rate of the thermal bath.
In the regime $\kappa \gg \freqmech$ of a broad cavity \footnote{We assumed $\freqmech\gg g^2/\kappa$ for the rotating wave approximation which imposes $\freqmech/\kappa \gg (g/\kappa)^2$. However, we also used $g/\kappa \ll 1$ to eliminate the cavity so $(g/\kappa)^2$ is a small parameter and the derivation holds for a range of linewidths from the sideband-resolved ($\kappa\ll \freqmech$) to the broad cavity regime ($\kappa \gg \freqmech$).}
and assuming $\kappa \gg \Delta_{\mrm{c}}$ the cooperativities reduce to the classical cooperativity, $C_{\pm}\approx C_{\mrm{cl}}$. As an example for the orders of magnitude involved here consider a recent experiment \cite{Rossi2018}, which realized $C \approx C_{\mrm{cl}} \sim 10^7$ and for $\bar{n}\sim 10^5$ a corresponding \textit{quantum cooperativity} \cite{Aspelmeyer2014}
\begin{align}
	C_{q} & := \frac{C}{\bar{n}+1} \sim 10^2.
\end{align}

To obtain a proper master equation from Eq.~\eqref{eq:AveragedConditionalMasterEquation} we still need to perform the integral over the measurement term, which depends on the choice of $\Delta_{\mrm{lo}}$. But Eq.~\eqref{eq:AveragedConditionalMasterEquation} already illustrates the point we made in Sec.~\ref{sec:OptomechanicalInteraction}: detuning the driving field affects the optomechanical interaction. Driving on resonance, $\Delta_{\mrm{c}} = 0$, TMS and BS interaction occur with equal strength which is reflected by $\Gamma_{+} = \Gamma_{-}$. A blue drive, $\Delta_{\mrm{c}}=\freqmech$, enhances TMS and causes $\Gamma_{+}>\Gamma_{-}$, while a red drive, $\Delta_{\mrm{c}}=-\freqmech$, enhances the BS interaction and causes $\Gamma_{-}>\Gamma_{+}$. Additionally, we can tune the local oscillator either to resonantly detect at the driving frequency, $\Delta_{\mrm{lo}} = 0$, or to either the blue or the red sideband, $\Delta_{\mrm{lo}} = \pm \freqeff$. We will explore these different dynamics step by step, starting with a resonant drive and resonant detection in the following section, then considering detection of the sidebands in Sec.~\ref{sec:ResoDriveDetectSBs}, and finally treating an off-resonant drive with sideband detection in Sec.~\ref{sec:OffResonantDrive}.

\subsection{Drive and detect on resonance}
We start by exploring a cavity driven on resonance, $\Delta_{\mrm{c}} = 0$, so we find equal rates $\Gamma_{+} = \Gamma_{-} =: \Gamma$, equal cooperativities $C := C_{+} = C_{-}$ with
\begin{align} \label{eq:Cooperativity}
	C = C_{\mrm{cl}} \frac{\kappa^2}{\kappa^2 + 4\freqmech^2},
\end{align}
and $\freqeff = \freqmech$. 
The first detection scheme we consider is homodyne detection on resonance, $\Delta_{\mrm{lo}} = 0$. Plugging this into Eq.~\eqref{eq:AveragedConditionalMasterEquation} yields the measurement operator
\begin{align}\label{eq:ResonantMeasurementOperator}
	\op{C}(\tau;\Delta_{\mrm{lo}} = 0) & := \sqrt{\Gamma}\bigl(\op{a}\expo{-i\freqmech\tau } + \op{a}^\dag\expo{i\freqmech\tau}\bigr)\\
	& = \sqrt{2\Gamma}\bigl(\op{x}\cos(\freqmech\tau) + \op{p}\sin(\freqmech\tau)\bigr),
\end{align}
with $\op{x} = (\op{a}+\op{a}^\dag)/\sqrt{2}$ and $\op{p} = -i(\op{a}-\op{a}^\dag)/\sqrt{2}$. Using again that we can pull $\oprho(t)$ out of the integrals we find the master equation
\begin{align}
	\begin{split}
		\ito\ \delta \oprho(t) & = \mcl{L}_{\mrm{th}}\oprho(t) + \Gamma\mcl{D}[\op{a}]\oprho(t)\delta t + \Gamma\mcl{D}[\op{a}^\dag]\oprho(t)\delta t \\
		& \qquad + \sqrt{\eta\Gamma}\mcl{H}[\op{x}]\oprho(t)\delta W_{\mrm{c}}(t)\\
		& \qquad + \sqrt{\eta\Gamma}\mcl{H}[\op{p}]\oprho(t)\delta W_{\mrm{s}}(t)
	\end{split}
\end{align}
with the coarse-grained Wiener increments
\begin{subequations} \label{eq:CosineSineWienerIncrements}
\begin{align}
	\ito\ \delta W_{\mrm{c}}(t) & := \sqrt{2}\int_{t}^{t+\delta t} \cos(\freqmech\tau) \diff W(\tau),\\
	\ito\ \delta W_{\mrm{s}}(t) & := \sqrt{2}\int_{t}^{t+\delta t} \sin(\freqmech\tau) \diff W(\tau).
\end{align}
\end{subequations}
It turns out that these are approximately normalized, $\delta W_{\mrm{c}}^2(t) = \delta t(1+\mcl{O}(\freqmech\delta t)^{-1})$ and $\delta W_{\mrm{s}}^2(t) = \delta t(1+\mcl{O}(\freqmech\delta t)^{-1})$, and independent $\delta W_{\mrm{c}}(t)\delta W_{\mrm{s}}(t) = \delta t \mcl{O}(\freqmech\delta t)^{-1}$. Thus we can replace $\delta t\to \diff t$, $\delta \oprho \to \diff \oprho$ and $\delta W_{\mrm{c/s}}\to \diff W_{\mrm{c/s}}$ to obtain the effective system dynamics
\begin{align}
	\begin{split}
		\ito\ \diff \oprho(t) & = \mcl{L}_{\mrm{th}}\oprho(t) + \Gamma\mcl{D}[\op{x}]\oprho(t)\diff t + \Gamma\mcl{D}[\op{p}]\oprho(t)\diff t \\
		& \quad + \sqrt{\eta\Gamma}\mcl{H}[\op{x}]\diff W_{\mrm{c}}(t) + \sqrt{\eta\Gamma}\mcl{H}[\op{p}]\diff W_{\mrm{s}}(t)
	\end{split}
\end{align}
with independent Wiener increments $\diff W_{\mrm{c}}(t)$ and $\diff W_{\mrm{s}}(t)$.

\subsubsection{Conditional state evolution}

\begin{figure}
	\centering
	\includegraphics[keepaspectratio=true,width=.8\columnwidth]{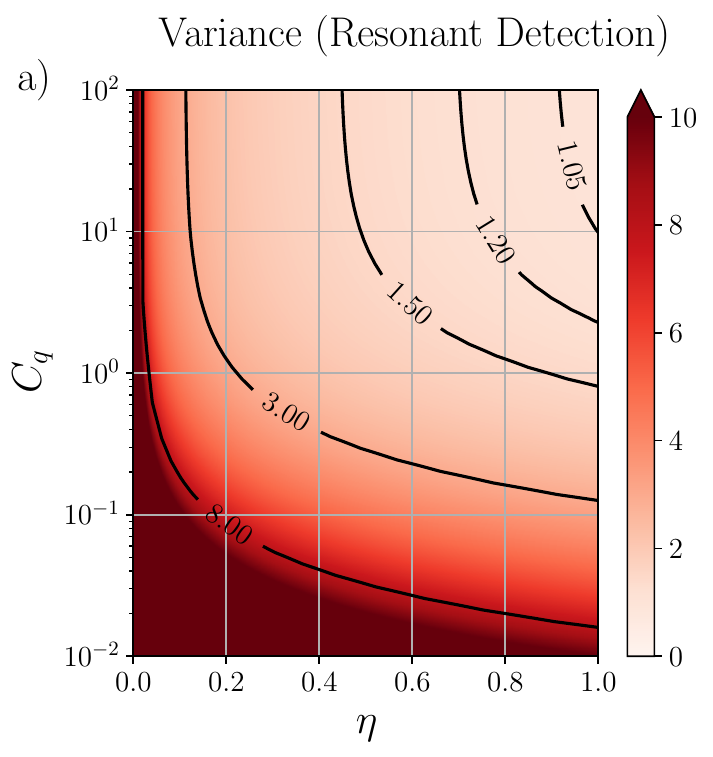}
	\includegraphics[keepaspectratio=true,width=.8\columnwidth]{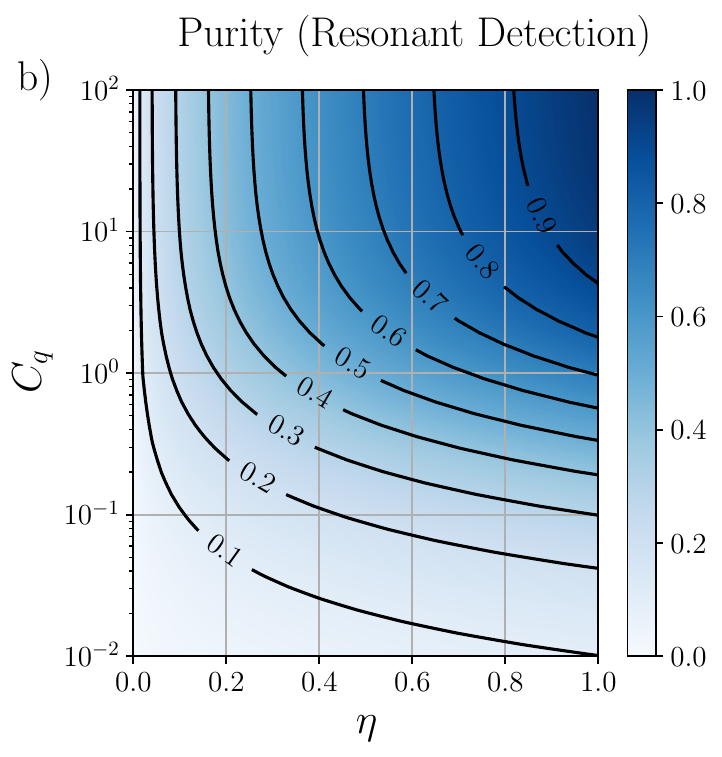}
	\caption{	 (a) Log-linear plot of the steady state variance $V_{\rho}^{\asympt}$ (same for $\hat{x}$ and $\hat{p}$) from Eq.~\eqref{eq:ExactFwdVarianceOnResonance} obtained for homodyne detection on resonance, plotted against detection efficiency $\eta$ and quantum cooperativity $C_{q} = C/(\bar{n} + 1)$. We chose a bath occupation of $\bar{n}\sim 10^5$ \cite{Rossi2018}, which entails $C\sim 10^{3}\dots 10^{7}$ in the plotted regime. The exact variance is virtually indistinguishable from its approximate value Eq.~\eqref{eq:ApproximateVarianceOnResonance} because the difference goes as $\sim 1/C$. The plot is also indistinguishable from the exact and approximate variances $V_{E}^{\asympt}$ of the effect operator in Eqs.~\eqref{eq:ExactBwdVarianceOnResonance} and Eq.~\eqref{eq:ApproximateBackwardVarianceOnResonance}.
	 (b) The purity of the covariance matrix corresponding to the variance in (a). \label{fig:VarianceAndPurityOnResonance}
	 }
\end{figure}

Using the notation of Sec.~\ref{sec:LinearDynamics} we find $\mat{H}=\mathbb{0}_2$, $\mat{\Delta} = (\Gamma+\frac{1}{2}\gamma(2\bar{n}+1))\mathbb{1}_{2}$ and $\mat{\Omega} = \frac{1}{2}\gamma\mat{\sigma}$, as well as the measurement matrices $\mat{A} = \sqrt{\eta\Gamma}\mathbb{1}_{2}$, and $\mat{B} = \mathbb{0}_{2}$. As before we solve $\dot{\mat{V}}_{\rho}=0$ to obtain the steady state covariance $V_{xp}^{\rho} = 0$ and equal variances
\begin{align}\label{eq:ExactFwdVarianceOnResonance}
	\begin{split}
		V_{\rho}^{\asympt} & := V_{xx}^{\rho} = V_{pp}^{\rho} \\
		& = \frac{1}{4\eta C}\Bigl(\sqrt{1 + 8\eta C (2C + 2\bar{n} + 1)} - 1\Bigr) 
	\end{split}
\end{align}
in terms of the cooperativity Eq.~\eqref{eq:Cooperativity}.
The purity is simply the inverse of the variance, $\purity (\oprho) = 1/V_{\rho}^{\asympt}$, 
so it suffices to consider $V_{\rho}^{\asympt}$. Note that as $\eta \to 0$ the variance approaches its thermal state value $V_{\rho}^{\asympt} \to 2\bar{n} + 1 + 2C$. 
From the covariance matrix we can compute the conditioned drift matrix from Eq.~\eqref{eq:FwdConditionalDriftMatrix} which turns out to be diagonal,
\begin{align}\label{eq:OMSetupDriftMatrix}
	\mat{M}_{\rho}^{\asympt} & = \lambda \mathbb{1}_{2}, &
	\lambda_{\rho} & = -\frac{\gamma}{2}\sqrt{1 + 8\eta C(2C + 2\bar{n} + 1)}.
\end{align}
The degenerate eigenvalue $\lambda_{\rho}$ is always real, and negative as long as $\gamma$ or $\eta\Gamma$ are non-zero and thus guarantees stable dynamics. We obtain the mode functions with which the cosine and sine components of the measurement current are It\^{o}-integrated in Eq.~\eqref{eq:FwdIntegratedMeans} by evaluating the kernel
\begin{align}
	\begin{bmatrix} f_{xc}(t) & f_{xs}(t) \\ f_{pc}(t) & f_{ps}(t) \end{bmatrix} & = \expo{\mat{M}_{\rho}^{\asympt}t}\bigl(\mat{V}_{\rho}^{\asympt}\mat{A}^\transpose - \mat{\sigma}\mat{B}^\transpose\bigr). 
\end{align}
We find that $f_{xs}(t) = f_{pc}(t) = 0$ and
\begin{align}\label{eq:ModeFunctionOnResonance}
	f_{\rho}(t) & := f_{xc}(t) = f_{ps}(t) = \sqrt{\eta\Gamma}V_{\rho}^{\asympt}\expo{-\lambda_{\rho} t},
\end{align}
which shows that the cosine and sine components of the photocurrent each only enter the corresponding ($\op{x}$ or $\op{p}$) quadrature.

In the following we assume that $\eta C\gg 1$ and $\bar{n}\gg 1$ so $\bar{n}+1\approx \bar{n}$. In terms of the quantum cooperativity \cite{Aspelmeyer2014}
\begin{align}
	C_{q} & = \frac{C}{\bar{n}+1} \approx \frac{C}{\bar{n}}
\end{align}
we find the variance
\begin{align} \label{eq:ApproximateVarianceOnResonance}
	V_{\rho}^{\asympt} & \approx \sqrt{\frac{C_{q}+1}{\eta C_{q}}} = \frac{1}{\sqrt{\eta}}\sqrt{1+\frac{1}{C_{q}}}, 
\end{align}
plotted in Fig.~\hyperref[fig:VarianceAndPurityOnResonance]{\ref{fig:VarianceAndPurityOnResonance}~(a)}, and the mode function damping rate is given by
\begin{align} \label{eq:ApproximateModeFunctionExponentOnResonance}
	\lambda_{\rho} & \approx -2\eta\Gamma\sqrt{\frac{C_{q}+1}{\eta C_{q}}}.
\end{align}

The equal variances in Eq.~\eqref{eq:ExactFwdVarianceOnResonance} and vanishing covariance indicate that we prepare a thermal steady state, which approaches a pure coherent state as $\eta\to 1$ and $C_{q}\to \infty$, as we see from the limiting expression Eq.~\eqref{eq:ApproximateVarianceOnResonance} and also from the purity plot in Fig.~\hyperref[fig:VarianceAndPurityOnResonance]{\ref{fig:VarianceAndPurityOnResonance}~(b)}. The exponent $\lambda_{\rho}$ in Eqs.~\eqref{eq:OMSetupDriftMatrix} and Eq.~\eqref{eq:ApproximateModeFunctionExponentOnResonance} determines how fast the mode functions Eq.~\eqref{eq:ModeFunctionOnResonance} decay, and thereby the ``memory time'' of the conditional state. In the regime where $C_{q}\gg 1\ \Leftrightarrow\ \Gamma \gg \gamma(\bar{n}+1)$ we find $\lambda_{\rho}\approx -2\sqrt{\eta}\Gamma$ so the mode function is only determined by the measurement rate. If $\Gamma$ is much larger than typical evolution time-scales it becomes sharply peaked at $t$, so the conditional state essentially follows the measurement current in real time. However, $\Gamma$ must stay well below $\freqmech$ or it violates the assumptions underlying our coarse-graining. In the opposite regime of $C_{q}\ll 1\ \Leftrightarrow\ \Gamma \ll \gamma(\bar{n}+1)$ the exponent is given by $\lambda_{\rho}\approx -2\sqrt{\Gamma\gamma(\bar{n}+1)/2}$. As $\Gamma \to 0$ the mode function becomes essentially flat but also goes to zero itself. In this limit the detection will yield mostly noise and only little signal, so the evolution becomes effectively unconditional.

\subsubsection{Retrodiction of POVM elements}
We obtain the asymptotic effect operator by solving the Riccati equation resulting from Eq.~\eqref{eq:BwdCovarianceEquationOfMotion}. Again $V_{xp}^{E} = 0$ and
\begin{align}
	V_{E}^{\asympt} & := V_{xx}^{E} = V_{pp}^{E} \notag \\
	& = \frac{1}{4\eta C}\Bigl(\sqrt{1 + 8\eta C (2C + 2\bar{n} + 1)} + 1\Bigr) \label{eq:ExactBwdVarianceOnResonance} \\ 
	& \approx \sqrt{\frac{C_{q}+1}{\eta C_{q}}}, \label{eq:ApproximateBackwardVarianceOnResonance} 
\end{align}
so we find effect operators with equal variance, which corresponds to a POVM realizing a heterodyne measurement.

The asymptotic variance of the retrodicted effect operator is strictly greater than the asymptotic variance of the conditional state, $V_{E}^{\asympt} - V_{\rho}^{\asympt} = 1/(2\eta C)$. 
The difference vanishes as $C \to \infty$ so the limits Eq.~\eqref{eq:ApproximateVarianceOnResonance} and Eq.~\eqref{eq:ApproximateBackwardVarianceOnResonance} are the same, and the plot in Fig.~\hyperref[fig:VarianceAndPurityOnResonance]{\ref{fig:VarianceAndPurityOnResonance}~(a)} also holds for $V_{E}^{\asympt}$. As expected, the exact $V_{E}^{\asympt}$ in Eq.~\eqref{eq:ExactBwdVarianceOnResonance} diverges without observations: $V_{E}^{\asympt} \sim 1/(2\eta C)$ 
as $\eta\to 0$. Otherwise the forward and backward dynamics are very similar: we find the same drift matrix as in Eq.~\eqref{eq:OMSetupDriftMatrix} with a degenerate negative eigenvalue $\lambda_{E}=\lambda_{\rho}$, and the mode function takes the same form as before,
\begin{align}
	f_{E}(t) & = \sqrt{\eta\Gamma}V_{E}^{\asympt}\expo{-\lambda_{E} t}, 
\end{align}
with the strictly greater variance $V_{E}^{\asympt}$ placing more weight on the backward optical mode compared to the evolution of the conditional state. Assuming $C,\bar{n}\gg 1$, forward and backward mode functions become identical.

For both preparation and retrodiction we see that we can never measure or prepare states with sub-shot noise resolution. In fact, in the ideal limit of perfect detection, $\eta \to 1$, and large cooperativity, $C_{q} \to \infty$,
both $V_{\rho}^{\asympt}$ and $V_{E}^{\asympt}$ approach $1$ 
so we can at best measure and prepare coherent states. This symmetry is not surprising since detecting on resonance means both TMS and BS interaction contribute equally to the observed light. The situation is different when the local oscillator is resonant with either of the sidebands.

\subsection{Drive on resonance, detect sidebands}\label{sec:ResoDriveDetectSBs}
We now detune the local oscillator with respect to the driving laser, $\Delta_{\mrm{lo}}= \pm \freqmech$, to resolve the information contained in the sidebands located at $\freqcav \pm \freqmech$. Recalling the general equation Eq.~\eqref{eq:AveragedConditionalMasterEquation}, we see that detecting the red sideband, $\Delta_{\mrm{lo}} = -\freqmech$, makes $\op{a}^\dag$ resonant while $\op{a}$ oscillates at $-2\freqmech$, and yields the measurement operator
\begin{align}\label{eq:RedMeasurementOperator}
	\op{C}(\tau;\Delta_{\mrm{lo}} = -\freqmech) & = \sqrt{\Gamma}\bigl(\op{a}\expo{-2i\freqmech\tau} + \op{a}^\dag\bigr).
\end{align}
Resonant detection of the blue sideband with $\Delta_{\mrm{lo}} = \freqmech$ analogously makes $\op{a}$ resonant and results in
\begin{align}\label{eq:BlueMeasurementOperator}
	\op{C}(\tau;\Delta_{\mrm{lo}} = \freqmech) & = \sqrt{\Gamma}\bigl(\op{a} + \op{a}^\dag\expo{2i\freqmech\tau}\bigr).
\end{align}
Thus, after coarse-graining we expect to better see an effect of the TMS interaction on the red sideband, and of the BS interaction on the blue sideband. 
To evaluate the integrals in Eq.~\eqref{eq:AveragedConditionalMasterEquation} we introduce
\begin{subequations}
\begin{align}
	\ito\ \delta W_{0}(t) & := \int_{t}^{t+\delta t} \diff W(\tau), \\
	\ito\ \delta W_{\mrm{c},2}(t) & := \sqrt{2}\int_{t}^{t+\delta t} \cos(2\freqmech\tau) \diff W(\tau), \\
	\ito\ \delta W_{\mrm{s},2}(t) & := \sqrt{2}\int_{t}^{t+\delta t} \sin(2\freqmech\tau) \diff W(\tau),
\end{align}
\end{subequations}
analogous to Eqs.~\eqref{eq:CosineSineWienerIncrements} which separate the photocurrent oscillating at twice the mechanical frequency from its DC component (at the given sideband frequency). As before these are approximately normalized and independent of one another (up to $\mcl{O}(\freqmech\delta t)^{-1}$) so we treat them as independent Wiener increments. Making the replacements $\delta t\to \diff t$ and $\delta W_{\alpha}\to \diff W_{\alpha}$ we obtain two coarse-grained master equations depending on the choice of $\Delta_{\mrm{lo}} = \pm \freqmech$.

\begin{figure}
	\centering
	\includegraphics[keepaspectratio=true,width=.8\columnwidth]{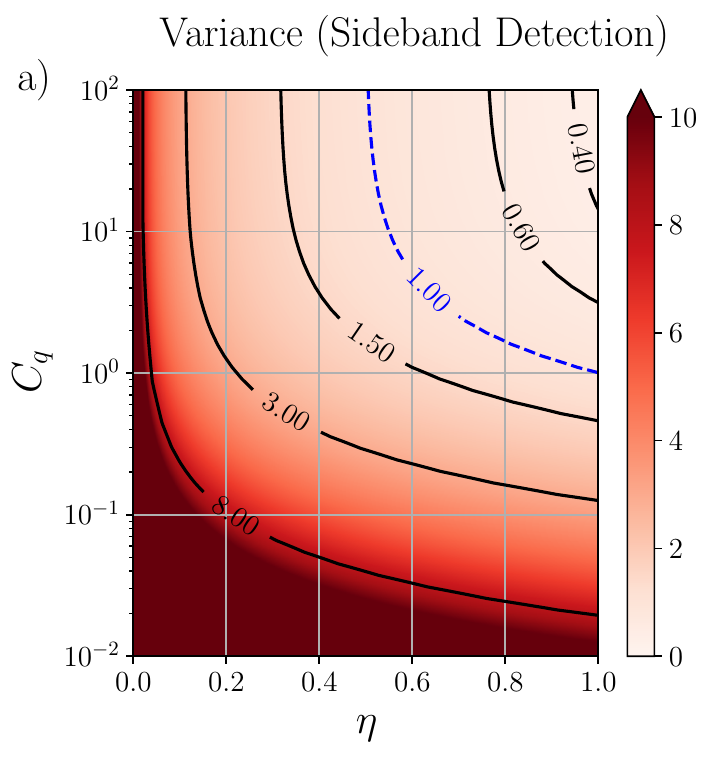}
	\includegraphics[keepaspectratio=true,width=.8\columnwidth]{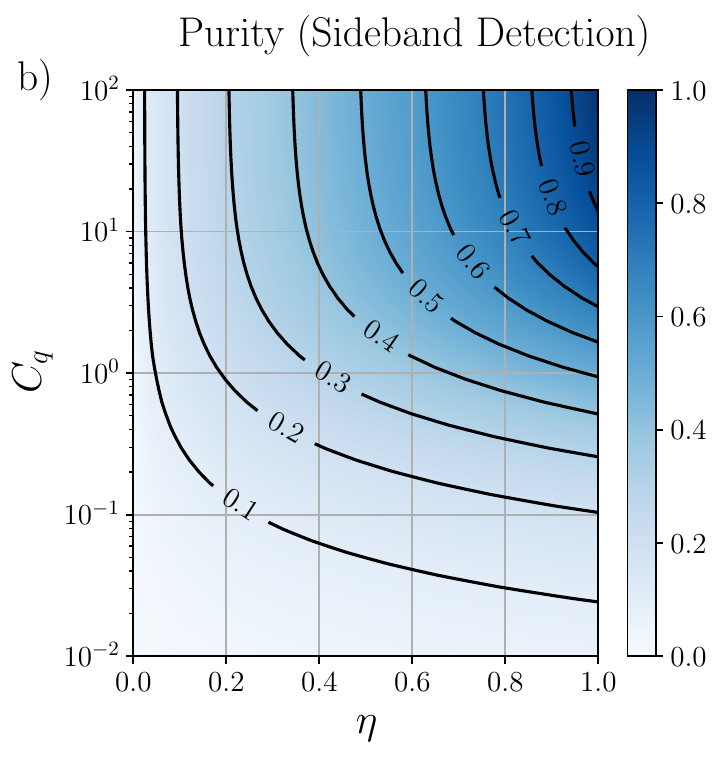}
	\caption{(a) Linear-logarithmic plot of the approximate steady state variances $V_{xx}^{\rho}$ and $V_{xx}^{E}$ from Eqs.~\eqref{eq:ApproximateVarianceOnSideband} and Eq.~\eqref{eq:ApproximateBackwardVarianceOnSideband}, plotted against detection efficiency $\eta$ and quantum cooperativity $C_{q}$ in the limit of large cooperativity $C\gg 1$. The dashed line denotes the shot noise-limited variance of the vacuum state at $V_{xx} = 1.$ (b) The purity of the covariance matrices corresponding to the variances plotted in (a).
	\label{fig:VarianceOnSideband}}
\end{figure}

\subsubsection{Detecting the red sideband}
We first consider the local oscillator tuned to the red sideband, $\Delta_{\mrm{lo}} = -\freqmech$.
This yields the coarse-grained master equation
\begin{align}
	\begin{split}
		\ito\ \diff \oprho(t) & = \mcl{L}_{\mrm{th}}\oprho(t) + \Gamma\mcl{D}[\op{x}]\oprho(t)\diff t + \Gamma\mcl{D}[\op{p}]\oprho(t)\diff t \\
		& \quad + \sqrt{\eta\Gamma}\mcl{H}[\op{a}]\oprho(t)\diff W_{\mrm{c},2}(t) \\
		& \quad - \sqrt{\eta\Gamma}\mcl{H}[i\op{a}]\oprho(t)\diff W_{\mrm{s},2}(t) \\
		& \quad + \sqrt{\eta\Gamma}\mcl{H}[\op{a}^\dag]\oprho(t)\diff W_{0}(t).
	\end{split}
\end{align}
Analogously to the case of resonant detection we can use the Gaussian formalism to compute the conditional steady state variances,
\begin{align}
	V_{xx}^{\rho} & = \frac{1}{3\eta C}\mleft( \sqrt{1 + 4\eta C ( (3-2\eta)C + 3\bar{n} + 2)} - 1 \mright) - \frac{1}{3},\\ 
	V_{pp}^{\rho} & = \frac{1}{\eta C}\mleft( \sqrt{1 + 4\eta C ( C + \bar{n} )} - 1 \mright) + 1, 
\end{align}
which for $C,\bar{n}\gg 1$ become approximately
\begin{align}
	V_{xx}^{\rho} & \approx \frac{2}{3}\sqrt{\frac{(3-2\eta)C_{q} + 3}{\eta C_{q}}} - \frac{1}{3}, \label{eq:ApproximateVarianceOnSideband} \\ 
	V_{pp}^{\rho} & \approx 2\sqrt{\frac{C_{q} + 1}{\eta C_{q}}} + 1. \label{eq:ApproximatePVarianceOnSideband} 
\end{align}

To find the corresponding Gaussian effect operators realizable through retrodiction we could translate the full master equation above to an effect equation and then apply the Gaussian formalism as before. Instead we take the shortcut of directly reading off the Riccati equation Eq.~\eqref{eq:BwdCovarianceEquationOfMotion} from the corresponding Riccati equation of the conditional state. Solving it yields the asymptotic variances
\begin{align}
	V_{xx}^{E} & = \frac{1}{3\eta C}\mleft( \sqrt{1 + 4\eta C ( (3-2\eta)C + 3\bar{n} + 2)} + 1 \mright) + \frac{1}{3} \\ 
	V_{pp}^{E} & = \frac{1}{\eta C}\mleft( \sqrt{1 + 4\eta C ( C + \bar{n} )} + 1 \mright) - 1, 
\end{align}
which for $C,\bar{n}\gg 1$ approach

\begin{align}
	V_{xx}^{E} & \approx \frac{2}{3}\sqrt{\frac{(3-2\eta)C_{q} + 3}{\eta C_{q}}} + \frac{1}{3}, \\ 
	V_{pp}^{E} & \approx 2\sqrt{ \frac{C_{q} + 1}{\eta C_{q}}} - 1. 
\end{align}
Considering the ideal limit $\eta \to 1$ and $C_{q} \to \infty$ we find
\begin{align}
	V_{xx}^{E} & \to 1, & V_{pp}^{E} & \to 1
\end{align}
for the effect operator, so at best we retrodict POVMs that project onto coherent states. On the other hand, we find
\begin{align}
	V_{xx}^{\rho} & \to \frac{1}{3}, & V_{pp}^{\rho} & \to 3
\end{align}
for the conditional steady state, showing that we can in principle prepare squeezed states. Necessary conditions for going below shot noise in the preparation are $C_{q} > 1$ and $\eta > 1/2$ since
\begin{align}
	V_{xx}^{\rho} & < 1 \quad \Leftrightarrow \quad \eta > \frac{C + \bar{n}}{2C} \approx \frac{1}{2}\biggl( 1 + \frac{1}{C_{q}} \biggr),
\end{align}
which is confirmed by the plot of $V_{xx}^{\rho}$ in Fig.~\hyperref[fig:VarianceOnSideband]{\ref{fig:VarianceOnSideband}~(a)}. However, even with one quadrature below shot noise the prepared state will never be entirely pure as seen in Fig.~\hyperref[fig:VarianceOnSideband]{\ref{fig:VarianceOnSideband}~(b)}.

\subsubsection{Detecting the blue sideband}
Tuning the local oscillator to the blue sideband, for $\Delta_{\mrm{lo}} = +\freqmech$, we find the master equation
\begin{align}
	\begin{split}
		\ito\ \diff \oprho(t) & = \mcl{L}_{\mrm{th}}\oprho(t) + \Gamma\mcl{D}[\op{x}]\oprho(t)\diff t + \Gamma\mcl{D}[\op{p}]\oprho(t)\diff t \\
		& \quad + \sqrt{\eta\Gamma}\mcl{H}[\op{a}^\dag]\oprho(t)\diff W_{\mrm{c},2}(t) \\
		& \quad + \sqrt{\eta\Gamma}\mcl{H}[i\op{a}^\dag]\oprho(t)\diff W_{\mrm{s},2}(t) \\
		& \quad + \sqrt{\eta\Gamma}\mcl{H}[\op{a}]\oprho(t)\diff W_{0}(t),
	\end{split}
\end{align}
which asymptotically results in a conditional state with variances
\begin{align}
	V_{xx}^{\rho} & = \frac{1}{3\eta C}\mleft( \sqrt{1 + 4\eta C ( (3-2\eta)C + 3\bar{n} + 1)} - 1 \mright) + \frac{1}{3} \\ 
	V_{pp}^{\rho} & = \frac{1}{\eta C}\mleft( \sqrt{1 + 4\eta C ( C + \bar{n} + 1)} - 1 \mright) - 1, 
\end{align}
which for $C,\bar{n}\gg 1$ become approximately
\begin{align}
	V_{xx}^{\rho} & \approx \frac{1}{3}\sqrt{\frac{(3-2\eta)C_{q} + 3}{\eta C_{q}}} + \frac{1}{6}, \\ 
	V_{pp}^{\rho} & \approx \sqrt{\frac{C_{q} + 1}{\eta C_{q}}} - \frac{1}{2}. 
\end{align}
We see here that in the limit of $\eta \to 1$ and $C_{q}\to \infty$ the variances approach
\begin{align}
	V_{xx}^{\rho} & \to 1, & V_{pp}^{\rho} & \to 1,
\end{align}
so we can at best prepare coherent states.

To find the corresponding effect operators we again translate the forward Riccati equation directly to a corresponding backward equation. This yields the asymptotic variances
\begin{align}
	V_{xx}^{E} & = \frac{1}{3\eta C}\mleft( \sqrt{1 + 4\eta C ( (3-2\eta)C + 3\bar{n} + 1)} + 1 \mright) - \frac{1}{3} \\ 
	V_{pp}^{E} & = \frac{1}{\eta C}\mleft( \sqrt{1 + 4\eta C ( C + \bar{n} + 1)} + 1 \mright) + 1, 
\end{align}
which for $C,\bar{n}\gg 1$ become approximately
\begin{align}
	V_{xx}^{E} & \approx \frac{2}{3} \sqrt{\frac{(3-2\eta)C_{q} + 3}{\eta C_{q}}} - \frac{1}{3}, \label{eq:ApproximateBackwardVarianceOnSideband} \\ 
	V_{pp}^{E} & \approx 2\sqrt{\frac{C_{q}+1}{\eta C_{q}}} + 1. \label{eq:ApproximateBackwardPVarianceOnSideband} 
\end{align}
Considering the ideal limit $\eta \to 1$ and $C_{q} \to \infty$ we see that the asymptotic effect operators can in principle project onto squeezed states,
\begin{align}
	V_{xx}^{E} & \to \frac{1}{3}, & V_{pp}^{E} & \to 3, 
\end{align}
provided $C_{q}>1$ and $\eta > 1/2$ since
\begin{align}
	V_{xx}^{E} < 1 \quad \Leftrightarrow \quad \eta > \frac{C + \bar{n} + 1}{2C} = \frac{1}{2}\biggl( 1 + \frac{1}{C_{q}} \biggr). 
\end{align}
Since both the limiting $\op{x}$-variances Eq.~\eqref{eq:ApproximateVarianceOnSideband} and Eq.~\eqref{eq:ApproximateBackwardVarianceOnSideband} and corresponding $\op{p}$-variances agree, the plots in Fig.~\ref{fig:VarianceOnSideband} also hold for the effect operators retrodicted on the blue sideband.

\begin{center}
\begin{table}
	\begin{tabular}{|c||c|c|}
	\hline
	Detected Sideband	&  Prediction $\oprho$	& Retrodiction $\opE$	\\\hhline{|=#=|=|}
	Blue 	$\Delta_{\mrm{lo}} = \freqmech$							& Coherent				& Squeezed	\\\hline
	Red	$\Delta_{\mrm{lo}} = -\freqmech$							& Squeezed				& Coherent \\\hline
	\end{tabular}
	\caption{Conditional states and retrodictive POVMs generated by resonant drive and homodyne detection of the blue or red sideband.}\label{table2}
	\end{table}
\end{center}

These results are summarized in a Table~\ref{table2}: For large quantum cooperativity and resonant drive, homodyne detection of the blue (red) sideband generates coherent (squeezed) conditional states and squeezed (coherent) retrodictive POVMs. This conforms with the expectation that blue (red) sideband photons have been generated via a beam splitter (two-mode squeezing) interaction, as discussed in Sec.~\ref{sec:OptomechanicalInteraction}. Thus, these to cases perform qualitatively similar to the basic examples studied in Sec.~\ref{sec:DecayingCavity} and Table~\ref{table}. There is, of course, a significant quantitative difference as e.g. the squeezed POVM realized by resonant drive exhibits a noise reduction by $66\%$ only. A perfect quadrature measurement, such as found in Sec.~\ref{sec:DecayingCavity}, would require infinite squeezing. In order to achieve this, the driving field has to be detuned from cavity resonance, as will be discussed next.

\subsection{Off-resonant drive}\label{sec:OffResonantDrive}

Very relevant in experiments is also the case of an off-resonant drive, $\Delta_{\mrm{c}} \neq 0$, for example to perform sideband cooling or to prepare squeezed mechanical states in pulsed schemes \cite{Hofer2017}. Detuning also enables richer retrodictive dynamics since it allows to selectively enhance and suppress the Stokes and anti-Stokes rates $\Gamma_{\pm}$, and thus the BS and TMS components of the optomechanical interaction.

To analyze the effects of non-zero detuning we need to return to the original coarse-grained master equation Eq.~\eqref{eq:AveragedConditionalMasterEquation}. Evaluating the integral over the measurement term for homodyne detection of the carrier or sideband frequencies proceeds analogously to the previous sections. We only need to remember that the sidebands are now located at $\freqcav \pm \freqeff$ with the effective frequency $\freqeff$ from Eq.~\eqref{eq:EffectiveFrequency}. The Stokes and anti-Stokes rates $\Gamma_{\pm}$ from Eq.~\eqref{eq:StokesRates} are no longer equal to a single rate
\begin{align}
	\Gamma & = \frac{g^2 \kappa}{(\kappa/2)^2 + \freqmech^2},
\end{align}
but can be written as
\begin{subequations}
\begin{align}
	\Gamma_{\pm} & = \Gamma f_{\pm},\\
	f_{\pm} & := \frac{1 + 4(\freqmech/\kappa)^2}{1 + 4(-\Delta_{\mrm{c}} \pm \freqmech)^2/\kappa^2}.
\end{align}
\end{subequations}

For a blue-detuned drive, $\Delta_{\mrm{c}} > 0$, such that $\Gamma_{+} \geq \Gamma_{-} + \gamma$ the mechanical dynamics are unstable. Since we are interested in stationary states obtained through continuous driving and observation, we will thus consider only a red-detuned drive, $\Delta_{\mrm{c}} < 0$, in the following. 
We see that with $\Delta_{\mrm{c}}=-\freqmech$ we can enhance $\Gamma_{-}$ by a factor $f_{-} = 1+4(\freqmech/\kappa)^2 > 1$ while suppressing $\Gamma_{+}$ by $f_{+} = (1+4(\freqmech/\kappa)^2)/(1+16(\freqmech/\kappa)^2) < 1$. In the broad cavity regime ($\freqmech/\kappa\ll 1$) this imbalance becomes negligible, so we do not expect any benefit from a detuned drive, but whenever $\freqmech/\kappa > 1$ the enhancement of $\Gamma_{-}$ greatly enhances our ability to retrodict POVMs with sub-shot noise resolution, as we will now show.

\begin{figure}
	\centering
	\includegraphics[keepaspectratio=true,width=\columnwidth]{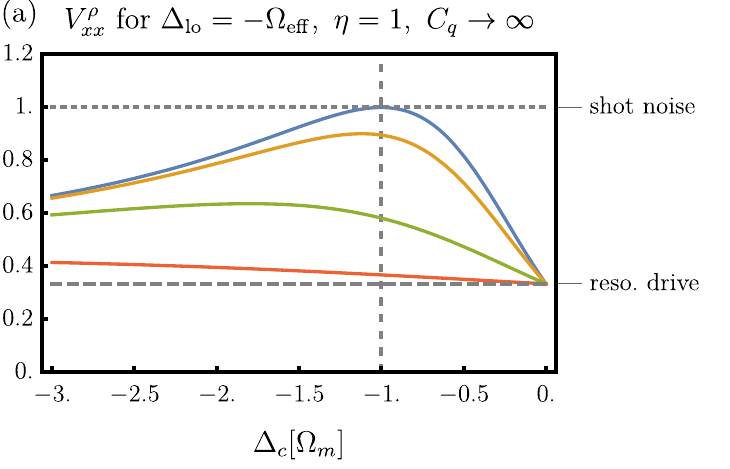}
	\vspace{0.5cm}
	\includegraphics[keepaspectratio=true,width=\columnwidth]{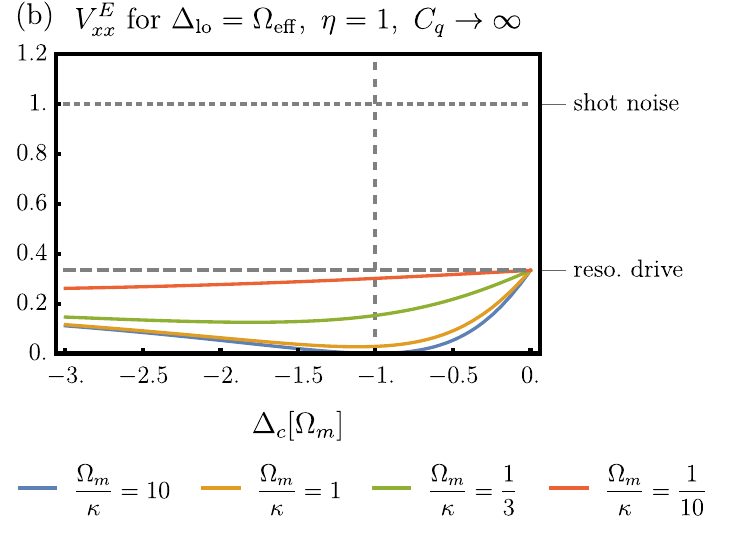}
	\caption{(a) Conditional steady state variance $V_{xx}^{\rho}$ from Eqs.~\eqref{eq:FwdSteadyStateVariance} and (b) asymptotic variance $V_{xx}^{E}$ of retrodicted effect operators from Eqs.~\eqref{eq:BwdAsymptoticVariance} in the ideal limit of $\eta= 1$ and $C_{q}\to \infty$, plotted against detuning of the drive $\Delta_{\mrm{c}}$ in units of the mechanical frequency $\freqmech$. Different curves correspond to different values of $\freqmech/\kappa$ ranging from the broad cavity ($\freqmech/\kappa\ll 1$) to the sideband-resolved regime ($\freqmech/\kappa\gg 1$). The upper dotted line is the shot noise limit, $V_{xx}^{\rho} = V_{xx}^{E} = 1$, 
	and the lower dashed line the limit obtainable with a resonant drive ($V_{xx}^{\rho} = V_{xx}^{E} = 1/3$
	).
	\label{fig:FwdVarianceVSDetuningIdeal}
	}
\end{figure}

Analogously to the previous sections we solve the Riccati equations for the asymptotic covariance matrices of filtered Gaussian states and retrodicted POVM elements. We will only consider $\op{x}$-variances since the results can be applied to any other quadrature by changing the local oscillator phase. Also, since we are interested in fundamental limits we consider only detection of the sidebands that are optimal for preparation and retrodiction, respectively, in the sense that they minimize the stationary variance: the red sideband, $\Delta_{\mrm{lo}} = -\freqeff$, for preparation and the blue sideband, $\Delta_{\mrm{lo}} = \freqeff$, for retrodiction.
The solutions are conveniently expressed in terms of the classical and quantum cooperativities
\begin{align}
	C_{\pm} & := \frac{\Gamma_{\pm}}{\gamma} = C f_{\pm},\\
	C_{q}^{\pm} & := \frac{C_{\pm}}{\bar{n}+1} = C_{q} f_{\pm},
\end{align}
where the ``bare'' cooperativities $C$ and $C_{q}$ are the same as for a resonant drive considered in the previous sections. The solution for a conditional Gaussian steady state prepared by observing the red sideband, $\Delta_{\mrm{lo}} = -\freqeff$, then reads
\begin{subequations}\label{eq:FwdSteadyStateVariance}
\begin{align}
	\begin{split}
		V_{xx}^{\rho} & = \frac{1}{\eta(C_{-} + 2C_{+})}\biggl( {- 1} - (1-\eta)C_{-}\\
		& \qquad\qquad\qquad\qquad\quad + (1-2\eta)C_{+} + \sqrt{r} \biggr), 
	\end{split}\\
	\begin{split}
		r & := (C_{-} - C_{+} + 1)^2 + 4\eta (3-2\eta)C_{-}C_{+}\\
		& \qquad  + 8\eta C_{+}(\bar{n} + 1) + 4 \bar{n}\eta C_{-}.
	\end{split}
\end{align}
\end{subequations}
Here we see that in the broad cavity regime, $\freqmech/\kappa \ll 1$, where $C_{-}\approx C_{+} \approx C$, the variance is just given by what one finds by driving on resonance. Thus the minimal variance obtained for $\eta = 1$ and $C_{q}\to\infty$ will be given by $V_{xx}^{\rho} \to 1/3 < 1$, 
and thus corresponds to a squeezed state. On the other hand, when $\freqmech/\kappa > 1$ we find that $C_{q}^{-} \gg C_{q}^{+}$, and $V_{xx}^{\rho} \to 1$ 
so we can at best prepare coherent states. The effect of different cavity linewidths is also depicted in Fig.~\ref{fig:FwdVarianceVSDetuningIdeal}, where we see that a red-detuned drive does not help preparation as expected.

We can compare these results to the asymptotic variance of a Gaussian effect operator retrodicted by observing the blue sideband, $\Delta_{\mrm{lo}} = \freqeff$, which reads
\begin{subequations}\label{eq:BwdAsymptoticVariance}
\begin{align}
\begin{split}
	V_{xx}^{E} & = \frac{1}{\eta(2C_{-} + C_{+})}\biggl( 1 - (1-\eta)C_{+} \\ 
	& \qquad\qquad\qquad\qquad\quad + (1-2\eta)C_{-} + \sqrt{s} \biggr),
\end{split}\\
\begin{split}
	s & := (C_{-} - C_{+} + 1)^2 + 4\eta (3-2\eta)C_{-}C_{+}\\
	& \qquad + 4\eta C_{+}(\bar{n} + 1) + 8 \bar{n}\eta C_{-}
\end{split}
\end{align}
\end{subequations}

\begin{figure}
	\centering
	\includegraphics[keepaspectratio=true,width=\columnwidth]{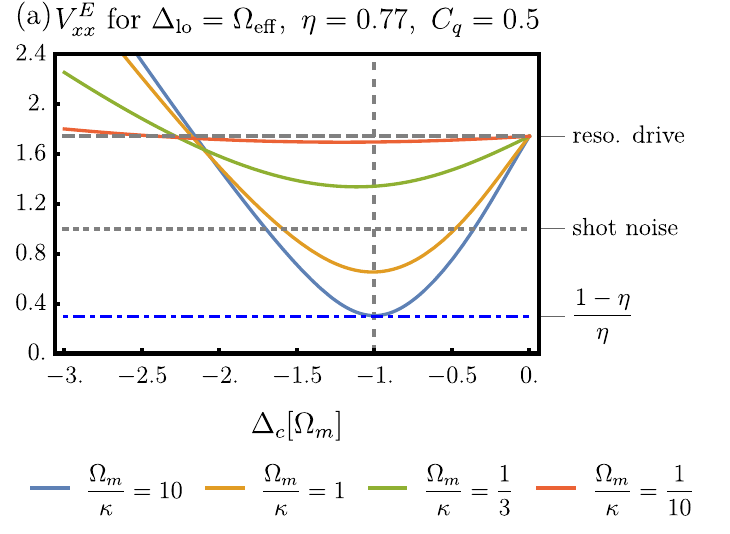}
	\vspace{.5cm}
	\includegraphics[keepaspectratio=true,width=\columnwidth]{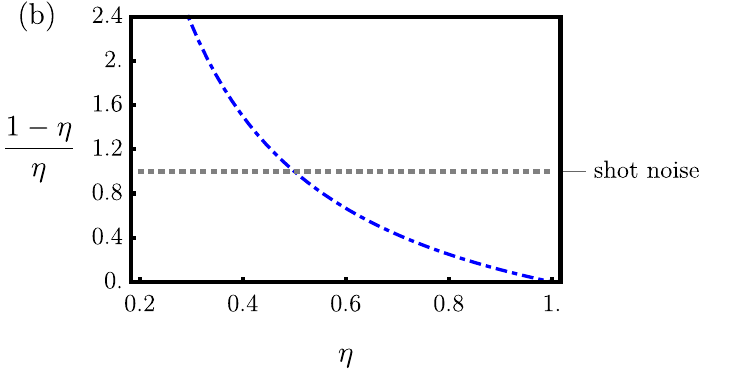}
	\caption{
	(a) The asymptotic variance $V_{xx}^{E}$ of retrodicted effect operators from Eqs.~\eqref{eq:BwdAsymptoticVariance} plotted against detuning of the drive $\Delta_{\mrm{c}}$ in units of the mechanical frequency $\freqmech$. Instead of the ideal limit $\eta = 1$ and $C_{q}\to\infty$ used to create Fig.~\hyperref[fig:FwdVarianceVSDetuningIdeal]{\ref{fig:FwdVarianceVSDetuningIdeal}~(b)} we consider a realistic efficiency $\eta = 0.77$ and sub-unit cooperativity $C_{q} = 1/2$. As the cavity enters the sideband-resolved regime where $\freqmech/\kappa > 1$  the variance approaches $V_{xx}^{E} \to (1-\eta)/\eta$ (dash-dotted line). 
	(b) This limiting value of the variance is plotted in the bottom figure against $\eta$.
	\label{fig:VarianceVSDetuningRealistic}
	}
\end{figure}

Here we find that to retrodict POVMs with sub-shot noise resolution, $V_{xx}^{E} < 1$, 
the detection efficiency must satisfy
\begin{align}
	\eta > \frac{1}{2}\biggl(1 + \frac{1}{C_{q}^{-}}\biggr),
\end{align}
and thus necessarily $\eta > 1/2$, but also $C_{q}^{-} > 1$. This is interesting because it means that with detuning $\Delta_{\mrm{c}} = -\freqmech$ we no longer require a large ``bare'' cooperativity $C_{q}>1$ to measure with sub-shot noise resolution, but only a large product $C_{q}(1+4(\freqmech/\kappa)^2) > 1$, which can be rewritten as
\begin{align}
	\biggl(\frac{\freqmech}{\kappa}\biggr)^2 & > \frac{1-C_{q}}{4C_{q}}.
\end{align}
Thus in the sideband-resolved regime a detuned drive allows to retrodict POVMs that beat the shot noise limit even for sub-unit quantum cooperativities. In fact, whenever $\freqmech/\kappa\gg 1$ such that $C_{q}^{-} \gg 1$ and $C_{q}^{-}\gg C_{q}^{+}$, the minimal variance will approach
\begin{align}
	V_{xx}^{E} & \to \frac{1-\eta}{\eta} 
\end{align}
as can also be seen in Fig.~\ref{fig:VarianceVSDetuningRealistic}, where we plot the achievable variances for conservative values of $C_{q} = 1/2$ and $\eta = 0.77$.
These results show that with an off-resonant (red-detuned) drive, and using only continuous measurements, it is possible to measure with sub-shot noise variance limited only by the detection efficiency $\eta$.

In summary, it is possible to perform quadrature measurements of the mechanical state with sub-shot noise variance through continuous monitoring of the cavity output. By using a red-detuned cavity drive and sufficiently efficient homodyne detection of the blue sideband of the output one achieves a squeezed retrodictive POVM realizing a quadrature measurement for the past mechanical state. In the resolved sideband limit, the quality of the quadrature measurement is essentially limited by the detection efficiency only and does not require a quantum cooperativity larger than one.


\section{Conclusion \& Outlook}\label{sec:Conclusion}

We have given here a self-contained introduction to the theory of retrodictive POVMs, demonstrating the potential to retrieve information about the initial quantum state of a system based on the outcomes of a continuous measurement process. The general formalism has been illustrated in detail for linear quantum systems and applied to realistic models of optomechanical systems. 

The application of our theoretical framework to optomechanics has revealed promising avenues for achieving retrodictive state analysis. By characterizing achievable retrodictive POVMs in various optomechanical operating modes, such as resonant and off-resonant driving fields, we have illustrated the potential for precise retrodictive measurements of mechanical oscillators. Notably, our findings unveil the possibility of nearly ideal quadrature measurements, offering direct access to the position or momentum distribution of mechanical oscillators at specific time instances. This advancement opens doors to novel possibilities in quantum state tomography, also of Non-Gaussian states, albeit with the caveat of being inherently destructive.

We hope that this presentation will facilitate and advance the use of retrodictive POVMs also in other linear quantum systems beyond optomechanics. Extending the formalism to more complex and nonlinear systems presents an intriguing challenge. As quantum technology continues to advance, the insights gained from this work will contribute to the expanding toolkit of quantum state analysis and manipulation.

We thank Klaus M\o lmer, Albert Schließer, Stefan Danilishin, Sebastian Hofer, David Reeb, Reinhard Werner und Lars Dammeier for discussions on this topic. We acknowledge funding  by the Deutsche Forschungsgemeinschaft (DFG, German Research Foundation)
through Project-ID 274200144 – SFB 1227 (projects A06) and Project-ID 390837967 - EXC 2123.

\appendix


\section{Derivation of the effect equation}
\label{sec:Appendix:DerivationEffectEquation}

\subsection{General retrodiction}
We now illustrate the claim made in Sec.~\ref{sec:BackwardEffectEquation} that effect operators are themselves dynamical objects \cite{Barnett2000,Barnett2001,Pegg2002,Tsang2009a,Tsang2010,Tsang2010a,Gammelmark2013,Zhang2017} whose time evolution is determined by that of the quantum system. Assume the system evolves according to some completely positive map $\tilde{\mcl{N}}$ \cite{Nielsen2010} from $t_{0}$ to $t$ (\eg, the solution to a master equation) such that
\begin{align}
	\tilde{\oprho}(t) & = \tilde{\mcl{N}}_{t_{0},t}[\oprho(t_{0})].
\end{align}
This may include closed unitary as well as open dissipative evolution. 
Performing continuous measurements leads to a conditional map,
\begin{align}
	\tilde{\oprho}_{\mcl{I}}(t) & = \tilde{\mcl{N}}_{t_{0},t,\mcl{I}}[\oprho(t_{0})],
\end{align}
which depends on the particular measurement record $\mcl{I}:=\{ Y(s),t_{0}\leq s < t \}$ obtained during a single run of the experiment. 
Now we let the state evolve further until some time $t_{1}>t$ so we find analogously
\begin{align}\label{eq:ConditionalFinalState}
	\tilde{\oprho}_{\mcl{I},\mcl{J}}(t_{1}) & = \tilde{\mcl{N}}_{t,t_{1},\mcl{J}}[\tilde{\oprho}_{\mcl{I}}(t)]\\
	& \equiv \tilde{\oprho}_{\mcl{Y}}(t_{1}),
\end{align}
with records $\mcl{J}:=\{ Y(s),t\leq s < t_{1} \}$ and $\mcl{Y} := \mcl{I}\cup\mcl{J} = \{ Y(s),t_{0} \leq s < t_{1} \}$. Performing a \textit{positive-operator valued measure (POVM)} measurement $\{\op{E}_{x}|x\in \mcl{X}\}$ with effect operators $\opE_{x}$ at time $t_{1}$ we expect outcome $x$ to occur with probability
\begin{align}
	P(x|\tilde{\oprho}_{\mcl{Y}}(t_{1})) & = \trace\{\opE_{x} \tilde{\oprho}_{\mcl{Y}}(t_{1})\}.
\end{align}
Plugging \eqref{eq:ConditionalFinalState} into this expression, and writing $\opE_{x}\equiv \opE_{x}(t_{1})$, yields
\begin{align}
	\trace\{\opE_{x}(t_{1}) \tilde{\oprho}_{\mcl{Y}}(t_{1})\} & = \trace\{\opE_{x}(t_{1}) \tilde{\oprho}_{\mcl{I},\mcl{J}}(t_{1})\} \\
	& = \trace\{\opE_{x}(t_{1}) \tilde{\mcl{N}}_{t,t_{1},\mcl{J}}[\tilde{\oprho}_{\mcl{I}}(t)]\}\\
	& = \trace\{\tilde{\mcl{N}}_{t,t_{1},\mcl{J}}^\dag[\opE_{x}(t_{1})] \tilde{\oprho}_{\mcl{I}}(t) \}\\
	& = \trace\{\opE_{x,\mcl{J}}(t) \tilde{\oprho}_{\mcl{I}}(t) \}, \label{eq:Appendix:ConditionalProbability}
\end{align}
where $\tilde{\mcl{N}}^\dag$ denotes the Hilbert-Schmidt adjoint of $\tilde{\mcl{N}}$, and $\opE_{x,\mcl{J}}(t) := \tilde{\mcl{N}}_{t,t_{1},\mcl{J}}^\dag[\opE_{x}(t_{1})]$. This shows that a POVM $\{\op{E}_{x}(t_{1})|x\in \mcl{X}\}$ at $t_{1}$ is equivalent to a different POVM $\{\opE_{x,\mcl{J}}(t)|x\in \mcl{X},\mcl{J}\in\mathfrak{J}\}$ at previous time $t$, where $\mathfrak{J}$ denotes the set of all possible observation records $\mcl{J}$ from $t$ to $t_{1}$. The dynamics of individual effect operators are given by 
\begin{align}
	\opE_{x,\mcl{J}}(t) & = \tilde{\mcl{N}}_{t,t_{1},\mcl{J}}^\dag[\opE_{x}(t_{1})].
\end{align}
This POVM backpropagation is what we call retrodiction. One could now use the first part $\mcl{I}$ of the whole record $\mcl{Y}$ to obtain $\tilde{\oprho}_{\mcl{I}}(t)$ through filtering, and then use the second part $\mcl{J}$ from $t$ to $t_{1}$ to effect a POVM measurement of $\opE_{x,\mcl{J}}$ on $\tilde{\oprho}_{\mcl{I}}(t)$, performing state preparation and verification with the same setup, but using disjoint sets of data. Of particular relevance is thus the case where no additional measurement is performed at the final time $t_{1}$ so we start with the trivial POVM $\opE(t_{1})=\op{1}$. In that case the retrodicted effect operators at $t$ will depend entirely on the continuous observations.

For simplicity we drop the subscripts $\mcl{I}$ and $\mcl{J}$ and remember that $\tilde{\oprho}$ and $\opE$ depend on respective parts of the measurement record.

\subsection{Conditional master equation}
We now consider the special case of a system governed by conditional master equation \eqref{eq:ItoSimpleMasterEquation},
\begin{align}
	\begin{split} \label{eq:Appendix:ItoSimpleMasterEquation}
		\ito\ \diff \tilde{\oprho} (t) & = -i[\op{H},\tilde{\oprho}(t)]\diff t \Dop{\op{L}}\tilde{\oprho}(t)\diff t \\
		& \qquad + \bigl(\op{C}\tilde{\oprho}(t) + \tilde{\oprho}(t)\op{C}\bigr) \diff Y(t).
	\end{split}
\end{align}
To derive the effect equation adjoint to this master equation consider again Eq.~\eqref{eq:Appendix:ConditionalProbability} for the probability to measure some particular value $x$ given a conditional state $\tilde{\oprho}(t_{1})$,
\begin{align}
	\trace\{\opE_{x} \tilde{\oprho}(t_{1})\} & = \trace\{\opE_{x}(t) \tilde{\oprho}(t)\}.
\end{align}
Obviously the left-hand side does not depend on the arbitrary parameter $t$, so when we take a variation with respect to $t$ \cite{Gammelmark2013} we find
\begin{align}
	0 & = \diff_t \trace\{\opE_{x}(t) \tilde{\oprho}(t) \} \\
	& = \trace\{\opE_{x}(t+\diff t) \tilde{\oprho}(t+\diff t) - \opE_{x}(t) \tilde{\oprho}(t) \}.
\end{align}
We know that $\tilde{\oprho}(t+\diff t) = \tilde{\oprho}(t) + \diff\tilde{\oprho}(t)$ with $\diff \tilde{\oprho}(t)$ given by Eq.~\eqref{eq:Appendix:ItoSimpleMasterEquation}, and we similarly assume we can write $\opE_{x}(t) = \opE_{x}(t+\diff t) - \diff \opE_{x}(t+\diff t)$. We can determine $\diff \opE_{x}(t+\diff t)$ from inserting these relation into the equation above,
\begin{align}
	0 & = \trace\{\opE_{x}(t+\diff t) \diff \tilde{\oprho}(t) + \diff\opE_{x}(t+\diff t) \tilde{\oprho}(t)\}.
\end{align}
Looking at the first term in conjunction with \eqref{eq:Appendix:ItoSimpleMasterEquation}, and suppressing the time-dependence of $\opE_{x}$ for the moment, we see the trace decomposes into three parts,
\begin{align}
	& \trace\{\opE_{x} \diff \tilde{\oprho}(t) \} = (A)\diff t + (B)\diff t + (C)\diff Y(t),
\end{align}
for the Hamiltonian, jump, and measurement operators respectively. For example
\begin{align}
	(A) & = -i\trace\{\opE_{x} [\op{H},\tilde{\oprho}(t)] \} \\
	& = -i\trace\{\opE_{x} \bigl( \op{H} \tilde{\oprho}(t) - \tilde{\oprho}(t)\op{H} \bigr) \} \\
	& = -i\trace\{\opE_{x} \op{H} \tilde{\oprho}(t) - \op{H} \opE_{x} \tilde{\oprho}(t) \} \\
	& = i\trace\{[\op{H},\opE_{x}] \tilde{\oprho}(t) \},
\end{align}
where from the second to third line we made use of the cyclic property of the trace. Similarly we find for the jump operator with $\Dop{\op{L}}\oprho = \op{L} \oprho \op{L}^\dag - (\op{L}^\dag \op{L}\oprho + \oprho \op{L}^\dag \op{L})/2$ that
\begin{align}
	(B) & = \trace\{\opE_{x} \left(\Dop{\op{L}}\tilde{\oprho}(t)\right) \} \\
	& = \trace\left\{\left(\Dopdag{\op{L}}\opE_{x}\right) \tilde{\oprho}(t) \right\}, 
\end{align}
with $\Dopdag{\op{L}}\opE = \op{L}^\dag \opE \op{L} - (\op{L}^\dag \op{L}\opE + \opE \op{L}^\dag \op{L})/2$, and for the measurement term
\begin{align}
	(C) & = \trace\{\opE_{x} \bigl(\op{C}\tilde{\oprho}(t) + \tilde{\oprho}(t)\op{C}\bigr) \}\\
	& = \trace\{\bigl(\op{C}^\dag\opE_{x} + \opE_{x}\op{C}\bigr) \tilde{\oprho}(t) \}.
\end{align}
Putting all three contributions together we find
\begin{align}
\begin{split}
	0 = \trace\Bigl\{\Bigl( & \diff\opE_{x}
	+ i[\op{H},\opE_{x}]\diff t
	+ \Dopdag{\op{L}}\opE_{x}\diff t\\
	& + \bigl(\op{C}^\dag\opE_{x} + \opE_{x}\op{C}\bigr) \diff Y(t)
	\Bigr) \tilde{\oprho}(t) \Bigr\}.
\end{split}
\end{align}
Note that we did not specify $\tilde{\oprho}$ so this equation has to hold for arbitrary density operators. Recalling that $\opE_{x}\equiv \opE_{x}(t+\diff t)$ we shift the time argument and conclude that
\begin{align}
	\begin{split}
		{-\diff}\opE_{x}(t) & = i[\op{H},\opE_{x}(t)]\diff t	+ \Dopdag{\op{L}}\opE_{x}(t)\diff t\\
		& \qquad + \bigl(\op{C}^\dag\opE_{x}(t) + \opE_{x}(t)\op{C}\bigr) \diff Y(t-\diff t).
	\end{split}
\end{align}

Our derivation swept a few things under the rug as the unusual argument of $\diff Y(t-\diff t)$ suggests. In particular, we started with a stochastic It\^{o} equation denoted by the $\ito$ in Eq.~\eqref{eq:Appendix:ItoSimpleMasterEquation}, but we did not say how to interpret the effect equation we just obtained. It turns out that it is a stochastic \textit{backward It\^{o} equation}, which is explained in the next section. This means when expressing the integral as a Riemann sum the stochastic increment needs to be evaluated at the \textit{upper} limit of each subinterval.


\section{Forward and backward It\^{o} integration}
\label{sec:Appendix:BackwardItoIntegration}
Deterministic integrals of integrable functions defined as limits of Riemann-Stieltjes sums do not depend on whether their integrands are evaluated at the lower or upper end of their subintervals. The upper and lower Riemann sums converge in the limit of vanishing subinterval length. For stochastic integrals this is not the case as the integrand may fluctuate rapidly \cite{Mikosch1998,Gardiner2009,Gardiner2010}. Thus there are different ways to integrate a random process depending on where one evaluates the integrand.

The type mostly used in this article is the It\^{o} stochastic integral with the integrand evaluated at the lower end of each subinterval. Another well-known type is the Stratonovich integral with evaluation performed at the mid-point \cite{Mikosch1998,Gardiner2009}. A third, lesser known type is the backward It\^{o} integral with the integrand evaluated at the upper limit of each subinterval \cite{Kuznetsov2017}.

We denote It\^{o} integrals and differentials by a prepended $\ito$ and backward It\^{o} integrals by $\bito$. Everything to the right of $\ito$ (or $\bito$) is an It\^{o} (or backward It\^{o}) integral. It is possible to mix different integral types \cite{Kuznetsov2017}. But since we never do this it should always be clear which type of integral is being used.

To clarify the distinction between $\ito$ and $\bito$ let us recall the definition of It\^{o} integrals \cite{Mikosch1998,Gardiner2009,Wiseman2010,Jacobs2014}. Consider an interval $[t_0,t_{1}]$ and partitions $P_{n}=\{\tau_{j}:j=0,\dots,n\}$ such that
\begin{align}
	t_{0} = \tau_{0} < \tau_{1} < \dots < \tau_{n} = t_{1}.
\end{align}
We consider only sequences of $P_n$ such that
\begin{align}
	\mrm{mesh}(P_{n}) & := \max_{j=1,\dots,n} (\tau_{j} - \tau_{j-1}) \to 0
\end{align}
as $n\to \infty$. Provided it exists and is independent of the partition sequence, the It\^{o} integral of some function (or stochastic process) $f(t)$ with respect to a white noise process $W_{t}(\equiv W(t))$ is defined as the mean-square limit
\begin{align}
	\ito\int_{t_{0}}^{t_{1}} f(\tau)\diff W_{\tau} := \lim_{n\to\infty} \sum_{j=1}^{n} f(\tau_{j-1}) (W_{\tau_{j}} - W_{\tau_{j-1}}).
\end{align}
The corresponding backward It\^{o} integral is defined as
\begin{align}
	\bito\int_{t_{0}}^{t_{1}} f(\tau)\diff W_{\tau} := \lim_{n\to\infty} \sum_{j=1}^{n} f(\tau_{j}) (W_{\tau_{j}} - W_{\tau_{j-1}}).
\end{align}
Given a stochastic process $X(t)$ with It\^{o} equation 
\begin{align}
	\ito\ \diff X_{t} & = f(X_{t},t) \diff t + g(X_{t},t)\diff W_{t},
\end{align}
the corresponding backward It\^{o} equation reads
\begin{align}
	\begin{split}
		\bito\ \diff X_{t} & = \left(f(X_{t},t) - g(X_{t},t) g'(X_{t},t) \right)\diff t + g(X_{t},t)\diff W_{t}
	\end{split}
\end{align}
with $g'(X_t,t) = (\partial g(x,t)/\partial x)|_{x=X_{t}}$.

We need backward It\^{o} integrals because they naturally govern the evolution of the effect operator. Let us illustrate this by considering a simple It\^{o} stochastic process driven only by white noise,
\begin{align}\label{eq:BackwardItoIntegration:ItoEquation}
	\ito\ \diff X(t) & = M_{t}X(t)\diff W_{t},
\end{align}
with Wiener increment $\diff W_{t}$ and some linear map $M_{t}$ acting on $X(t)$. Integrating both sides we obtain an equivalent integro-differential equation,
\begin{align}
	X(t) & = X(t_{0}) + \ito\int_{t_{0}}^{t} M_{\tau}X(\tau) \diff W_{\tau}.
\end{align}
If we continue replacing $X(\tau)$ on the right-hand side by this expression we find
\begin{align}
	X(t) & = \sum_{n=0}^{\infty} \mcl{M}_{t,t_{0}}^{(n)} X(t_{0})
\end{align}
with the operators $\mcl{M}_{t,t_{0}}^{(n)}$ defined recursively via
\begin{align}
	\mcl{M}_{t,t_{0}}^{(0)} & = \mathbb{1},\\
	\mcl{M}_{t,t_{0}}^{(n)} & = \ito\int_{t_{0}}^{t} M_{\tau} \mcl{M}_{\tau,t_{0}}^{(n-1)} \diff W_{\tau}.
\end{align}
We retrieve the It\^{o} equation of $X(t)$ through variation with respect to $t$ since
\begin{align}
	\diff_{t} \mcl{M}_{t,t_{0}}^{(0)} & = 0,\\
	\begin{split}
		\diff_{t} \mcl{M}_{t,t_{0}}^{(n)} & = \ito\ \diff_{t}\int_{t_{0}}^{t} M_{\tau_{1}} \diff W_{\tau_{1}} \int_{t_{0}}^{\tau_{1}} M_{\tau_{2}} \diff W_{\tau_{2}} \dots\\
		& \qquad\qquad \dots \int_{t_{0}}^{\tau_{n-1}} M_{\tau_{n}} \diff W_{\tau_{n}}
	\end{split} \\
	\begin{split}
		& = \ito\ M_{t} \diff W_{t} \int_{t_{0}}^{t} M_{\tau_{2}} \diff W_{\tau_{2}} \dots\\
		& \qquad\qquad \dots \int_{t_{0}}^{\tau_{n-1}} M_{\tau_{n}} \diff W_{\tau_{n}}
	\end{split} \\
	& = \ito\ M_{t} \mcl{M}_{t,t_{0}}^{(n-1)} \diff W_{t},
\end{align}
so as expected we find
\begin{align}
	\diff_{t} X(t) & = \sum_{n=0}^{\infty} \diff_{t}\mcl{M}_{t,t_{0}}^{(n)} X(t_{0})\\
	& = \ito\ M_{t} \sum_{n=1}^{\infty} \mcl{M}_{t,t_{0}}^{(n-1)} X(t_{0}) \diff W_{t}\\
	& = \ito\ M_{t} X(t) \diff W_{t}.
\end{align}
Now consider a second process $E(t)$ which starts at $t_1>t$ and evolves as the adjoint of $X(t)$, such that
\begin{align}
	\braket{E(t_{1})}{X(t_{1})} & = \braket{E(t_{1})}{\sum_{n=0}^{\infty} \mcl{M}_{t_{1},t_{0}}^{(n)} X(t_{0})}\\
	& = \braket{\sum_{n=0}^{\infty} (\mcl{M}_{t_{1},t_{0}}^{(n)})^\dag E(t_{1})}{ X(t_{0})}\\
	& = \braket{E(t_{0})}{ X(t_{0})}
\end{align}
independent of $t_{0}$, so
\begin{align}
	E(t) & = \sum_{n=0}^{\infty} (\mcl{M}_{t_{1},t}^{(n)})^\dag E(t_{1}).
\end{align}
To obtain a differential equation analogous to \eqref{eq:BackwardItoIntegration:ItoEquation} for $E(t)$ we need to take a derivative with respect to $t$. It is not immediately clear how to do this from
\begin{align}
	\begin{split}
		(\mcl{M}_{t_{1},t}^{(n)})^\dag E(t_{1}) & = \ito\int_{t}^{t_{1}} \diff W_{\tau_{1}} \int_{t}^{\tau_{1}}\diff W_{\tau_{2}} \dots\\
		& \quad\dots \int_{t}^{\tau_{n-1}} \diff W_{\tau_{n}} M_{\tau_{n}}^\dag \dots M_{\tau_{2}}^\dag M_{\tau_{1}}^\dag E(t_{1}),
	\end{split}
\end{align}
since $t$ appears in every integral.

If the integrals were regular deterministic integrals we could simply re-order the integration boundaries. For example, for an integrable deterministic function $f(t_1,t_2)$ one finds
\begin{align}
	\int_{t}^{t_{1}} \diff \tau_{1} \int_{t}^{\tau_{1}} \diff \tau_{2} f(\tau_{1}, \tau_{2}) & = \int_{t}^{t_{1}} \diff \tau_{2} \int_{\tau_{2}}^{t_{1}} \diff \tau_{2} f(\tau_{1}, \tau_{2}).
\end{align}
An equivalent result for stochastic integrals was proven by Kuznetsov \cite[Ch.~7]{Kuznetsov2017}.
He showed that one can swap the order of integration provided one simultaneously changes from regular It\^{o} to backward It\^{o} integrals, so proceeding inductively we find

\begin{align*}
		&(\mcl{M}_{t_{1},t}^{(n)})^\dag E(t_{1})  =\\
		&=\ito\int_{t}^{t_{1}} \diff W_{\tau_{1}} \int_{t}^{\tau_{1}}\diff W_{\tau_{2}} \dots \\
		&\hspace{3cm}\cdots\int_{t}^{\tau_{n-1}} \diff W_{\tau_{n}} M_{\tau_{n}}^\dag \dots M_{\tau_{2}}^\dag M_{\tau_{1}}^\dag E(t_{1}) \nonumber\\
		& = \bito\int_{t}^{t_{1}} \diff W_{\tau_{n}} \dots \\
		&\hspace{1.9cm}\cdots\int_{\tau_{3}}^{t_{1}}\diff W_{\tau_{2}} \int_{\tau_{2}}^{t_{1}} \diff W_{\tau_{1}} M_{\tau_{n}}^\dag \dots M_{\tau_{2}}^\dag M_{\tau_{1}}^\dag E(t_{1})\\
		& = \bito\int_{t}^{t_{1}} M_{\tau_{n}}^\dag \diff W_{\tau_{n}} \dots \int_{\tau_{3}}^{t_{1}} M_{\tau_{2}}^\dag \diff W_{\tau_{2}} \int_{\tau_{2}}^{t_{1}} M_{\tau_{1}}^\dag \diff W_{\tau_{1}} E(t_{1})\\
		& = \bito\int_{t}^{t_{1}} M_{\tau}^{\dag} (\mcl{M}_{t_{1},\tau}^{(n-1)})^\dag \diff W_{\tau} E(t_{1}),
\end{align*}

where all nested integrals in the last three lines are backward It\^{o} integrals. Taking a variation with respect to the lower limit of an integral yields a negative sign, so we find
\begin{align}
	{-\diff}_{t} (\mcl{M}_{t_{1},t}^{(n)})^\dag & = \bito\ M_{t}^{\dag} (\mcl{M}_{t_{1},t}^{(n-1)})^\dag \diff W_{t},
\end{align}
and consequently
\begin{align}
	{-\diff}_{t} E(t) & = \bito\ M_{t}^{\dag} E(t) \diff W_{t}.
\end{align}


\section{Cumulant equations of motion}
\label{sec:Appendix:CumulantEoMs}

We will now derive the cumulant equations of motion for the quantum state of an arbitrary linear system as described in Secs.~\ref{sec:StatePreparation} and \ref{sec:StateVerification}. We write $N$\textsuperscript{th}-order cumulants, \ie, of products of $N$ canonical operators, as
\begin{align}
	\cmlnt_{m_{1},\dots,m_{N}}^{(N)} & := \langle \op{r}_{m_{1}}\dots \op{r}_{m_{N}} \rangle_{\rho}^{\mrm{c}}
\end{align}
where the superscript $\mrm{c}$ stands for cumulant, and the indices $1\leq m_{k} \leq 2M$ indicate that operator $\op{r}_{m_{k}}$ is at position $k$ in the cumulant. We consider only symmetrically ordered cumulants, so the operator position does not matter, and the $V$ are symmetric under permutations of indices. These general cumulants relate to the means and covariance matrix from the main text as
\begin{align}
	r_{j} & = \langle \op{r}_{j} \rangle_{\rho} = \cmlnt_{j}^{(1)},\\
	V_{jk} & = \langle \{ \op{r}_{j} - r_{j}, \op{r}_{k} - r_{k} \} \rangle_{\rho} = 2\cmlnt_{jk}^{(2)}. 
\end{align}
If one is only interested in Gaussian states all cumulants of order $N\geq 3$ vanish identically. But for any non-Gaussian state \textit{all} cumulants have to be taken into account \cite{Ivan2012}.

Cumulants are obtained from the quantum characteristic function $\chi(\vect{\xi})$, which is defined as the expectation value of the symmetric Weyl operator $\weyl(\vect{\xi}) = \exp(-i\vect{\xi}^\transpose \mat{\sigma} \vect{\op{r}})$ \cite{Eisert2005,Wang2008,Heinosaari2010},
\begin{align}\label{eq:CharacteristicFunctionDefinitionAppendix}
	\chi(\vect{\xi}) := \langle \weyl(\vect{\xi}) \rangle_{\rho} = \trace\{\weyl(\vect{\xi})\oprho\},
\end{align}
with phase space variables $\vect{\xi}\in\mathbb{R}^{2M}$. The matrix $\mat{\sigma}$ comprises the canonical commutation relations \eqref{eq:SigmaDefinition}.
Let us introduce the twisted derivative
\begin{align}
	\tilde{\partial}_{j} & := \sum_{k=1}^{2M} \sigma_{jk} \frac{\partial}{\partial\xi_{k}}, &
	\vect{\tilde{\nabla}} & := \mat{\sigma}\vect{\nabla}, \label{eq:TwistedDerivative}
\end{align}
then $\chi$ serves as cumulant-generating function via
\begin{subequations}\label{eq:CumulantsInTermsOfChi}
\begin{align}
	\cmlnt_{m_1,\dots,m_{N}}^{(N)} & = (-i\tilde{\partial}_{m_{1}})\dots (-i\tilde{\partial}_{m_{N}}) \ln[\chi(\vect{\xi})]|_{\vect{\xi}=0}\\
	& = (-i\tilde{\partial}_{m_{1}})\dots (-i\tilde{\partial}_{m_{N}}) G(\vect{\xi})|_{\vect{\xi}=0},
\end{align}
\end{subequations}
with $G:=\ln[\chi]$ or $\chi = \exp(G)$. The usual approach \cite[Appendix 12]{Barnett1997} is to translate the master (or effect) equation into partial differential equations for the cumulants via \eqref{eq:CumulantsInTermsOfChi}.

\subsection{From operator to partial differential equation}
To find the correspondence between quantum and partial differential operators, consider the action of $\vect{\tilde{\nabla}}$ on the Weyl operator. We single out one $\xi_{i}$,
\begin{align}
	-i\vect{\xi}^\transpose \mat{\sigma} \vect{\op{r}} & = -i\xi_{i}\sum_{k}\sigma_{ik}\op{r}_{k} -i\sum_{\substack{j,k\\ j\neq i}}\xi_{j}\sigma_{jk}\op{r}_{k} =: \op{A}_{i} + \op{B}_{\neq i}
\end{align}
so we can write
\begin{align}
	\weyl(\vect{\xi}) & := \exp\mleft( \op{A}_{i} + \op{B}_{\neq i} \mright)\\
	& = \exp\mleft( \op{A}_{i} \mright) \exp\mleft(\op{B}_{\neq i} \mright) \exp\mleft( -\frac{1}{2}[\op{A}_{i},\op{B}_{\neq i}] \mright)\\
	& = \exp\mleft(\op{B}_{\neq i} \mright) \exp\mleft( \op{A}_{i} \mright) \exp\mleft( +\frac{1}{2}[\op{A}_{i},\op{B}_{\neq i}] \mright).
\end{align}
We find
\begin{align}
	[\op{A}_{i},\op{B}_{\neq i}] & = -i\xi_{i}\sum_{\substack{k\\ k\neq i}}\sigma_{ik} \xi_{k} = -i\xi_{i}(\mat{\sigma}\vect{\xi})_{i},
\end{align}
where we used that $\mat{\sigma}$ is skew-symmetric and thus zero on the diagonal.
Using this result to apply $\vect{\tilde{\nabla}}$ to $\weyl$ yields
\begin{align}
	-i\mat{\sigma}\vect{\nabla}\weyl(\vect{\xi}) & = \mleft(\vect{\op{r}} - \frac{1}{2}\vect{\xi}\mright) \weyl(\vect{\xi}) = \weyl(\vect{\xi}) \mleft(\vect{\op{r}} + \frac{1}{2}\vect{\xi}\mright),
\end{align}
This gives us the important relations
\begin{subequations}
\begin{align}
	\vect{\op{r}} \weyl(\vect{\xi}) & = \mleft(-i\mat{\sigma}\vect{\nabla} + \frac{1}{2}\vect{\xi}\mright) \weyl(\vect{\xi}),\\
	\weyl(\vect{\xi}) \vect{\op{r}} & = \mleft(-i\mat{\sigma}\vect{\nabla} - \frac{1}{2}\vect{\xi}\mright) \weyl(\vect{\xi}),
\end{align}
\end{subequations}
from which we read off the replacement rules
\begin{subequations}\label{eq:ReplacementRules}
\begin{align}
	\oprho\vect{\op{r}} & \to \mleft(-i\vect{\tilde{\nabla}} + \frac{1}{2}\vect{\xi}\mright) \chi(\vect{\xi}),\\
	\vect{\op{r}}\oprho & \to \mleft(-i\vect{\tilde{\nabla}} - \frac{1}{2}\vect{\xi}\mright) \chi(\vect{\xi}).
\end{align}
\end{subequations}
The following combinations frequently appear in master equations
\begin{align}
	[\vect{\op{r}},\oprho] & \to -\vect{\xi}\chi(\vect{\xi}), & [\vect{\op{r}}^\transpose,\oprho] & \to -\vect{\xi}^\transpose\chi(\vect{\xi}), \\
	\{\vect{\op{r}},\oprho\} & \to -2i\vect{\tilde{\nabla}}\chi(\vect{\xi}), & \{\vect{\op{r}}^\transpose,\oprho\} & \to -2i\vect{\tilde{\nabla}}^\transpose\chi(\vect{\xi}).
\end{align}

\subsection{Hamiltonian}
For a Hamiltonian with quadratic and linear terms,
\begin{align}
	\op{H} & = \frac{1}{2}\vect{\op{r}}^\transpose \mat{H} \vect{\op{r}} + \vect{h}^\transpose \vect{\op{r}}
\end{align}
with symmetric $\mat{H}\in\mathbb{R}^{2M\times 2M}$ and $\vect{h}\in\mathbb{R}^{2M}$ we find
\begin{align}
	-i[\op{H},\oprho] & = -i\mleft(\frac{1}{2}\vect{\op{r}}^\transpose\mat{H}[\vect{\op{r}},\oprho] + \frac{1}{2}[\vect{\op{r}}^\transpose,\oprho]\mat{H}\vect{\op{r}} + \vect{h}^\transpose[\vect{\op{r}},\oprho]\mright).
\end{align}
The operator nature of $\vect{\op{r}}$ and $\oprho$ prohibits us from commuting them, but otherwise we can treat $\mat{H}$ as a matrix, $\vect{\op{r}}$ as a vector, and $\oprho$ as a scalar, which greatly simplifies the notation.

Using the replacement rules from the previous section we obtain
\begin{align}
	\begin{split}
		[\dot{\chi}]_{H} & = -\frac{i}{2}\mleft(-i\vect{\tilde{\nabla}} - \frac{1}{2}\vect{\xi}\mright)^\transpose\mat{H}(-\vect{\xi}) \\
		& \quad - \frac{i}{2}(-\vect{\xi})^\transpose\mat{H}\mleft(-i\vect{\tilde{\nabla}} + \frac{1}{2}\vect{\xi}\mright) - i\vect{h}^\transpose(-\vect{\xi}) \chi(\vect{\xi})
	\end{split}\\
	& = \frac{1}{2}\mleft(\vect{\tilde{\nabla}}^\transpose\mat{H}\vect{\xi} + \vect{\xi}^\transpose\mat{H}\vect{\tilde{\nabla}} + 2i\vect{h}^\transpose\vect{\xi}\mright) \chi(\vect{\xi}).
\end{align}
Note that the first term requires use of the product rule since $\vect{\tilde{\nabla}}$ acts on both $\vect{\xi}$ and $\chi$, so to make this explicit we write
\begin{align}\label{eq:ProductRule}
	\mleft(\vect{\tilde{\nabla}}^\transpose\mat{H}\vect{\xi}\mright) \chi(\vect{\xi}) & = \chi(\vect{\xi}) \mleft(\vect{\tilde{\nabla}}^\transpose\mat{H}\vect{\xi}\mright) + \vect{\xi}^\transpose\mat{H}(\vect{\tilde{\nabla}} \chi(\vect{\xi})).
\end{align}
Here we see
\begin{align}
	\vect{\tilde{\nabla}}^\transpose\mat{H}\vect{\xi} & = \sum_{j,k}\partial_{j} (\mat{\sigma}^\transpose \mat{H})_{jk} \xi_{k} = \sum_{j}(\mat{\sigma}^\transpose \mat{H})_{jj}\\
	& = \trace\mleft[ \mat{\sigma}^\transpose\mat{H} \mright] = 0,
\end{align}
so the final expression reads
\begin{align}
	-i[\op{H},\oprho] & \to \mleft(\vect{\xi}^\transpose\mat{H}\vect{\tilde{\nabla}} + i\vect{h}^\transpose\vect{\xi}\mright) \chi(\vect{\xi}).
\end{align}

To obtain the equations of motion of the cumulants, one has to take the derivative of Eq.~\eqref{eq:CumulantsInTermsOfChi}, so
\begin{align}
	& [\diff \cmlnt_{m_1,\dots,m_{N}}^{(N)}]_{H} = (-i\tilde{\partial}_{m_{1}})\dots (-i\tilde{\partial}_{m_{N}}) \diff \ln[\chi(\vect{\xi})]|_{\vect{\xi}=0}\\
	&\ \ = (-i\tilde{\partial}_{m_{1}})\dots (-i\tilde{\partial}_{m_{N}}) \frac{1}{\chi}\diff\chi|_{\vect{\xi}=0}\\
	\begin{split}
		&\ \ = (-i\tilde{\partial}_{m_{1}})\dots (-i\tilde{\partial}_{m_{N}})\times \\
		&\qquad\qquad \times \bigl[ \expo{-G(\vect{\xi})}\bigl(\vect{\xi}^\transpose\mat{H}\vect{\tilde{\nabla}} + i\vect{h}^\transpose\vect{\xi}\bigr) \expo{G(\vect{\xi})}\bigr]|_{\vect{\xi}=0}\diff t
	\end{split}\\
	\begin{split}
		&\ \ = (-i\tilde{\partial}_{m_{1}})\dots (-i\tilde{\partial}_{m_{N}}) \times\\
		&\qquad\qquad \times \bigl[\vect{\xi}^\transpose\mat{H}(\vect{\tilde{\nabla}}G(\vect{\xi})) + i\vect{h}^\transpose\vect{\xi}\bigr] |_{\vect{\xi}=0}\diff t.
	\end{split}
\end{align}
The twisted derivatives act on $\vect{\xi}$ as $\tilde{\partial}_{j} \xi_{k} = \sum\nolimits_{l} \sigma_{jl}\partial_{l}\xi_{k} = \sigma_{jk}$.
We thus find, using the Einstein sum convention,
\begin{align}\label{eq:HamiltonianCumulantContribution}
\begin{split}
	[\diff \cmlnt_{m_1,\dots,m_{N}}^{(N)}]_{H} & = \sum_{\tau\in \mcl{S}^{\text{cycl}}(N)}(\mat{\sigma}\mat{H})^{\phantom{(N)}}_{m_{\tau(1)},k} \cmlnt_{k,m_{\tau(2)},\dots,m_{\tau(N)}}^{(N)}\diff t \\
	& \qquad + (\mat{\sigma}\vect{h})_{m_1}\delta_{N,1}\diff t
\end{split}
\end{align}
where $\tau$ runs through all cyclic permutations of $1,\dots,N$.

\subsection{Lindblad operators}
For linear jump operators $\vect{\op{L}} = \mat{\Lambda}\vect{\op{r}}$ with $\mat{\Lambda}^\dag \mat{\Lambda} =: \mat{\Delta} + i\mat{\Omega}$ we find the Lindblad operators
\begin{align}
	\sum_{j} \Dop{\op{L}_{j}} \oprho & = \sum_{j,k,l}\Lambda_{jk}^* \Lambda_{jl} \mleft( \op{r}_{l}\oprho\op{r}_{k} - \frac{1}{2}\{\op{r}_{k}\op{r}_{l},\oprho\} \mright)\\
	& = \vect{\op{r}}^\transpose \oprho (\mat{\Lambda}^\dagger\mat{\Lambda})^\transpose\vect{\op{r}} - \frac{1}{2}\{ \vect{\op{r}}^\transpose \mat{\Lambda}^\dagger\mat{\Lambda} \vect{\op{r}},\oprho \}\\
	\begin{split}
		& = \frac{1}{2}\vect{\op{r}}^\transpose\mat{\Delta} [\oprho ,\vect{\op{r}}] + \frac{1}{2}[\vect{\op{r}}^\transpose,\oprho]\mat{\Delta}\vect{\op{r}} \\
		& \quad - \frac{i}{2}\mleft( \vect{\op{r}}^\transpose\mat{\Omega} \{\oprho ,\vect{\op{r}}\} + \{\vect{\op{r}}^\transpose,\oprho\}\mat{\Omega} \vect{\op{r}} \mright)
	\end{split}
\end{align}
which becomes
\begin{widetext}
\begin{align}
	\begin{split}
		[\dot{\chi}]_{L} & = \mleft[ \frac{1}{2}\mleft(-i\vect{\tilde{\nabla}} - \frac{1}{2}\vect{\xi}\mright)^\transpose\mat{\Delta} \vect{\xi} + \frac{1}{2}(-\vect{\xi})^\transpose\mat{\Delta}\mleft(-i\vect{\tilde{\nabla}} + \frac{1}{2}\vect{\xi}\mright) - \mleft(-i\vect{\tilde{\nabla}} - \frac{1}{2}\vect{\xi}\mright)^\transpose\mat{\Omega} \vect{\tilde{\nabla}} - \vect{\tilde{\nabla}}^\transpose\mat{\Omega} \mleft(-i\vect{\tilde{\nabla}} + \frac{1}{2}\vect{\xi}\mright) \mright] \chi(\vect{\xi})
	\end{split}\\
	\begin{split}
		& = \mleft[ -\frac{1}{2}\vect{\xi}^\transpose\mat{\Delta}\vect{\xi} - \frac{i}{2}\vect{\tilde{\nabla}}^\transpose\mat{\Delta}\vect{\xi} + \frac{i}{2}\vect{\xi}^\transpose\mat{\Delta}\vect{\tilde{\nabla}} + 2i\vect{\tilde{\nabla}}^\transpose\mat{\Omega}\vect{\tilde{\nabla}} + \frac{1}{2}\vect{\xi}^\transpose\mat{\Omega}\vect{\tilde{\nabla}} - \frac{1}{2}\vect{\tilde{\nabla}}^\transpose\mat{\Omega}\vect{\xi} \mright]\chi(\vect{\xi})
	\end{split}\\
	& = \mleft[-\frac{1}{2} \vect{\xi}^\transpose\mat{\Delta}\vect{\xi} + \vect{\xi}^\transpose\mat{\Omega}\vect{\tilde{\nabla}} + \frac{1}{2}\trace[\mat{\sigma}\mat{\Omega}]\mright]\chi(\vect{\xi})
\end{align}
where we used the product rule as in \eqref{eq:ProductRule} to get
\begin{align}\label{eq:ProductRuleOmega}
	\vect{\tilde{\nabla}}^\transpose\mat{\Omega}\vect{\xi} & = - \vect{\xi}^\transpose\mat{\Omega}\vect{\tilde{\nabla}} + \trace[\mat{\sigma}\mat{\Omega}].
\end{align}
We also used that $\mat{\Delta}$ is a symmetric matrix so $-\vect{\tilde{\nabla}}^\transpose\mat{\Delta}\vect{\xi} + \vect{\xi}^\transpose\mat{\Delta}\vect{\tilde{\nabla}} = \trace[\mat{\sigma}\mat{\Delta}] = 0$, and $\mat{\Omega}$ is skew-symmetric so $\vect{v}^\transpose\mat{\Omega}\vect{v}=0$ for any vector $\vect{v}$; in particular
\begin{align}\label{eq:SkewSymmetricScalarProduct}
	\vect{\tilde{\nabla}}^\transpose\mat{\Omega}\vect{\tilde{\nabla}} & = 0.
\end{align}
The contribution to the cumulant evolution will be given by
\begin{align}
	\begin{split}
		[\diff & \cmlnt_{m_1,\dots,m_{N}}^{(N)}]_{L} = (-i\tilde{\partial}_{m_{1}})\dots (-i\tilde{\partial}_{m_{N}}) \expo{-G(\vect{\xi})} \mleft[-\frac{1}{2} \vect{\xi}^\transpose\mat{\Delta}\vect{\xi} + \vect{\xi}^\transpose\mat{\Omega}\vect{\tilde{\nabla}} + \frac{1}{2}\trace[\mat{\sigma}\mat{\Omega}]\mright]\expo{G(\vect{\xi})}\diff t 
	\end{split}\\
	\begin{split}
		& = (-i\tilde{\partial}_{m_{1}})\dots (-i\tilde{\partial}_{m_{N}}) \mleft[ -\frac{1}{2} \vect{\xi}^\transpose\mat{\Delta}\vect{\xi} + (\vect{\xi}^\transpose\mat{\Omega}\vect{\tilde{\nabla}}G(\vect{\xi})) + \frac{1}{2}\trace[\mat{\sigma}\mat{\Omega}]\mright] |_{\vect{\xi}=0}\diff t
	\end{split}\\
	\begin{split}
		& = (\mat{\sigma}\mat{\Delta}\mat{\sigma}^\transpose)_{m_1 m_2}\delta_{N,2} \diff t + \frac{1}{2}\trace[\mat{\sigma}\mat{\Omega}]\delta_{N,0} \diff t + \sum_{\tau\in \mcl{S}^{\text{cycl}}(N)} (\mat{\sigma}\mat{\Omega})_{m_{\tau(1)},k}\cmlnt_{k,m_{\tau(2)},\dots,m_{\tau(N)}}^{(N)} \diff t.
	\end{split}
\end{align}

\subsection{Measurement terms}
We assume linear measurement operators, $\vect{\op{C}} = (\mat{A}+i\mat{B})\vect{\op{r}}$, that generate terms
\begin{align}
		\ito\ \sum_{k} & (\op{C}_{k}\oprho + \oprho\op{C}_{k}^\dag)\diff Y_{k} 
		 = (\vect{\op{C}}\oprho + \oprho\vect{\op{C}^\dag})^\transpose\diff \vect{Y}
	 = \mleft(\mat{A}(\vect{\op{r}}\oprho + \oprho\vect{\op{r}}) + i \mat{B}(\vect{\op{r}}\oprho - \oprho\vect{\op{r}})\mright)^\transpose\diff \vect{Y}
	 = \mleft(\mat{A}\{\vect{\op{r}},\oprho\} + i \mat{B}[\vect{\op{r}},\oprho] \mright)^\transpose\diff \vect{Y},
\end{align}
which yield a contribution
\begin{align}
	\ito\  [\diff \chi]_{C} & = \mleft(-2i\mat{A}\vect{\tilde{\nabla}} - i \mat{B}\vect{\xi} \mright)^\transpose\diff \vect{Y} \chi(\vect{\xi}).
\end{align}
Note that due to the stochastic nature of the It\^{o} increment we need to implement the It\^{o} table $\diff Y_{j}\diff Y_{k} = \delta_{jk}\diff t$ so we keep terms of up to second order,
\begin{align}
	\ito\  [\diff G]_{C} & = [\diff \ln[\chi]]_{C}
	= \frac{1}{\chi}[\diff \chi]_{C} + \frac{1}{2}\mleft(-\frac{1}{\chi^2}\mright)[\diff \chi]_{C}^2
	=: [\diff G]_{Y} + [\diff G]_{\text{It\^{o}}}.
\end{align}
The linear term yields
\begin{align}
	\ito\  [\diff G]_{Y} & = \expo{-G(\vect{\xi})} \mleft(-2i\mat{A}\vect{\tilde{\nabla}} - i \mat{B}\vect{\xi} \mright)^\transpose\expo{G(\vect{\xi})} \diff \vect{Y}
	 = \mleft(-2i\mat{A}\vect{\tilde{\nabla}}G(\vect{\xi}) - i \mat{B}\vect{\xi} \mright)^\transpose \diff \vect{Y},
\end{align}
which contributes to the cumulants
\begin{align}
	\begin{split}
		\ito\  [\diff \cmlnt_{m_1,\dots,m_{N}}^{(N)}]_{Y} &= (-i\tilde{\partial}_{m_{1}})\dots (-i\tilde{\partial}_{m_{N}}) 
		\times \mleft(-2i\mat{A}\vect{\tilde{\nabla}}G(\vect{\xi}) - i \mat{B}\vect{\xi} \mright)^\transpose \diff \vect{Y}|_{\vect{\xi}=0}
	\end{split} \\
	& = \mleft( 2A_{jk}\cmlnt_{k,m_1,\dots,m_N}^{(N+1)} - B_{jk}\sigma_{m_1 k}\delta_{N,1} \mright) \diff Y_{j} \\
	& = \mleft( 2\cmlnt_{m_1,\dots,m_N,k}^{(N+1)}\mat{A}_{kj}^\transpose - (\mat{\sigma}\mat{B}^\transpose)_{m_1 j}\delta_{N,1} \mright) \diff Y_{j}.
\end{align}
If we replace the measurement signal with a Wiener increment by extracting the mean,
\begin{align}
	\ito\  \diff \vect{Y} & = \langle \vect{\op{C}} + \vect{\op{C}}^\dag \rangle \diff t + \diff \vect{W}
	 = 2 \mat{A} \vect{r} \diff t + \diff \vect{W}
	 = 2 \mat{A} \mat{\cmlnt}^{(1)} \diff t + \diff \vect{W} \label{eq:ExtractingMeanFromWiener}
\end{align}
we find
\begin{align}\label{eq:MeasurementTermWienerIncrement}
	\begin{split}
		\ito\ [\diff \cmlnt_{m_1,\dots,m_{N}}^{(N)}]_{Y} = \mleft( 2\cmlnt_{m_1,\dots,m_N,k}^{(N+1)}\mat{A}_{kj}^\transpose - (\mat{\sigma}\mat{B}^\transpose)_{m_1 j}\delta_{N,1} \mright)\diff W_{j} + 2\mleft( 2\cmlnt_{m_1,\dots,m_N,k}^{(N+1)}(\mat{A}^\transpose\mat{A}\vect{r})_{k} - (\mat{\sigma}\mat{B}^\transpose\mat{A}\vect{r})_{m_1} \delta_{N,1} \mright) \diff t.
	\end{split}
\end{align}

The quadratic It\^{o} correction contributes terms
\begin{align}
	\begin{split}
		[\diff G]_{\text{It\^{o}}} & = -\frac{1}{2}\mleft(-2i\mat{A}\vect{\tilde{\nabla}}G(\vect{\xi}) - i \mat{B}\vect{\xi} \mright)^\transpose \mleft(-2i\mat{A}\vect{\tilde{\nabla}}G(\vect{\xi}) - i \mat{B}\vect{\xi} \mright)\diff t
	\end{split}\\
	\begin{split}
		& = \frac{1}{2}\Bigl(4 (\vect{\tilde{\nabla}}G(\vect{\xi}))^\transpose \mat{A}^\transpose\mat{A}(\vect{\tilde{\nabla}}G(\vect{\xi})) + \vect{\xi}^\transpose \mat{B}^\transpose\mat{B}\vect{\xi} + 4 \vect{\xi}^\transpose\mat{B}^\transpose\mat{A}(\vect{\tilde{\nabla}}G(\vect{\xi})) \Bigl)\diff t,
	\end{split}
\end{align}
where the derivatives act only on the $G$ right next to them. Evaluating the last two terms is straightforward and one finds
\begin{align}
	(-i\tilde{\partial}_{m_{1}})\dots (-i\tilde{\partial}_{m_{N}}) [\vect{\xi}^\transpose \mat{B}^\transpose\mat{B}\vect{\xi}]|_{\vect{\xi}=0} = -2(\mat{\sigma}\mat{B}^\transpose\mat{B}\mat{\sigma}^\transpose)_{m_1 m_2}\delta_{N,2},
\end{align}
and
\begin{align}
	(-i\tilde{\partial}_{m_{1}})\dots (-i\tilde{\partial}_{m_{N}}) [\vect{\xi}^\transpose\mat{B}^\transpose\mat{A}(\vect{\tilde{\nabla}}G(\vect{\xi}))]|_{\vect{\xi}=0} = \sum_{\tau\in \mcl{S}^{\text{cycl}}(N)} (\mat{\sigma}\mat{B}^\transpose\mat{A})_{m_{\tau(1)},k}\cmlnt_{k,m_{\tau(2)},\dots,m_{\tau(N)}}^{(N)}.
\end{align}
To compute the remaining term we have to apply the product rule multiple times. This results in a sum over all possible sequences of derivatives acting either on the right or on the left $\vect{\tilde{\nabla}}G$, so
\begin{align}
\begin{split}
	& (-i\tilde{\partial}_{m_{1}})\dots (-i\tilde{\partial}_{m_{N}}) 2[(\vect{\tilde{\nabla}}G(\vect{\xi}))^\transpose \mat{A}^\transpose\mat{A}(\vect{\tilde{\nabla}}G(\vect{\xi}))]_{\vect{\xi}=0} = -2\sum_{n=0}^{N}\sum_{\sigma\in S(N)} \cmlnt_{m_{\sigma(1)},\dots,m_{\sigma(n)},j}^{(n+1)} (\mat{A}^\transpose\mat{A})_{jk} \cmlnt_{k,m_{\sigma(n+1)},\dots,m_{\sigma(N)}}^{(N-n+1)}.
\end{split}
\end{align}
where $\sigma\in S(N)$ runs through all possible permutations of $\{1,\dots,N\}$, not just cyclic ones. We extract the terms with $n=0$ and $n=N$ from the sum which each contain only one permutation,
\begin{align}
\begin{split}
	& - 2\mleft( \cmlnt_{j}^{(1)} (\mat{A}^\transpose\mat{A})_{jk} \cmlnt_{k,m_{1},\dots,m_{N}}^{(N+1)} + \cmlnt_{m_{1},\dots,m_{N},j}^{(N+1)} (\mat{A}^\transpose\mat{A})_{jk} \cmlnt_{k}^{(1)} \mright) = -4 \cmlnt_{m_{1},\dots,m_{N},k}^{(N+1)} (\mat{A}^\transpose\mat{A}\vect{r})_{k},
\end{split}
\end{align}
and we see that this cancels exactly a term from \eqref{eq:MeasurementTermWienerIncrement}.
Combining the remaining measurement contributions we find
\begin{align}
	\begin{split}
		\ito\ & [\diff \cmlnt_{m_1,\dots,m_{N}}^{(N)}]_{C} = \mleft( 2\cmlnt_{m_1,\dots,m_N,k}^{(N+1)}\mat{A}_{kj}^\transpose - (\mat{\sigma}\mat{B}^\transpose)_{m_1 j}\delta_{N,1} \mright)\diff W_{j} - 2(\mat{\sigma}\mat{B}^\transpose\mat{A}\vect{r})_{m_1} \delta_{N,1} \diff t - (\mat{\sigma}\mat{B}^\transpose\mat{B}\mat{\sigma}^\transpose)_{m_1 m_2}\delta_{N,2}\diff t\\
		& \quad + 2\sum_{\mathclap{\tau\in \mcl{S}^{\text{cycl}}(N)}} (\mat{\sigma}\mat{B}^\transpose\mat{A})_{m_{\tau(1)},k}\cmlnt_{k,m_{\tau(2)},\dots,m_{\tau(N)}}^{(N)} \diff t - 2\sum_{n=1}^{N-1}\sum_{\sigma\in S(N)} \cmlnt_{m_{\sigma(1)},\dots,m_{\sigma(n)},j}^{(n+1)} (\mat{A}^\transpose\mat{A})_{jk} \cmlnt_{k,m_{\sigma(n+1)},\dots,m_{\sigma(N)}}^{(N-n+1)}.
	\end{split}
\end{align}

\subsection{Combined evolution}\label{sec:EvolutionOfGeneralQuantumStates}
We put all terms from the previous sections together and find for cumulants of first order
\begin{align}
	\begin{split}
		\ito\  \diff r_{m} = \diff \cmlnt_{m}^{(1)} & = \mat{Q}_{m,k} \cmlnt_{k}^{(1)}\diff t + 2\cmlnt_{m,k}^{(2)}(\mat{A}^\transpose\diff\vect{W})_{k} + (\mat{\sigma}\vect{h}\diff t - \mat{\sigma}\mat{B}^\transpose\diff\vect{W})_{m}
	\end{split}
\end{align}
with implicit sums over the repeated index $k$, and
\begin{align}
	\mat{Q} & := \mat{\sigma}(\mat{H} + \mat{\Omega}), &
	\mat{\tilde{\Delta}} & := \mat{\Delta} - \mat{B}^\transpose\mat{B}.
\end{align}
In vector form and with $\mat{V}_{\rho} = 2\cmlnt_{\rho}^{(2)}$ 
these equations read
\begin{align}
	 \ito\  \diff\vect{r}_{\rho} & = \mat{Q}\vect{r}_{\rho}\diff t + \mat{\sigma}\vect{h}\diff t + (2\cmlnt_{\rho}^{(2)}\mat{A}^\transpose - \mat{\sigma}\mat{B}^\transpose)\diff \vect{W} 
	  = \mat{M}_{\rho}\vect{r}_{\rho}\diff t + \mat{\sigma}\vect{h}\diff t + (2\cmlnt_{\rho}^{(2)}\mat{A}^\transpose - \mat{\sigma}\mat{B}^\transpose)\diff \vect{Y}, 
\end{align}
where we replaced the Wiener increment by the measurement current $\diff \vect{Y}$ as in Eq.~\eqref{eq:ExtractingMeanFromWiener}, and introduced the drift matrix from Eq.~\eqref{eq:FwdConditionalDriftMatrix},
\begin{align}
	\mat{M}_{\rho}(t) & = \mat{Q}+2\mat{\sigma}\mat{B}^\transpose\mat{A} - 4\cmlnt_{\rho}^{(2)}(t)\mat{A}^\transpose\mat{A}. 
\end{align}
At second order we find for the (co)variances
\begin{align}
	\begin{split}
		\ito\  \diff \cmlnt_{m_1,m_2}^{(2)} & = \sum_{\tau\in \mcl{S}^{\text{cycl}}(2)}(\mat{Q} + 2\mat{\sigma}\mat{B}^\transpose\mat{A})_{m_{\tau(1)},k} \cmlnt_{k,m_{\tau(2)}}^{(2)}\diff t + (\mat{\sigma}\mat{\tilde{\Delta}}\mat{\sigma}^\transpose)_{m_1 m_2}\diff t + 2\cmlnt_{m_1,m_2,k}^{(3)}(\mat{A}^\transpose\diff\vect{W})_{k}\\
		& \quad - 2\sum_{\sigma\in \mcl{S}(2)} \cmlnt_{m_{\sigma(1)},j}^{(2)} (\mat{A}^\transpose\mat{A})_{jk} \cmlnt_{k,m_{\sigma(2)}}^{(2)} \diff t
	\end{split}\\
	\begin{split}
		& = (\mat{Q} + 2\mat{\sigma}\mat{B}^\transpose\mat{A})_{m_1,k} \cmlnt_{k,m_2}^{(2)}\diff t + (\mat{Q} + 2\mat{\sigma}\mat{B}^\transpose\mat{A})_{m_2,k} \cmlnt_{k,m_1}^{(2)}\diff t + (\mat{\sigma}\mat{\tilde{\Delta}}\mat{\sigma}^\transpose)_{m_1 m_2}\diff t \\
		& \quad + 2\cmlnt_{m_1,m_2,k}^{(3)}(\mat{A}^\transpose\diff\vect{W})_{k} - 2 \cmlnt_{m_{1},j}^{(2)} (\mat{A}^\transpose\mat{A})_{jk} \cmlnt_{k,m_{2}}^{(2)} \diff t - 2 \cmlnt_{m_{2},j}^{(2)} (\mat{A}^\transpose\mat{A})_{jk} \cmlnt_{k,m_{1}}^{(2)} \diff t,
	\end{split}
\end{align}
which can be written more concisely as
\begin{align}\label{eq:ConditionalSecondOrderEoMs}
	\begin{split}
		\ito\  \diff\mat{\cmlnt}^{(2)}_{m_{1},m_{2}} & = \Bigl[ \mat{M}_{\rho}\mat{\cmlnt}_{\rho}^{(2)} +  \mat{\cmlnt}_{\rho}^{(2)}\mat{M}_{\rho}^\transpose + \mat{\sigma}\mat{\tilde{\Delta}}\mat{\sigma}^\transpose + 4\mat{\cmlnt}_{\rho}^{(2)}\mat{A}^\transpose\mat{A}\mat{\cmlnt}_{\rho}^{(2)}\Bigr]_{m_{1},m_{2}} \diff t  + 2\cmlnt_{m_{1},m_{2},k}^{(3)}(\mat{A}^\transpose\diff\vect{W})_{k},
	\end{split}
\end{align}
with an implicit sum over $k$ in the last term, turning $\mat{\cmlnt}_{\rho}^{(3)}$ from a tensor into a matrix with the free indices $m_{1}$, $m_{2}$. The evolution of cumulants of order $N\geq 3$ is given by
\begin{align}
	\begin{split}
		\ito\  \diff \cmlnt_{m_1,\dots,m_{N}}^{(N)} & = \sum_{\tau\in \mcl{S}^{\text{cycl}}(N)}(\mat{Q} + 2\mat{\sigma}\mat{B}^\transpose\mat{A})_{m_{\tau(1)},k} \cmlnt_{k,m_{\tau(2)},\dots,m_{\tau(N)}}^{(N)}\diff t \\
		& \quad - 2\sum_{n=1}^{N-1}\sum_{\sigma\in S(N)} \cmlnt_{m_{\sigma(1)},\dots,m_{\sigma(n)},j}^{(n+1)} (\mat{A}^\transpose\mat{A})_{jk} \cmlnt_{k,m_{\sigma(n+1)},\dots,m_{\sigma(N)}}^{(N-n+1)} \diff t + 2\cmlnt_{m_1,\dots,m_N,k}^{(N+1)}(\mat{A}^\transpose\diff\vect{W})_{k}
	\end{split}\\
	\begin{split}\label{eq:HigherOrderCumulantsEoMs}
		& = \sum_{\tau\in \mcl{S}^{\text{cycl}}(N)}M_{m_{\tau(1)},k}^{\rho}(t) \cmlnt_{k,m_{\tau(2)},\dots,m_{\tau(N)}}^{(N)}\diff t\\
		& \quad - 2\sum_{n=2}^{N-2}\sum_{\sigma\in S(N)} \cmlnt_{m_{\sigma(1)},\dots,m_{\sigma(n)},j}^{(n+1)} (\mat{A}^\transpose\mat{A})_{jk} \cmlnt_{k,m_{\sigma(n+1)},\dots,m_{\sigma(N)}}^{(N-n+1)} \diff t + 2\cmlnt_{m_1,\dots,m_N,k}^{(N+1)}(\mat{A}^\transpose\diff\vect{W})_{k},
	\end{split}
\end{align}
where the terms with $n=1$ and $n=N-1$ were used to introduce the drift matrix $\mat{M}_{\rho}(t)$. It is worthy to note that generally all cumulants couple to the next higher order through the stochastic term.

\subsection{Stability analysis}
\label{sec:Appendix:StabilityAnalysis}
A common restriction in stable linear systems is to consider only Gaussian states, which collapse to a steady state with fixed covariance matrix (second cumulants) and bounded or decaying means. 
This restriction is justified if any non-Gaussian state also becomes Gaussian asymptotically, which means that all higher-order cumulants decay over time. Proving this result rigorously is beyond the scope of this article but we make some qualitative observations, which lead us to conjecture that this is indeed the case.

First, note that assuming stable dynamics implies that the second cumulants of a Gaussian state asymptotically satisfy
\begin{align}
	0 & = \mat{M}_{\rho}^{\asympt}\cmlnt_{\asympt}^{(2)} + \mat{\cmlnt}_{\asympt}^{(2)}(\mat{M}_{\rho}^{\asympt})^\transpose + \mat{\sigma}\mat{\tilde{\Delta}}\mat{\sigma}^\transpose + 4\mat{\cmlnt}_{\asympt}^{(2)}\mat{A}^\transpose\mat{A}\mat{\cmlnt}_{\asympt}^{(2)},
\end{align}
with $\mat{M}_{\asympt} := \mat{Q}+2\mat{\sigma}\mat{B}^\transpose\mat{A} - 4\mat{\cmlnt}_{\asympt}^{(2)}\mat{A}^\transpose\mat{A}$ having only eigenvalues with negative real part. The collapse to this particular $\mat{\cmlnt}_{\asympt}^{(2)}$ is a consequence of the deterministic part of the evolution (\ie, the bracket) in Eq.~\eqref{eq:ConditionalSecondOrderEoMs}. If we consider some non-Gaussian state with $\cmlnt_{\rho}^{(3)}\neq 0$, then $\cmlnt_{\rho}^{(2)}(t)$ itself becomes a random process, which in addition to the deterministic drift will experience some diffusion induced by the additive white noise dependent on $\cmlnt_{\rho}^{(3)}$. Looking at Eq.~\eqref{eq:HigherOrderCumulantsEoMs} we see that these third-order cumulants (as well as all higher orders) themselves also experience deterministic decay induced by all lower orders through matrices $\mat{M}_{\rho}(t)$ and $\mat{A}^\transpose\mat{A}$, and white-noise diffusion through the next higher order only.
This leads us to conjecture that in stable systems asymptotically all higher-order cumulants vanish on average, but experience stochastic fluctuations about zero (diffusion) induced by the Wiener increment, which leads to some residual variance. Thus also any initial state of the system will asymptotically collapse to a Gaussian on average.

\subsection{Backward addition}\label{sec:Appendix:BackwardAddition}
To obtain the evolution equations of the cumulants associated with general effect operators we can compare the unnormalized master equation for $\oprho$,
\begin{align}
	\begin{split}
		\ito\  \diff \oprho (t) & = -i[\op{H},\oprho(t)]\diff t + \sum_{j=1}^{N_L}\Dop{\op{L}_{j}}\oprho(t)\diff t + \sum_{k=1}^{N_C} \mleft(\op{C}_{k}\oprho(t) + \oprho(t)\op{C}_{k}^\dag \mright) \diff Y_k(t),
	\end{split}
\end{align}
to the corresponding adjoint equation for $\opE$,
\begin{align}
	\begin{split}
		\bito\ & {-\diff} \opE(t) = i[\op{H},\opE(t)]\diff t + \sum_{j=1}^{N_L}\Dopdag{\op{L}_{j}}\opE(t)\diff t + \sum_{k=1}^{N_C} \mleft(\op{C}_{k}^\dag\opE(t) + \opE(t)\op{C}_{k} \mright) \diff Y_k(t)
	\end{split}\\
	\begin{split}
		& \quad = -i[-\op{H},\opE(t)]\diff t + \sum_{j=1}^{N_{L}}\mcl{D}[\op{L}_{j}]\opE(t)\diff t + \sum_{k=1}^{N_C} \mleft(\op{C}_{k}^\dag\opE(t) + \opE(t)\op{C}_{k} \mright) \diff Y_k(t) + \sum_{j=1}^{N_L}\mleft( \op{L}_{j}^\dag \opE(t)\op{L}_{j} - \op{L}_{j} \opE(t)\op{L}_{j}^\dag \mright)\diff t,
	\end{split}
\end{align}
where the last sum compensates for replacing $\mcl{D}^\dag\mapsto \mcl{D}$. We can immediately read off the following changes. The sign flip of $\op{H}$ causes $\mat{H}\to -\mat{H}$ and replacing the measurement operators $\op{C}_{k}$ by their adjoint entails $\mat{B} \to -\mat{B}$. We only need to work out the change  $[\diff\chi_{E}]_{\text{bwd}}$ stemming from the sandwich terms in the last sum. Using the relations \eqref{eq:ReplacementRules}, \eqref{eq:ProductRuleOmega} and \eqref{eq:SkewSymmetricScalarProduct} we obtain
\begin{align}
	{-[\diff\opE]_{\text{bwd}}} & = 2i\vect{\op{r}}^\transpose\mat{\Omega}\opE \vect{\op{r}}\diff t
\end{align}
which turns into
\begin{align}
	{-[\diff\chi_{E}]_{\text{bwd}}} & = 2i\mleft(-i\vect{\tilde{\nabla}}^\transpose - \frac{1}{2}\vect{\xi}\mright)^\transpose \mat{\Omega} \mleft(-i\vect{\tilde{\nabla}}^\transpose + \frac{1}{2}\vect{\xi}\mright)\chi_{E}
	= -\trace[\mat{\sigma}\mat{\Omega}]\chi_{E}\diff t-2\vect{\xi}^\transpose\mat{\Omega}(\vect{\tilde{\nabla}}\chi_{E})\diff t.
\end{align}
So the backward cumulants with $N\geq 1$ will have the additional term
\begin{align}
\begin{split}
	{-[\diff \cmlnt_{m_1,\dots,m_{N}}^{(N)}]_{\text{bwd}}} & = -2\sum_{\tau\in \mcl{S}^{\text{cycl}}(N)}\mat{\Omega}_{m_{\tau(1)},k} \cmlnt_{k,m_{\tau(2)},\dots,m_{\tau(N)}}^{(N)}\diff t.
\end{split}
\end{align}
Implementing all these changes we find that the difference between forward and backward equation amounts to replacing the drift matrix $\mat{M}_{\rho}(t)$ with the backward drift matrix from Eq.~\eqref{eq:BwdConditionalDriftMatrix},
\begin{align}
	\mat{M}_{E}(t) & := -\mat{Q}-2\mat{\sigma}\mat{B}^\transpose\mat{A} - 4\mat{\cmlnt}_{E}(t)\mat{A}^\transpose\mat{A},
\end{align}
and changing the sign of the constant term in the equations of the means. Spelling this out we find that the means satisfy
\begin{align}
	\bito\  {-\diff}\vect{r}_{E} & = -\mat{Q}\vect{r}_{E}\diff t - \mat{\sigma}\vect{h}\diff t +(2\mat{\cmlnt}_{E}^{(2)}\mat{A}^\transpose + \mat{\sigma}\mat{B}^\transpose)\diff \vect{W}\\
	& = \mat{M}_{E}\vect{r}_{E}\diff t - \mat{\sigma}\vect{h}\diff t +(2\mat{\cmlnt}_{E}^{(2)}\mat{A}^\transpose + \mat{\sigma}\mat{B}^\transpose)\diff \vect{Y}.
\end{align}
For the covariance matrix we find
\begin{align}
	\begin{split}
		\bito\  {-\diff}\mat{\cmlnt}_{m_{1},m_{2}}^{(2)} & = \Bigl[ \mat{M}_{E}\mat{\cmlnt}_{E}^{(2)} + \mat{\cmlnt}_{E}^{(2)}\mat{M}_{E}^\transpose + \mat{\sigma}\mat{\tilde{\Delta}}\mat{\sigma}^\transpose + 4\mat{\cmlnt}_{E}^{(2)}\mat{A}^\transpose\mat{A}\mat{\cmlnt}_{E}^{(2)}\Bigr] \diff t + 2\cmlnt_{m_{1},m_{2},k}^{(3)}(\mat{A}^\transpose\diff\vect{W})_{k},
	\end{split}
\end{align}
Higher order ($N\geq 3$) cumulants evolve as
\begin{align}
	\begin{split}
		\bito\ & {-\diff} \cmlnt_{m_1,\dots,m_{N}}^{(N)} \\
		& = -\sum_{\tau\in \mcl{S}^{\text{cycl}}(N)}(\mat{Q} + 2\mat{\sigma}\mat{B}^\transpose\mat{A})_{m_{\tau(1)},k} \cmlnt_{k,m_{\tau(2)},\dots,m_{\tau(N)}}^{(N)}\diff t\\
		& \quad - 2\sum_{n=1}^{N-1}\sum_{\sigma\in S(N)} \cmlnt_{m_{\sigma(1)},\dots,m_{\sigma(n)},j}^{(n+1)} (\mat{A}^\transpose\mat{A})_{jk} \cmlnt_{k,m_{\sigma(n+1)},\dots,m_{\sigma(N)}}^{(N-n+1)} \diff t + 2\cmlnt_{m_1,\dots,m_N,k}^{(N+1)}(\mat{A}^\transpose\diff\vect{W})_{k}.
	\end{split}\\
	\begin{split}
		& = \sum_{\tau\in \mcl{S}^{\text{cycl}}(N)}M_{m_{\tau(1)},k}^{E} \cmlnt_{k,m_{\tau(2)},\dots,m_{\tau(N)}}^{(N)}\diff t\\
		& \quad - 2\sum_{n=2}^{N-2}\sum_{\sigma\in S(N)} \cmlnt_{m_{\sigma(1)},\dots,m_{\sigma(n)},j}^{(n+1)} (\mat{A}^\transpose\mat{A})_{jk} \cmlnt_{k,m_{\sigma(n+1)},\dots,m_{\sigma(N)}}^{(N-n+1)} \diff t + 2\cmlnt_{m_1,\dots,m_N,k}^{(N+1)}(\mat{A}^\transpose\diff\vect{W})_{k}.
	\end{split}
\end{align}
With regards to stability we can apply the same reasoning to the backward dynamics as to the forward dynamics, which, \textit{mutatis mutandis}, leads to the same conjecture as in \ref{sec:Appendix:StabilityAnalysis}. Backward dynamics are stable whenever any Gaussian effect operator with arbitrary final covariance matrix $\mat{V}_{E}(t_{1})$ approaches an asymptotic covariance matrix $\mat{V}_{E}^{\asympt}$ as $t\to -\infty$ such that $\mat{M}_{E}^{\asympt}$ has only eigenvalues with negative real parts. In that case we conjecture that also any non-Gaussian effect operator $\opE(t_{1})$ will collapse to a Gaussian one with $\mat{V}_{E}^{\asympt}$ as $t\to -\infty$, up to stochastic fluctuations of higher-order cumulants about zero.
\end{widetext}


%

\end{document}